\documentclass[oupdraft,pdftex]{bio}
\usepackage{url}
\usepackage{comment}
\usepackage{bm,mathtools}
\usepackage{algorithm}
\usepackage{subcaption}
\usepackage[noend]{algpseudocode}
\usepackage{hyperref}

\begin{document}

\title{Efficient Bayesian inference for two-stage models in environmental epidemiology}

\author{Konstantin Larin$^{1,\ast}$, Daniel R. Kowal$^2$ \\[4pt]
\text{$^1$Department of Statistics, 
Rice University,
6100 Main St, Houston, TX 77005, United States} \\
\text{
$^2$Department of Statistics and Data Science, Cornell University,
127 Hoy Rd, Ithaca, NY 14853, United States}
\\[2pt]
{konstantin.larin@rice.edu}}

\markboth%
{K. Larin and D.R. Kowal}
{Efficient Bayesian inference for two-stage models}

\maketitle

\footnotetext{To whom correspondence should be addressed.}

\begin{abstract}
{Statistical models often require inputs that are not completely known. This can occur when  inputs are measured with error, indirectly, or when they are predicted using another model. In environmental epidemiology, air pollution exposure is a key determinant of health, yet typically must be estimated for each observational unit by a complex model. Bayesian two-stage models  combine this stage-one model with a stage-two model for the health outcome given the exposure. However, analysts usually only have access to the stage-one model output without all of its specifications or input data, making joint Bayesian inference apparently intractable. We show that two prominent workarounds---using a point estimate or using the posterior from the stage-one model without feedback from the stage-two model---lead to miscalibrated inference. Instead, we propose efficient algorithms to facilitate joint Bayesian inference and provide more accurate estimates and well-calibrated uncertainties. Comparing different approaches, we investigate the association between PM$_{2.5}$ exposure and county-level mortality rates in the South-Central USA.}{air pollution; environmental health; importance sampling; multi-stage models; uncertainty propagation.}
\end{abstract}

\section{Introduction}
\label{sec:Intro}


When analyzing modern data, it is often the case that some variables are either not completely known or not observed directly. Our primary focus is on air pollution exposure and its adverse effects on human health   (e.g., \citealp{schwartz_is_1996,jin_comparison_2019,wu_evaluating_2020,dong_impacts_2025}). However, air pollution measurements are typically only available from select monitoring stations in fixed locations, so relying on those direct measurements as inputs can lead to  spatial misalignment.  Thus, it is typically necessary to estimate air pollution exposure across many spatial locations using advanced statistical models (\citealp{chang_estimating_2011}). 
Critically, these estimates are imperfect, with errors and uncertainties that vary in space and time. Failure to account for this can lead to biased effect and uncertainty estimates (\citealp{gryparis_measurement_2009}). Similar challenges occur in other contexts: modeling the effects of water contaminant exposure on pregnancy (\citealp{bove_drinking_2002}), using ocean circulation models to estimate the effects of increasing seawater temperature on marine life (\citealp{santana_increasing_2025}), and classical  measurement error scenarios (\citealp{carroll_measurement_2006}), among many others.

\begin{figure}[h]
    \centering
    \includegraphics[width=0.8\textwidth]{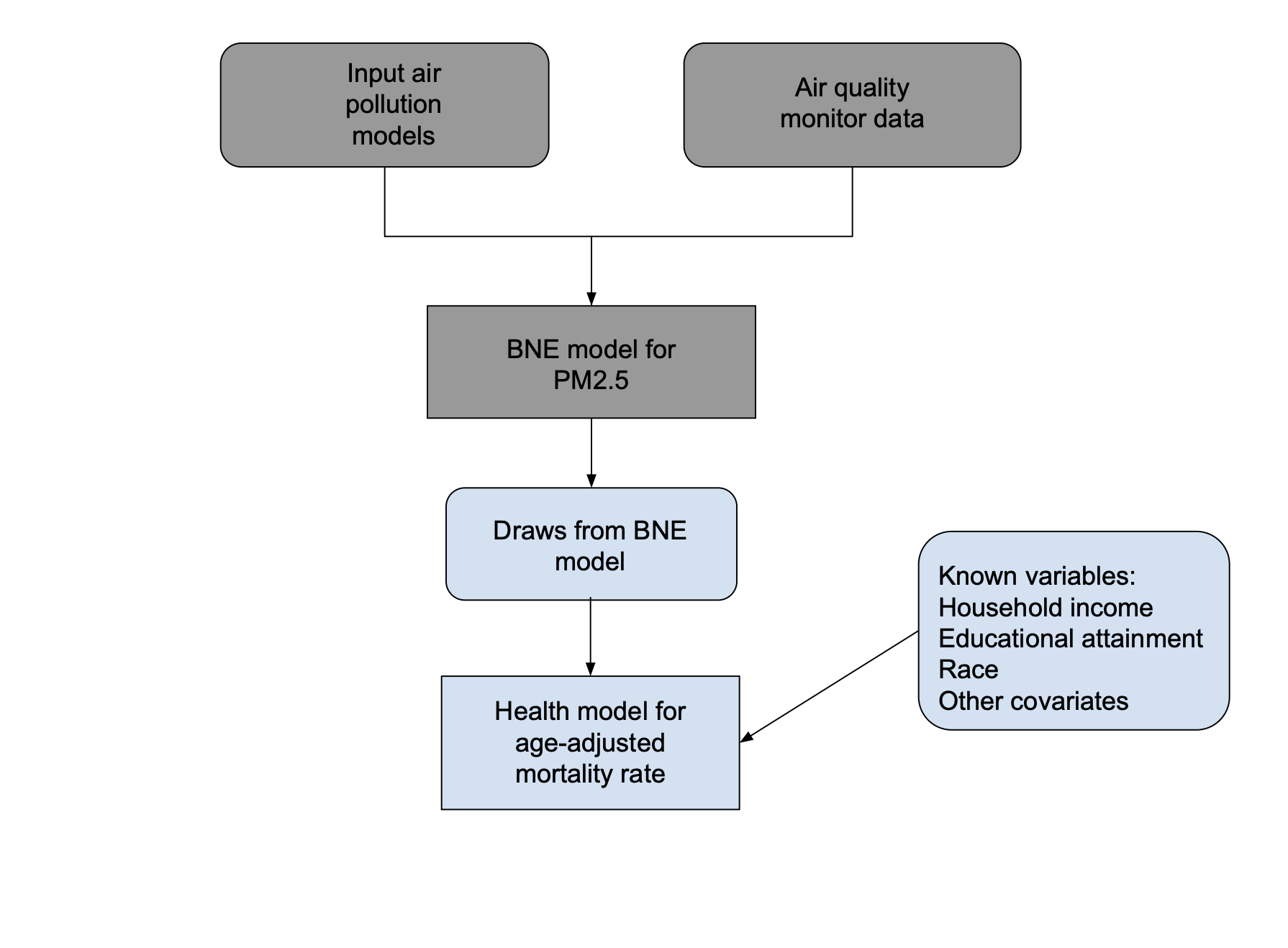}
    \caption{Two-stage modeling framework with potentially unknown quantities (gray) and known inputs (blue). In the stage-one  Bayesian nonparametric ensemble (BNE) model, PM$_{2.5}$ is modeled and estimated across space and time using various input models and data sources.  These data  and modeling details may not be fully available for downstream users. The stage-two model uses the output from the stage-one model along with known covariates to model a health outcome.} 
    \label{fig:model_diag}
\end{figure}

The central statistical challenge is how to integrate these ``stage-one" model estimates and uncertainties---i.e., the air pollution exposures---into a subsequent ``stage-two" regression model that relates  health outcomes to these exposures and other (known) covariates. 
We illustrate this setting in Figure~\ref{fig:model_diag}. Models for air pollution exposure are well-designed, highly complex, computationally intensive to fit  (across massive spatial domains), and leverage various data sources that may not be publicly available or easy to access and store. Thus, it is typically the case that these models are fit in isolation without regard for any particular downstream regression or inference task.  A prominent example is the Bayesian spatiotemporal downscaler (DS) model (\citealp{berrocal_space-time_2012}), whose output is published by the USA Environmental Protection Agency  (\citealp{epa_faqsd_data}). This output is used by academic, government, and industry communities for a variety of tasks that require air pollution estimates. Recently, \cite{benavides_bner_2025} proposed a Bayesian nonparametric ensemble (BNE) model that aggregates multiple air pollution models and incorporates other monitoring data to generate posterior distributions of air pollution exposures across space and time. We adopt the BNE model in our data analysis (Section~\ref{sec:5_empmain}).

Two-stage estimation procedures are the overwhelming default: the output from a stage-one air pollution model is fed into a stage-two regression model with a health outcome. Crucially, it is typically impossible to fit these models jointly using standard approaches. A joint Bayesian model would seemingly require the full model parameterization and data inputs for the stage-one model: this information would be combined with the stage-two model and data inputs in order to develop and implement an algorithm that targets the joint posterior distribution   (e.g., MCMC or variational inference). However, this  would require re-fitting the stage-one (air pollution) model for every downstream analysis. That is not only computationally infeasible, but also  impossible without full knowledge of the stage-one model specifications and input datasets. Thus, in our setting, we only allow for \emph{access} to the stage-one model output---specifically, in the form of posterior  draws---but do not assume any knowledge of the stage-one model specifications or input datasets.

Most commonly, two-stage estimation procedures disregard the uncertainty in the stage-one model estimates and instead treat the air pollution estimates as a fixed and known covariate  (e.g., \citealp{lim_long-term_2019,schulz_independent_2020,huang_effects_2021,kowal_bayesian_2021,ahmed_explainable_2021,warren_critical_2022}). The primary advantages of this ``plug-in" approach are its simplicity and its computational efficiency: it  reduces the two-stage joint estimation problem to require only estimation of the stage-two model. Yet critically, ignoring the uncertainty from the stage-one estimates creates problems for the stage-two model: the parameter estimates can be biased and the inferences miscalibrated  (see Sections~\ref{sec:Insuf_Back2} and \ref{sec:4_Sim}). 

In an attempt to incorporate the uncertainties from the stage-one model, a ``partial posterior" approach propagates the stage-one model posterior distribution into the stage-two model. In effect, this occurs by feeding the posterior draws from the stage-one model into the stage-two model   (\citealp{blangiardo_two-stage_2016}). However, despite the intuitiveness of this approach, it fails to target the specified posterior distribution under the joint two-stage model. Crucially, we show that this partial posterior approach creates further inferential miscalibration issues and does not correct the bias from the plug-in approach (see Sections~\ref{sec:Insuf_Back2} and \ref{sec:4_Sim}).

There are several approaches that more directly target the joint posterior of the two-stage model. The general strategy is to construct an approximation of the stage-one posterior and then take this approximation as a prior for the air pollution covariate in the stage-two model. A multivariate normal (MVN) distribution is a common choice \citep{warren_spatial-temporal_2012,lee_rigorous_2016}, but requires estimation of the  covariance matrix and thus faces computational and statistical challenges as the number of observations grows large. For more scalable analyses, \citet{huang_multivariate_2018,cameletti_bayesian_2019} instead use a product of normal distributions, which removes the need to estimate a covariance matrix but ignores the (e.g., spatial) correlations among the stage-one outputs. Alternatively, \citet{lee_scalable_2024} retain the MVN approximation but use a sparse estimate of the inverse-covariance based on a Vecchia approximation. In each case, the MVN distribution is central; yet in actuality, the stage-one posteriors may be skewed, heavy-tailed, multi-modal, or exhibit nonlinear correlation patterns. One approach that does not rely on the MVN approximation is \citet{comess_bayesian_2022}, which instead uses kernel density estimation (KDE) to approximate the stage-one posterior. However, this approach is not conducive to multivariate cases, so the authors recommend using a product of KDEs and thus impose a strict independence approximation for the stage-one posterior. 

In this paper, we prioritize joint posterior inference that  propagates uncertainty and feedback across the stage-one and stage-two models. First, we document the inadequacies of aforementioned plug-in and partial posterior approaches using a tractable two-stage model. Then we propose an algorithm that effectively targets the joint two-stage model posterior (predictive) distribution. Building upon importance sampling---which is not effective in our context due to weight degeneracy---our approach has minimal requirements for the stage-one model: it \emph{only} requires posterior draws and no other information or data. Our approach is then compatible with virtually any stage-two regression model and can leverage accompanying MCMC algorithms for the case of known covariates. More specifically, we design two modifications to importance sampling: a streamlined version that is accurate when the stage-one model outputs are approximately independent and a corrected version for when that does not occur (Section~\ref{sec:IS_main}). 
We validate our approach through simulation studies that include cases for independent and dependent stage-one models and show how competing methods deteriorate for estimation, inference, or prediction in different settings (Section~\ref{sec:4_Sim}). We then apply our methods to evaluate the associations between PM$_{2.5}$ exposure and county-level mortality rates across the South-Central United States (Arkansas, Louisiana, Oklahoma, and Texas) using air pollution estimates from the BNE model (Section~\ref{sec:5_empmain}). Supplementary materials are available online.

\section{Background}
\label{sec:Back}

\subsection{Two-stage Bayesian models}
\label{sec:Back_intro}
Our goal is to associate health outcomes with model-based estimates of air pollution exposures and other (known) covariates (Figure~\ref{fig:model_diag}). The two-stage model setting is broader than this, but we continue with our environmental epidemiology setting for clarity. For a given unit of observation (e.g., county) $i=1,\ldots,n$, let $y_i$ be the health outcome, $\zeta_i$ the  (unknown) air pollution exposure, and $\bm x_i$ the (known) covariates. The central challenge is that $\zeta_i$ is not observed directly, but rather is estimated using a stage-one model. Concisely, we write $p(\bm \zeta \mid  \bm z)$ for the stage-one posterior  distribution, where $\bm \zeta = (\zeta_1,\ldots,\zeta_n)^\top$ contains all $n$ air pollution exposures and $\bm z$ refers to the auxiliary data used to train the stage-one model. For instance, the DS model   uses monitoring station data and output from the Community Multiscale Air Quality model (\citealp{epa_cmaq}) for $\bm z$, while the BNE model  uses both monitoring station data and an ensemble of air quality models.

This stage-one model $p(\bm \zeta \mid  \bm z)$ is fit completely distinct from any downstream analysis. In fact, we---the downstream analysts---may not have access to the data $\bm z$, or perhaps even the model specifications and algorithms used for $p(\bm \zeta \mid  \bm z)$. We also do not assume any special distribution (e.g., MVN) for $p(\bm \zeta \mid  \bm z)$. Instead, we only assume access to posterior  draws from the stage-one model:
\begin{align}
\{\bm \zeta^s\}^S_{s=1} \sim p(\bm \zeta\mid \bm z). \label{eq:ppost}
\end{align}
These draws can be computed, stored, and shared across many potential downstream analyses, e.g., with a variety of health outcomes or regression models. In other two-stage modeling contexts, $p(\bm \zeta\mid \bm z)$ might be a Bayesian machine learning or a deep learning model. 

Then we consider stage-two regression models of the form 
\begin{align}
y_i = f_{\bm\theta}(\bm{x}_i,\zeta_i) + \epsilon_i, \quad \epsilon_i \stackrel{iid}{\sim}N(0,\sigma_\epsilon^2), \quad i=1,\ldots,n\label{eq:stage_two}
\end{align}
so that $f_{\bm \theta}$ is a regression function with known inputs $\bm x_i$ and unknown inputs $\zeta_i$ and parameterized by $\bm \theta$. The goal is to infer $\bm \theta$, which  is sufficient for learning $f_{\bm \theta}$ (e.g., as in linear regression). In general, we assume prior independence between $\bm \theta$ and the stage-one outputs $\bm \zeta$. Modifications for non-Gaussian models or non-iid errors in \eqref{eq:stage_two} are  available. 

Given these two modeling stages, we now describe our target distribution for inference. Bayesian analysis requires a posterior distribution for $\bm \theta$ that jointly considers both models and conditions on both datasets: the auxiliary data $\bm z$ for the stage-one model and the  outcome data $\bm y = (y_1,\ldots,y_n)^\top$ for the stage-two model. Omitting dependence on $\bm x$ for notational simplicity, the two-stage model posterior for $\bm \theta$ can be written by marginalizing over the unknown $\bm \zeta$,
\begin{align}
    p(\bm \theta\mid  \bm y, \bm z) &= \int p(\bm \theta ,\bm \zeta\mid  \bm y,  \bm z) d \bm \zeta = \int p(\bm \theta\mid \bm y,\bm \zeta) \ p(\bm \zeta\mid \bm y,\bm z)\ d\bm\zeta. \label{eq:ideal}
  \end{align}
The first term, $p(\bm \theta\mid \bm y,\bm \zeta)$, is precisely the posterior distribution for $\bm \theta$ under the stage-two model \eqref{eq:stage_two} if $\bm \zeta$ were fixed and known. This a standard Bayesian regression problem, for which many estimation algorithms are available for a variety of choices for \eqref{eq:stage_two} (as well as non-Gaussian and non-iid generalizations). Throughout, we will  assume that we can sample from this conditional posterior distribution, for instance using MCMC. 

The second term, $p(\bm \zeta\mid \bm y,\bm z)$, is   \emph{not} simply the partial posterior in \eqref{eq:ppost}, but rather incorporates data and feedback from the stage-two model \eqref{eq:stage_two}. This ``full data" posterior is 
\begin{equation}\label{eq:fpost}
    p(\bm \zeta \mid \bm y, \bm z)   \propto  p(\bm y \mid \bm \zeta) \ p(\bm \zeta \mid \bm z)
\end{equation}
assuming that conditional on $\bm \zeta$, $\bm y$ is independent of $\bm z$. 
While the partial posterior in \eqref{eq:ppost} may be viewed as a shared term, the \emph{full} posterior in \eqref{eq:fpost} will be different for every downstream regression model \eqref{eq:stage_two} and dataset $\bm y$. 

To further elaborate on the challenges, Algorithm~\ref{alg:ideal} outlines an idealized two-step sampling algorithm based on \eqref{eq:ideal}. 
Given the ability to sample from the stage-two conditional posterior $p(\bm \theta\mid \bm y,\bm \zeta)$, we would simply draw from the full posterior $[\bm \zeta \mid \bm y, \bm z]$, then substitute these draws to sample from the conditional posterior $[\bm \theta \mid \bm y, \bm \zeta]$. However, Algorithm~\ref{alg:ideal} is simply infeasible for our problem: it is not obvious how to access or sample from $[\bm \zeta \mid \bm y, \bm z]$. Modifying Algorithm~\ref{alg:ideal} to be a Gibbs sampler---i.e., replacing the first step with the full conditional $[\bm \zeta \mid \bm y, \bm z, \bm \theta]$---does not resolve this issue. The limitation is that we 
\emph{only} have output (i.e., draws) from the partial posterior \eqref{eq:ppost}. Thus, we either cannot analytically solve \eqref{eq:fpost} to due a lack of stage-one model details, or we cannot compute \eqref{eq:fpost} because we lack access to the auxiliary data $\bm z$ or the computational capabilities to re-fit a model for $\bm \zeta$ that now accounts for both $\bm z$ and $\bm y$. 

\begin{algorithm}[ht]
\renewcommand{\baselinestretch}{1.2}\selectfont
\caption{Idealized Sampling Algorithm for $p(\bm \theta \mid \bm y, \bm z)$}\label{alg:ideal}
\begin{algorithmic}
\State \textbf{Input:} Data $\bm y$, $\bm z$
\For{$t = 1,..., T$}
\begin{enumerate}
\item Sample $\bm{\zeta}^t \sim p(\bm \zeta\mid \bm y,\bm z)$ 
\item Sample $\bm \theta^t \sim p(\bm \theta\mid \bm y,   \bm { \zeta}^t)$
\end{enumerate}
\EndFor
\State \textbf{Output:}  draws $\{\bm \theta^t\}_{t=1}^T$.
\end{algorithmic}
\end{algorithm}



These issue also arise for prediction of future or unobserved outcome variables $\tilde y$. Consistent with \eqref{eq:ppost}, suppose that we have access to draws from a stage-one partial posterior predictive distribution, now denoted $\bm{\tilde \zeta}$, which may also condition on new auxiliary data $\bm{\tilde z}$ in addition to $\bm z$, and we wish to predict $\tilde y$ at known covariates $\bm{\tilde x}$. The posterior predictive distribution can be expressed by taking the full joint posterior distribution over $(\tilde y, \bm{\tilde \zeta}, \bm \theta, \bm \zeta)$ and marginalizing over all but $\tilde y$. Generally, this full distribution is $p(\tilde y, \tilde {\bm \zeta}, \bm \theta,\bm \zeta \mid  \bm y,\bm z, \tilde {\bm z}) =  p(\tilde y \mid  \tilde {\bm \zeta}, \bm \theta,\bm \zeta,   \bm y,\bm z, \tilde {\bm z}) \ p(\tilde {\bm \zeta} \mid  \bm \theta,\bm \zeta,   \bm y,\bm z, \tilde {\bm z}) \ p(\bm \theta,\bm \zeta \mid  \bm y,\bm z, \tilde {\bm z})$. The first term simplifies to $p(\tilde y\mid \bm{\tilde \zeta}, \bm \theta)$ due to conditional independence assumed in the stage-two model. The second and third terms will simplify depending on the form of the stage-one model. Here, we will assume that the second term simplifies to  $p(\tilde {\bm \zeta} \mid  \bm z, \tilde {\bm z})$. This occurs when the new $\bm{\tilde \zeta}$ is 1) not linked to any observed $\bm y$ and 2) conditionally independent of $\bm{\zeta}$ given $(\bm{\tilde z}, \bm z)$. Under the same assumptions, the third term does not depend on  $\bm{\tilde z}$. 

Using these simplifications,  the posterior predictive distribution is
\begin{equation}\label{eq:postpred}
    p(\tilde y \mid \bm y, \bm z, \bm{\tilde z}) = \int \int p(\tilde y \mid \bm{\tilde \zeta}, \bm \theta) \  p(\tilde {\bm \zeta}\mid \bm z,   \tilde {\bm z} ) \ p(\bm \theta \mid \bm y, \bm z) \ d \bm{\tilde \zeta} \ d\bm \theta  
\end{equation}
where we have also marginalized over $\bm \zeta$, which appeared only in the third term. Now, if we can sample from the full data posterior distribution $p(\bm \theta \mid \bm y, \bm z)$---the primary statistical goal of this paper---and if we are given draws from the stage-one partial posterior predictive $p(\tilde {\bm \zeta}\mid \bm z,   \tilde {\bm z} )$, then our sampling algorithm for the posterior predictive distribution \eqref{eq:postpred} is completed by sampling from the  stage-two likelihood  \eqref{eq:stage_two} given the sampled parameters $\bm \theta$, the sampled inputs $\bm{\tilde \zeta}$, and the known covariates $\bm{\tilde x}$. This final step is the simplest one. Thus, the key to both posterior \emph{and} predictive inference is the full data posterior distribution $p(\bm \theta \mid \bm y, \bm z)$.

\subsection{The ineffectiveness of plug-in and partial posterior methods}
\label{sec:Insuf_Back2}

We now describe two common yet flawed approaches for estimation and inference with two-stage models. 
We then use a tractable two-stage model to illustrate their statistical flaws. These approaches are  revisited in our simulated (Section~\ref{sec:4_Sim}) and real data (Section~\ref{sec:5_empmain}) examples.


\subsubsection{Plug-in}
The plug-in approach (Algorithm~\ref{alg:plug}) simply fixes $\bm \zeta$ at a point estimate, say $\hat {\bm \zeta}$, and treats this quantity as fixed and known within the stage-two regression model \eqref{eq:stage_two}. A natural estimator given the partial posterior draws \eqref{eq:ppost} is the partial posterior mean, $\hat {\bm \zeta} = \mathbb{E}(\bm \zeta \mid \bm z) \approx S^{-1} \sum_{s=1}^S \bm \zeta^{s}$. 
Notably, this estimate is computed exclusively from the stage-one model output with no regard for the stage-two model. Perhaps more glaringly, it also does not account for any uncertainty about the air pollution exposure estimates---even though these exposures are unknown quantities in the model. We detail the implications for estimation and inference for $\bm \theta$ and prediction of $\tilde y$ both analytically (Section~\ref{sec-anexamp}) and empirically (Section~\ref{sec:4_Sim}). 
\begin{algorithm}
\renewcommand{\baselinestretch}{1.2}\selectfont
\caption{Plug-in Approach}\label{alg:plug}
\begin{algorithmic}
\State \textbf{Input:} Fixed point estimate $\hat{\bm \zeta}$, data $\bm y$
\For{$t = 1,..., T$} 

Sample $\bm \theta^t \sim p(\bm \theta\mid \bm y, \hat {\bm \zeta})$

\EndFor
\State \textbf{Output:}  draws $\{\bm \theta^t\}_{t=1}^T$.
\end{algorithmic}
\end{algorithm}


\subsubsection{Partial Posterior}
To rectify the lack of uncertainty quantification in the plug-in approach, the partial posterior approach (Algorithm~\ref{alg:ppost}) adopts the intuitive---yet ultimately misguided---strategy of feeding the draws from the stage-one partial posterior \eqref{eq:ppost} into the stage-two model \eqref{eq:stage_two}. 
More generally, this approach is closely related to cut posteriors (\citealp{plummer_cuts_2015}), which are primarily used to guard against misspecification in multi-stage models.  The uniform draw on the indices $\{1,\ldots,S\}$ is simply to accommodate cases where the number of partial posterior draws $S$ differs from the desired number of draws $T$ for $\bm \theta$.

\begin{algorithm}
\renewcommand{\baselinestretch}{1.2}\selectfont
\caption{Partial Posterior Approach}\label{alg:ppost}
\begin{algorithmic}
\State \textbf{Input:} Draws $\{\bm\zeta^s \}^{S}_{s=1}\sim p(\bm \zeta\mid \bm z)$, data $\bm y$
\For{$t = 1,..., T$}
\begin{enumerate}
\item Sample $j \sim \text{Uniform}(\{1,...,S\})$
\item Sample $\bm\theta^t \sim p(\bm \theta\mid \bm y,   \bm { \zeta}^j)$
\end{enumerate}
\EndFor
\State \textbf{Output:}  draws $\{\bm \theta^t\}_{t=1}^T$.
\end{algorithmic}
\end{algorithm}

At first glance, this approach is appealing: it leverages the partial posterior draws of \eqref{eq:ppost} to propagate uncertainty from the stage-one exposure estimates to the stage-two model. It has minimal requirements on the stage-one model and does not require any re-fitting to account for the stage-two regression model. However, it does not actually target the two-stage model posterior \eqref{eq:ideal} or the accompanying posterior predictive \eqref{eq:postpred}. Compare Algorithm~\ref{alg:ppost} with the idealized Algorithm~\ref{alg:ideal}: clearly, draws from the partial posterior  $p(\bm \zeta \mid \bm z)$ are not the same as draws from the full data posterior $p(\bm\zeta \mid \bm y, \bm z)$. While it may seem that the partial posterior may be a suitable approximation, we will show analytically (Section~\ref{sec-anexamp}) and empirically (Section~\ref{sec:4_Sim})  that this is not the case. More specifically, we show that Algorithm~\ref{alg:ppost} indeed provides additional uncertainty about $\bm \theta$, but it is substantially miscalibrated and often biased.

\subsubsection{Analytical example}\label{sec-anexamp}

To explore the statistical flaws in the plug-in and partial posterior approaches, we analyze a tractable  two-stage model.  The stage-one model is a Gaussian measurement error model,
\begin{equation}
      \bm z = \bm \zeta + \bm u, \quad \bm u \sim  MVN(\bm 0, \bm \Sigma_u)\label{eq:slr_one}
\end{equation}      
where each vector is $n$-dimensional, $\bm \Sigma_u$ is the $n \times n$ error covariance matrix, and $\bm \zeta \sim MVN(\bm 0, \bm \Sigma_\zeta)$ is the prior. 
The stage-two model is a Gaussian linear model,
\begin{equation}\label{eq:slr_two}
    y_i = \beta_0 + \theta_\zeta\zeta_i + \epsilon_i, \quad \epsilon_i \stackrel{iid}{\sim}N(0,\sigma_\epsilon^2), \quad i=1,\ldots,n
\end{equation}
For simplicity, we omit known covariates $\bm x$ and assume that $\bm \Sigma_u = \sigma_u^2 \bm I_n$ and $\bm \Sigma_\zeta = \sigma_\zeta^2 \bm I_n$, but allow for more general covariances in the simulation study (Section~\ref{sec:4_Sim}). 

Although \eqref{eq:slr_one} is clearly much simpler than the DS and BNE models used for air pollution estimation, the two-stage model \eqref{eq:slr_one}--\eqref{eq:slr_two} retains the main ingredients of our setting: 1) an unobservable quantity $\bm \zeta$ is a required input for our regression model with outcome $\bm y$ and 2) it is inferred based on a separate model with observable data $\bm z$. Unlike our more general context, here we have full knowledge of the stage-one model specification, rather than just draws \eqref{eq:ppost}, and the joint model is tractable. In particular, the partial posterior distribution is $[\bm \zeta \mid \bm z] \sim MVN(\lambda \bm z, \bm \Sigma_{\zeta \mid z})$, where $\lambda = \sigma_\zeta^2/(\sigma_\zeta^2 + \sigma_u^2)$ and $\bm \Sigma_{\zeta \mid z} = \lambda \sigma_u^2 \bm I_n$. The results presented below are supported by full derivations in Section~A of the supplementary materials. 

We consider two versions of the plug-in approach along with the partial posterior approach. First, Plug-in (Z) takes $\bm{\hat \zeta} = \bm z$ to use the corrupted observations in place of  $\bm \zeta$. This case is well-studied in the literature on measurement error models (e.g., \citealp{carroll_measurement_2006}). In the context of air pollution exposures, we may view Plug-in (Z) as a proxy for using the air pollution measurements from the nearest monitoring station. Second, Plug-in ($\hat\zeta$) takes $\hat{\bm\zeta}=\mathbb{E}(\bm\zeta\mid \bm z) = \lambda \bm z$, akin to using  model-based air pollution exposure estimates (e.g., from DS or BNE). Finally, the partial posterior substitutes $[\bm \zeta \mid \bm z]$, which is known analytically for model \eqref{eq:slr_one}--\eqref{eq:slr_two} but more generally would use sample averages of the partial posterior draws \eqref{eq:ppost} (e.g., from DS or BNE). 

In each case, a different ``covariate" is used in place of $\zeta_i$ in the stage-two regression model. First, we analyze the implied regression coefficients in each case. The regression coefficient is defined as the ratio of the covariance between $y$ and that covariate to the variance of that covariate. For Plug-in (Z), Plug-in ($\hat\zeta$), and partial posterior,  the estimands are
\begin{equation}\label{bias}
\theta_Z \coloneq \frac{\mbox{Cov}(y, z)}{\mbox{Var}(z)} = \lambda \theta_\zeta, \quad 
\theta_{\hat\zeta} \coloneq \frac{\mbox{Cov}(y, \lambda z)}{\mbox{Var}(\lambda z)} =  \theta_\zeta, \quad 
\theta_{\zeta \mid z} \coloneq \frac{\mbox{Cov}(y, [\zeta \mid z])}{\mbox{Var}(\zeta \mid z)} = \lambda \theta_\zeta
\end{equation}
respectively. Notably, both Plug-in (Z) \emph{and} the partial posterior approach suffer from attenuation bias: $\lambda < 1$, so the estimand is pulled toward zero relative to the  true coefficient $\theta_\zeta$. The attenuation bias increases as the measurement error uncertainty $\sigma_u$ grows. For Plug-in (Z), this phenomenon is a classical measurement error result. However, it is perhaps surprising for the partial posterior approach---especially because Plug-in ($\hat\zeta$), which simply uses the partial posterior mean, completely avoids attenuation bias.

Next, we consider the conditional variance of $y$ given its covariate. In the stage-two model, this is simply the error variance $\sigma_\epsilon^2$. But for these alternative approaches, we obtain much different estimands: 
\begin{equation}\label{variance}
\mbox{Var}( y\mid  z) = \sigma^2_\epsilon + \theta_\zeta^2  \lambda \sigma_u^2 =
\mbox{Var}(y\mid \lambda z), 
\quad \quad \mbox{Var}( y\mid [\zeta \mid z]) =  \sigma^2_\epsilon + \theta_\zeta^2(1+\lambda)\lambda\sigma_u^2
\end{equation}
for Plug-in (Z), Plug-in ($\hat\zeta$), and partial posterior, respectively. In each case, the variance is inflated by an additive term that depends on the signal strength $\theta_\zeta^2$, the attenuation factor $\lambda$, and the measurement error variance $\sigma_u^2$. Interestingly, Plug-in (Z) and the partial posterior agree on the attenuation bias in \eqref{bias}, while Plug-in (Z) and Plug-in ($\hat\zeta$) agree on the variance inflation in \eqref{variance}. What is clear is that each of these approaches is unsatisfactory for estimation, uncertainty quantification, or both. Despite its intuitive appeal, the partial posterior approach not only fails to correct the attenuation bias of Plug-in (Z), but also overinflates the variance. We confirm both \eqref{bias} and \eqref{variance}  empirically in Section~\ref{sec:4_Sim}. 




\section{Modified Importance Sampling for Two-stage  Bayesian Models}
\label{sec:IS_main}

\subsection{Importance sampling for two-stage Bayesian models}
\label{sec:IS_intro}

Despite the clear limitations of plug-in and partial posterior approaches, their popularity suggests that the general framework has potential: fitting a stage-one model separately, and then feeding its output into (one or more) stage-two model(s).  
However, joint Bayesian posterior (predictive) inference for two-stage models requires the distribution in \eqref{eq:ideal}. Given only output from the stage-one model in the form of partial posterior draws \eqref{eq:ppost}, the challenge remains to estimate the full data posterior $p(\bm \zeta \mid \bm y, \bm z)$. 

A natural way to map draws from one distribution to another is importance sampling (IS). Specifically, take the partial posterior distribution $p(\bm \zeta \mid \bm z)$ as the proposal distribution, generate draws  \eqref{eq:ppost}, and then reweight those draws to approximate the full data posterior distribution   (e.g., \citealp{tokdar_importance_2010}). Here, we target the \emph{conditional} full data distribution $p(\bm \zeta \mid \bm y, \bm z, \bm \theta)$, which leads to substantial simplifications. Substituting this sampling step into Algorithm~\ref{alg:ideal} is straightforward and converts that algorithm into a Gibbs sampler for $(\bm\theta, \bm \zeta)$. 

Now, for each $\bm \zeta^s  \sim p(\bm \zeta\mid \bm z)$, the IS weight is 
\begin{equation} \label{eq:is}
        w(\bm \zeta^s,\bm \theta) = \frac{p(\bm \zeta^s\mid \bm y,\bm z, \bm \theta)}{p(\bm \zeta^s \mid \bm z)} \propto 
    p(\bm y \mid \bm \theta, \bm \zeta^s)
\end{equation}
which is simply the likelihood under the stage-two model \eqref{eq:stage_two}. The IS weights can be used to approximate expectations via weighted averages (after renormalizing) or for sample importance resampling (SIR) which draws from $\{\bm\zeta^s\}_{s=1}^S$ with probabilities proportional to each $w(\bm \zeta^s,\bm \theta)$.

 IS has several advantages: it requires only the  draws \eqref{eq:ppost} from the stage-one model, provides an easy to compute adjustment for valid posterior (predictive) inference  under the two-stage model, and seamlessly integrates with (the Gibbs sampler version of) Algorithm~\ref{alg:ideal}. Unfortunately, IS is ineffective: it suffers from weight degeneracy as the dimension of the target distribution grows. Here, the dimensions of $\bm \zeta$ is $n$, so IS fails even for moderate sample sizes  (see Section~\ref{sec:4_Sim}). More specifically, one (or a small number of) draw(s), say $\bm \zeta^{s^*}$, accumulates by far the largest IS weight, so the IS approximation reduces to a point mass at $\bm \zeta^{s^*}$. Clearly, this is not an improvement over plug-in or partial posterior approaches.  Previous implementations of IS for two-stage models have appeared mainly as competing methods that performed poorly, such as in 
\citet{thijssen_approximating_2020} or the ``discrete uniform approach" in \citet{comess_bayesian_2022}.  

We introduce two algorithms that improve upon IS for two-stage models. 
These methods preserve the advantages of IS but are designed to avoid weight degeneracy. 

\subsection{Independent Importance Sampling}
\label{sec:IS2_IIS}

Independent importance sampling (IIS) replaces the $n$-dimensional proposal in IS with $n$ one-dimensional proposals. Specifically, consider the weights for each observation separately:
\begin{equation}\label{eq:iis}
    w_{\rm IIS}(\zeta_i^s, \bm \theta) = 
    p(y_i \mid \bm \theta, \zeta^s_i), \quad i=1,\ldots,n.
\end{equation}
For each draw from the partial posterior distribution $\bm \zeta^s \sim p(\bm \zeta \mid \bm z)$, we now have the $n$ weights $\{w_{\rm IIS}(\zeta_i^s, \bm \theta)\}_{i=1}^n$ rather than the single weight $w(\bm \zeta^s, \bm \theta)$. We illustrate how IIS integrates within a Gibbs sampler for $(\bm \zeta, \bm \theta)$ in Algorithm~\ref{alg:iis}. Note that Steps 1-2 are parallelizable.

\begin{algorithm}
\renewcommand{\baselinestretch}{1.2}\selectfont
\caption{A Gibbs Sampling Sweep with IIS}\label{alg:iis}
\begin{algorithmic}
\State \textbf{Input:} Draws $\{\bm\zeta^s \}^{S}_{s=1}\sim p(\bm \zeta \mid \bm z)$, data $\bm y$, current draw $\bm \theta^{t-1}$

\For{$i=1,\ldots,n$}
\begin{enumerate}
    \item Compute weights: $w_{{\rm IIS}, i}^s =  p(y_i \mid \bm \theta^{t-1}, \zeta_i^s)$ for each $s = 1, ..., S$.

 \item Sample $\zeta_i^t$ from $\{\zeta_i^s \}^{S}_{s=1}$ with  
  $\mathbb{P}(\zeta_i^t = \zeta_i^s) = w_{{\rm IIS}, i}^s/\sum_{k=1}^S w_{{\rm IIS}, i}^k$. 

\end{enumerate}
 \EndFor
 
 \State Sample $\bm \theta^{t} \sim p(\bm \theta \mid \bm y, \bm\zeta^t)$, where $\bm\zeta^t = (\zeta_1^t,..., \zeta_i^t,...\zeta_n^t)^\top$
 
\State \textbf{Output:}  draws $\bm{\zeta}^t, \bm \theta^t$
\end{algorithmic}
\end{algorithm}

The central advantages of IIS are 1) it inherits the advantages of IS but 2) avoids the  weight degeneracy via low-dimensional proposals. The main limitation is that IIS targets the distribution for $\bm \zeta$ proportional to
\begin{equation}\label{iis-target}
\prod_{i=1}^n p(y_i \mid \bm \theta, \zeta_i) \ p(\zeta_i \mid \bm z)
\end{equation}
which only equals the full data (conditional) posterior distribution $p(\bm \zeta \mid \bm y, \bm z, \bm \theta)$ when the partial posterior distribution is independent, $p(\bm \zeta \mid \bm z) = \prod_{i=1}^n p(\zeta_i \mid \bm z)$. Thus, we view IIS as a posterior approximation algorithm that will be most accurate when the partial posterior is approximately independent. This kind of independence approximation is not unique to IIS and has appeared previously for two-stage models  (\citealp{cameletti_bayesian_2019}; \citealp{comess_bayesian_2022}). We evaluate it empirically under both independent and dependent stage-one posteriors in Section~\ref{sec:4_Sim}. 



\subsection{Adjusted Importance Sampling (AIS)}
\label{sec:IS3_AIS}

Although IIS circumvents the weight degeneracy problems of IS, it introduces a new approximation error based on the dependencies in the partial posterior  $p(\bm \zeta \mid \bm z)$. Since our stage-one models correspond to air pollution exposures across space or time, such dependencies are expected. Here, we propose an adjusted importance sampling (AIS) algorithm to correct for that approximation error. The basic idea is to view IIS as a proposal distribution in an importance sampling algorithm. Notably, IIS empirically dominates the partial posterior approach (Section~\ref{sec:4_Sim}), which suggests that IIS yields a better proposal distribution than the partial posterior in \eqref{eq:is}. 

Leveraging the output from IIS, the proposal distribution is now given by the IIS target distribution proportional to \eqref{iis-target}. With this proposal, the IS weights are
\begin{equation}
w_{\rm AIS}(\bm \zeta^s) \propto \frac{p(\bm \zeta^s \mid \bm y,\bm z, \bm \theta)}{\prod_{i=1}^n p(y_i \mid \bm \theta, \zeta_i^s) \ p(\zeta_i^s \mid \bm z)} \propto \frac{p(\bm \zeta^s \mid \bm z)}{\prod_{i=1}^n  p(\zeta_i^s \mid \bm z)}
\label{eq:ais_weights}
\end{equation}
which first uses the ratio of target to proposal (up to constants) and then simplifies based on conditional independence in the stage-two model \eqref{eq:stage_two} and the assumed independence between $\bm \zeta$ and $\bm \theta$. The AIS weights directly measure the dependencies in the partial posterior and provide the necessary correction to IIS in order to target the joint two-stage model posterior   \eqref{eq:ideal}. Although $w_{\rm AIS}(\bm \zeta^s)$ does not depend on $\bm \theta$,  the target distribution remains conditional on $\bm \theta$ as before. 

We summarize AIS in Algorithm~\ref{alg:ais} for a single sweep of a Gibbs sampler. AIS starts similarly to IIS, but instead takes $R$ draws from this proposal distribution. Fortunately, many of these computations may be recycled or parallelized. 


\begin{algorithm}
\renewcommand{\baselinestretch}{1.2}\selectfont
\caption{A Gibbs Sampling Sweep with AIS}\label{alg:ais}
\begin{algorithmic}
\State \textbf{Input:} Draws $\{\bm\zeta^s \}^{S}_{s=1}\sim p(\bm \zeta \mid \bm z)$, data $\bm y$, current draw $\bm \theta^{t-1}$

\begin{enumerate}

      \item Compute the IIS weights (Step 1, Algorithm \ref{alg:iis}) for $i=1,\ldots,n$ and $s=1,\ldots,S$

      \For{$r=1,\ldots,R$}
        \begin{enumerate}
            \item Sample $\zeta_i^r$ from $\{\zeta_i^s \}^{S}_{s=1}$ with  
  $\mathbb{P}(\zeta_i^r = \zeta_i^s) = w_{{\rm IIS}, i}^s/\sum_{k=1}^S w_{{\rm IIS}, i}^k$ for $i=1,\ldots,n$
            \item Compute $w_{\rm AIS}^r = p(\bm \zeta^r \mid \bm z)/\prod_{i=1}^n  p(\zeta_i^r \mid \bm z)$
        \end{enumerate}
      \EndFor

    \item Sample $\bm \zeta^t$ from $\{\bm \zeta^r \}_{r=1}^{R}$ with  
  $\mathbb{P}(\bm \zeta^t = \bm \zeta^r) = w_{{\rm AIS}}^r/\sum_{k=1}^R w_{{\rm AIS}, r}^k$. 

   \item Sample $\bm \theta^{t} \sim p(\bm \theta \mid \bm y, \bm\zeta^t)$, where $\bm\zeta^t = (\zeta_1^t,..., \zeta_i^t,...\zeta_n^t)^\top$
\end{enumerate}

\State \textbf{Output:} draws $\bm{\zeta}^t, \bm \theta^t$

\end{algorithmic}
\end{algorithm}

The remaining challenge for AIS is how to compute the weights in \eqref{eq:ais_weights} using only the draws from the partial posterior \eqref{eq:ppost}. This quantity is the ratio of a joint distribution to the product of its marginals, so several options are available. We adopt a relatively simple, stable, and computationally efficient approach based on a MVN approximation to the ratio \eqref{eq:ais_weights}. This does \emph{not} imply that we are using MVN distributions for our partial posterior or our IS proposal. Rather, our proposals are still taken directly from the stage-one model draws \eqref{eq:ppost}, and we seek to improve upon the IIS weights to account for dependencies in the partial posterior---but without incurring the weight degeneracies of IS or the limitations of a MVN approximation for $p(\bm \zeta \mid \bm z)$.  

The use of MVN densities to approximate the ratio in \eqref{eq:ais_weights} is especially convenient because 1) approximating the joint distribution in the numerator  only requires estimates of the expectation and covariance matrix and 2) approximating the marginals in the denominator simply takes these expectations and the diagonals of the covariance matrix. Specifically, let $\hat {\bm \zeta} = \mathbb{E}(\bm \zeta \mid \bm z)$ and 
 $\bm{\hat \Sigma_{\zeta \mid z}} = \mbox{Cov}(\bm \zeta \mid \bm z)$ be the partial posterior mean and covariance, respectively, and let $\bm{\hat{D}_{\zeta \mid z}} = \mbox{diag}(\bm{\hat \Sigma_{\zeta \mid z}})$. 
 Now, the ratio in \eqref{eq:ais_weights} may be approximated 
 \begin{equation}\label{ais-approx}
     w_{\rm AIS}(\bm \zeta^s) \approx \sqrt{\frac{\det(\bm{\hat{D}_{\zeta \mid z}})}{\det(\bm{\hat \Sigma_{\zeta \mid z}})}} \exp\Big\{-\frac{1}{2}(\bm{\zeta}^s- \hat {\bm \zeta} )^\top(\bm{\hat \Sigma_{\zeta \mid z}}^{-1}-\bm{\hat{D}_{\zeta \mid z}}^{-1}) (\bm{\zeta}^s- \hat {\bm \zeta} )\Big\},
 \end{equation}
up to constants. By default, we estimate $\hat {\bm \zeta}$ and $\bm{\hat \Sigma_{\zeta \mid z}}$ using the sample mean and sample covariance, respectively, of the draws in \eqref{eq:ppost}. More sophisticated estimation techniques remain applicable within this framework, especially for the covariance matrix \citep{lee_scalable_2024}. Notably, the estimates needed for \eqref{ais-approx} are a one-time cost prior to running the MCMC sampling algorithm. 

\section{Simulations}
\label{sec:4_Sim}

\subsection{Simulation Design}
\label{sec:4_1_linmod}
We design a simulation study to compare the performance of the proposed IIS and AIS algorithms against popular and state-of-the-art competitors. We use the two-stage model \eqref{eq:slr_one}--\eqref{eq:slr_two} from Section~\ref{sec-anexamp}. Despite its simplicity, this two-stage model still includes the salient features in our setting. Crucially, it is sufficiently tractable that we may derive an ``Oracle" Gibbs sampler for the model (see Section~\ref{sec-compete}) to use as the ground truth for the target posterior distribution \eqref{eq:ideal} and predictive distribution \eqref{eq:postpred}. In addition, the simulated data allow us to verify our analytical results from Section~\ref{sec-anexamp}, which provide important critiques for plug-in and partial posterior approaches. The goal is to understand if and when IIS and AIS correct those significant flaws.

For the parameters in \eqref{eq:slr_one}--\eqref{eq:slr_two}, we set $\beta_0=0$, $\theta_\zeta=4$, and $\sigma^2_\epsilon=2$, and generate $n=200$ observations. We consider two scenarios for the covariance matrices $\bm \Sigma_u$ and $\bm \Sigma_\zeta$. In Example 1 (independent case), the covariances are $\bm \Sigma_u = \sigma_u^2 \bm I_n$ and $\bm \Sigma_\zeta = \sigma_\zeta^2 \bm I_n$ as in Section~\ref{sec-anexamp}. We set $\sigma_u^2 = \sigma_\zeta^2 = 1$. Here, the partial posterior $p(\bm \zeta \mid \bm z)$ satisfies independence, so we expect IIS to perform well. 
Then Example 2 (correlated case) uses the same diagonal entries but fixes the off-diagonal entries to $\rho = 0.3$. The partial posterior distribution no longer satisfies independence, so we expect IIS to perform worse. For all competing methods, $\bm \Sigma_u$ and $\bm \Sigma_\zeta$ are treated as known. We generate 100 simulated datasets for both Example~1 and Example~2. In Section D of the supplement, we include results for settings with different values of $\theta_\zeta$ and $\sigma^2_\epsilon$.

\subsection{Competing methods and evaluations}\label{sec-compete}

To mimic our general setting, we supply each competing method with $S = 500$ partial posterior draws \eqref{eq:ppost}. Under model \eqref{eq:slr_one}, the partial posterior distribution is $[\bm \zeta \mid \bm z] \sim MVN(\bm{\hat \zeta}, \bm{\hat \Sigma_{\zeta \mid z}})$
where $\bm{\hat \Sigma_{\zeta \mid z}} = (\bm \Sigma_u^{-1} + \bm \Sigma_\zeta^{-1})^{-1}$ and $\bm{\hat \zeta} = \bm{\hat \Sigma_{\zeta \mid z}}\bm\Sigma_u^{-1} \bm z$. 
In addition to IIS and AIS, we include each of the alternatives detailed previously: Plug-in (Z), Plug-in ($\hat\zeta$), Partial Posterior, and importance sampling. Note that Plug-in (Z) does not use the partial posterior draws, but we include it nonetheless as a reasonable competitor for this two-stage model. 
Finally, we add the univariate KDE (UKDE) approach from \cite{comess_bayesian_2022} implemented in the \texttt{R} package \texttt{KDExp}. We set $R=500$ for AIS and use 1000 draws for importance sampling. 

Each of these competing methods provides output from the stage-one model \eqref{eq:slr_one} or $\bm z$ to be fed into the stage-two regression model \eqref{eq:slr_two}. For all approaches, we adopt a Gibbs sampling version of Algorithm~\ref{alg:ideal} that alternates between that method's treatment of $\bm \zeta$  and sampling $ \bm\theta = (\beta_0, \theta_\zeta, \sigma_\epsilon)$. The common sampling steps for $p(\bm \theta \mid \bm y, \bm \zeta)$ are as follows. 
Let $\bm X = (\bm 1, \bm \zeta)$ and assume the priors $\sigma^2_\epsilon\sim IG(a, b)$ and $\beta_0, \theta_\zeta \sim N(0, \sigma_\theta^2)$. The full conditionals are $[(\beta_0, \theta_\zeta)^\top \mid \bm y, -] \sim MVN\big[ \{\bm X^\top \bm X + (\sigma^2_\epsilon/\sigma^2_\theta) \bm I_2\}^{-1}\bm X^\top \bm y, \sigma^2_\epsilon \{\bm X^\top \bm X + (\sigma^2_\epsilon/\sigma^2_\theta) \bm I_2\}^{-1}\big]$ and $[\sigma^2_\epsilon \mid \bm y, -] \sim IG\big\{a+ n/2,b + 0.5 \sum^n_{i=1}(y_i- \beta_0-\theta_\zeta\zeta_i)^2\big\}$.
The hyperparameters are fixed at $a = 3$, $b = 6$, and $\sigma_\theta = 1000$.


Finally, we add an ``Oracle Gibbs" competitor for benchmarking. For this particular two-stage model, it is straightforward to derive the \emph{exact} conditional full posterior distribution, $p(\bm \zeta \mid \bm y, \bm z, \theta)$. Thus, the Gibbs sampling version of our idealized Algorithm~\ref{alg:ideal} is feasible here. Oracle Gibbs is guaranteed to target the desired joint posterior distribution \eqref{eq:ideal} and its only approximation error is from MCMC sampling. As a result, Oracle Gibbs is our gold standard for estimation, inference, and prediction in this simulation study. We emphasize, however, that in our general setting and with real air pollution exposure models (Section~\ref{sec:5_empmain}), this option is not feasible.

Our evaluations are based on estimation and inference for a stage-two regression coefficient as well as prediction of new outcomes. First, we compute the posterior means of $\theta_\zeta$ and $\sigma^2_\epsilon$ to check for the statistical flaws identified in \eqref{bias} and \eqref{variance}. Next, we compute interval widths and empirical coverage of the 95\% credible intervals for $\theta_\zeta$, which measures the sharpness and calibration, respectively, of posterior inference for this key term. We also compute the 2-Wasserstein distances to the Oracle Gibbs posterior distribution for $\theta_\zeta$. Finally, we compute 95\% prediction interval widths and empirical coverages for $n_{test}=1000$ testing data points $\bm{\tilde y}$ generated iid to $\bm y$. Additional simulaltion results are available in the supplement.


\subsection{Parameter Estimation and Uncertainty Quantification}
\label{sec:4_1_simres_main}

\begin{figure}[h!] 
\begin{subfigure}{0.48\textwidth}
\includegraphics[width=\linewidth]{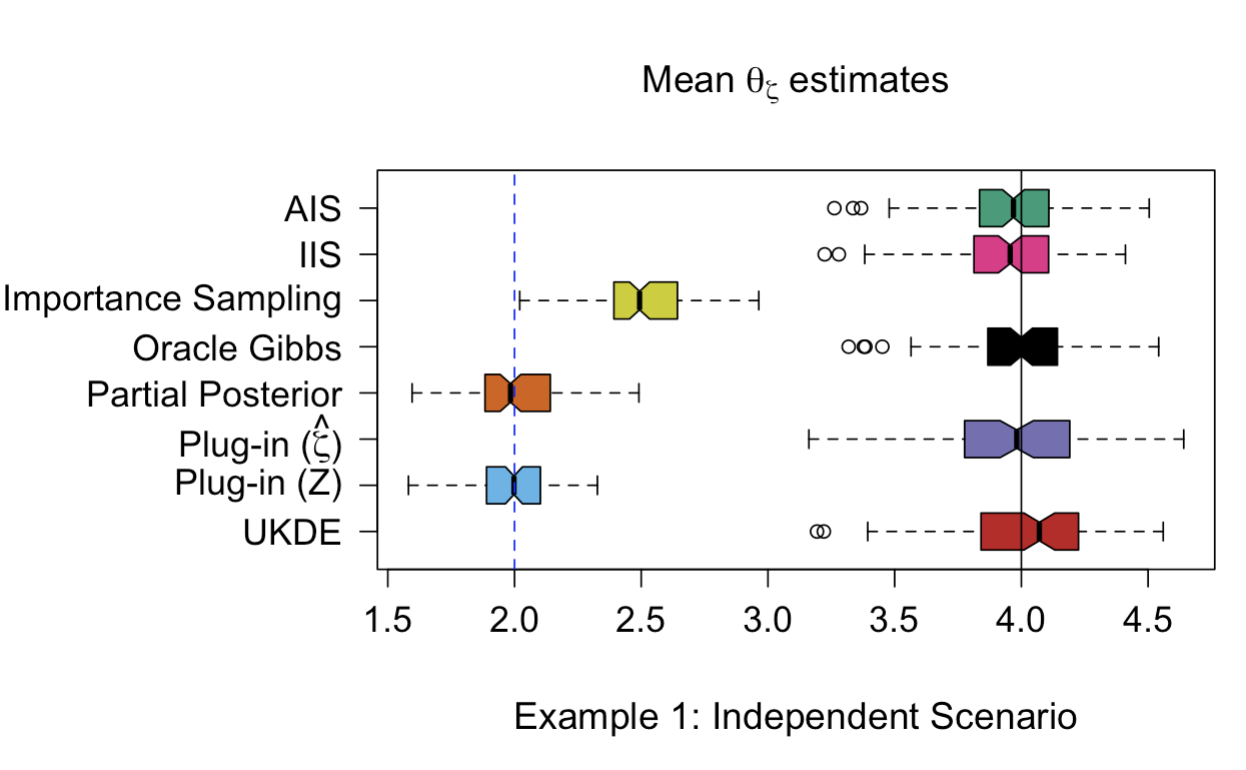}
\end{subfigure}
\begin{subfigure}{0.48\textwidth}
\includegraphics[width=\linewidth]{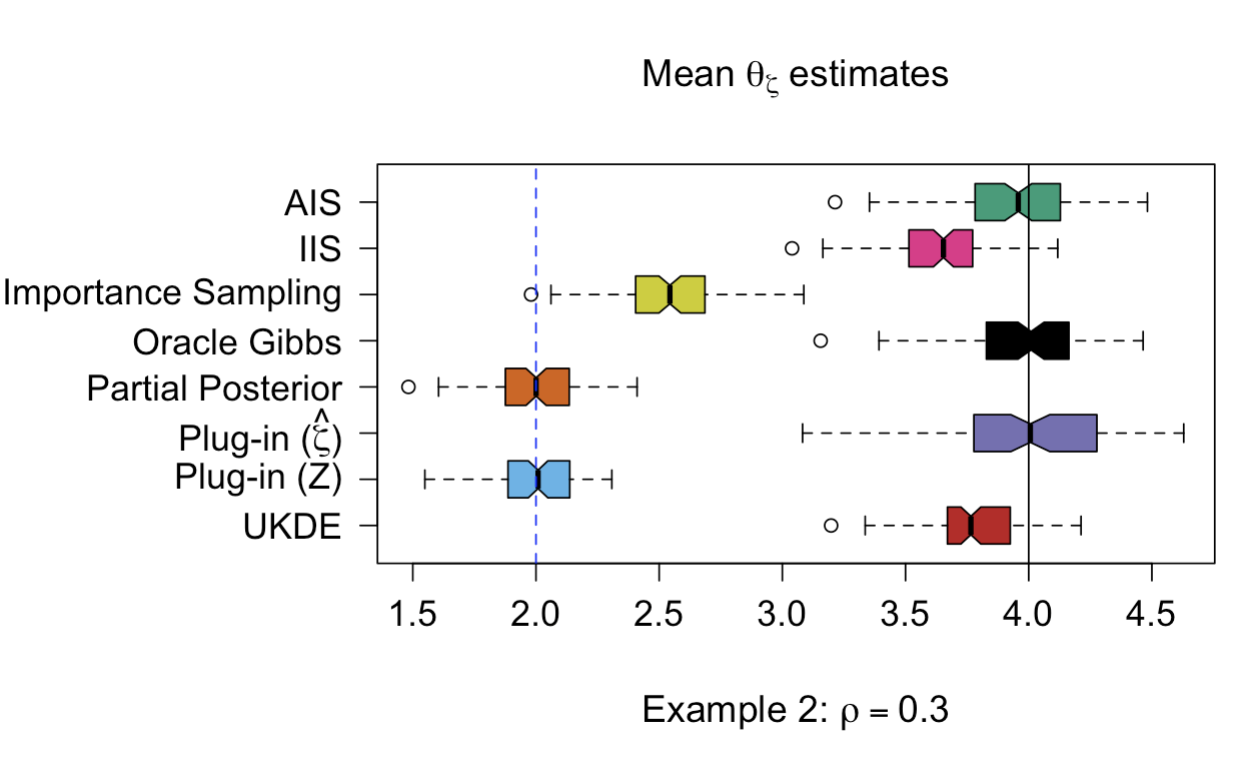}
\end{subfigure}
\begin{subfigure}{0.5\textwidth}
\includegraphics[width=\linewidth]{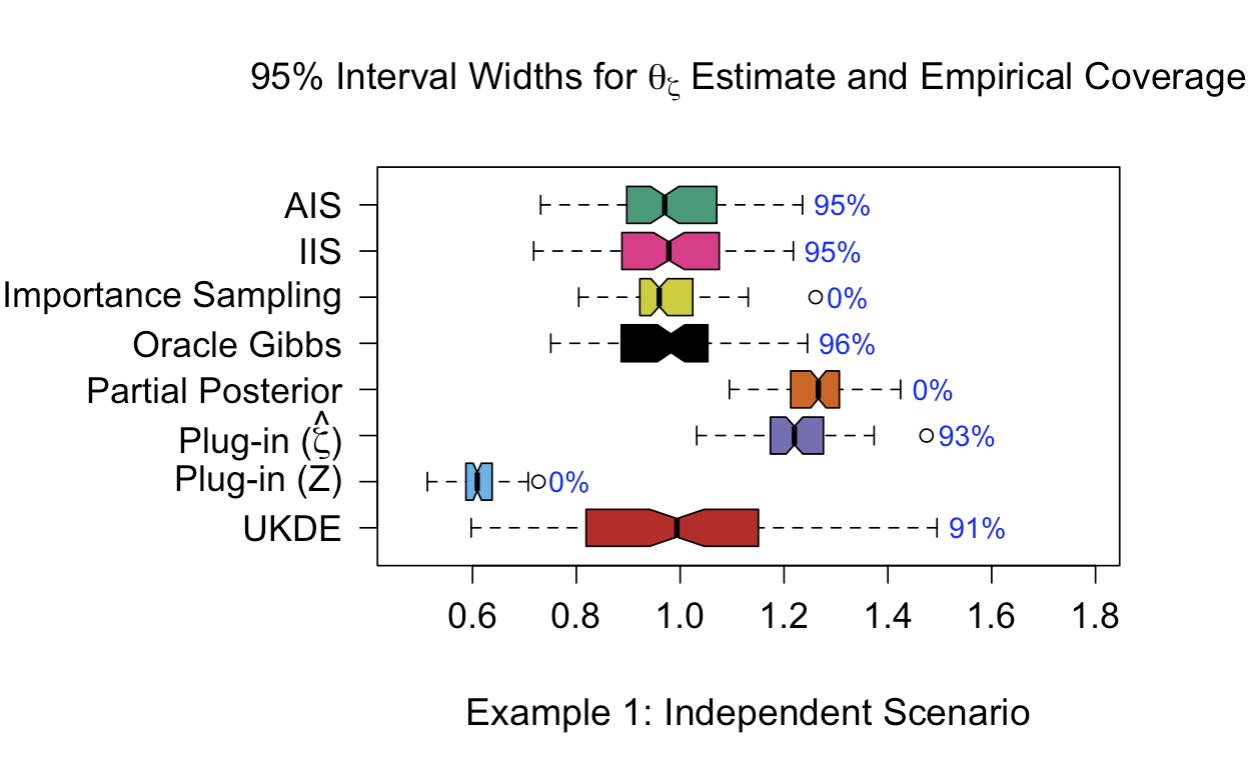}
\end{subfigure}
\begin{subfigure}{0.5\textwidth}
\includegraphics[width=\linewidth]{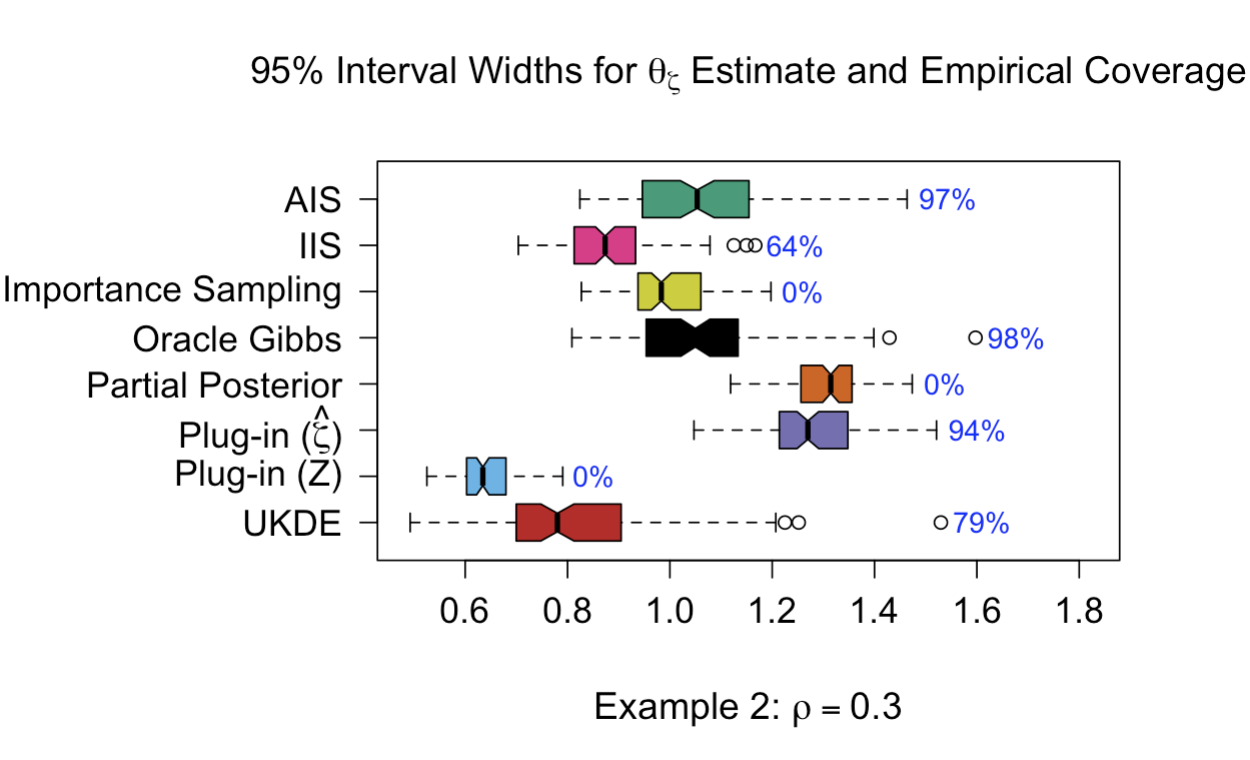}
\end{subfigure}
\caption{Posterior means (top) and 95\% credible interval widths and coverages (bottom) for $\theta_\zeta$ in independent (left) and correlated (right) settings. For estimation, we mark the true regression coefficient (solid black) and the attenuated version from \eqref{bias} (dashed blue); for inference, the annotations give the empirical coverages.  IIS  performs well in the independent case, while AIS performs comparably to Oracle Gibbs in all cases.} \label{fig:1}
\end{figure}

In Figure~\ref{fig:1}, we show the posterior means for $\theta_\zeta$ across all simulations (top) and summarize the 95\% credible interval widths and coverages for $\theta_\zeta$ (bottom) for each approach  in both examples. This confirms the bias results from \eqref{bias}: Plug-in (Z) and the partial posterior approach are both biased near the amount predicted by \eqref{bias}, while Plug-in ($\hat\zeta$) is nearly unbiased. Crucially, this bias implicates the credible intervals and drives the empirical coverages to zero. We also see that importance sampling is clearly ineffective both for point estimation and inference. Notably, IIS and AIS offer dramatic improvements. In particular, AIS closely matches Oracle Gibbs for both estimation and inference in both examples. Finally, in the correlated example, the methods that rely on partial posterior independence (IIS and UKDE) are less competitive.


\begin{figure}[t!] 
\begin{subfigure}{0.48\textwidth}
\includegraphics[width=\linewidth]{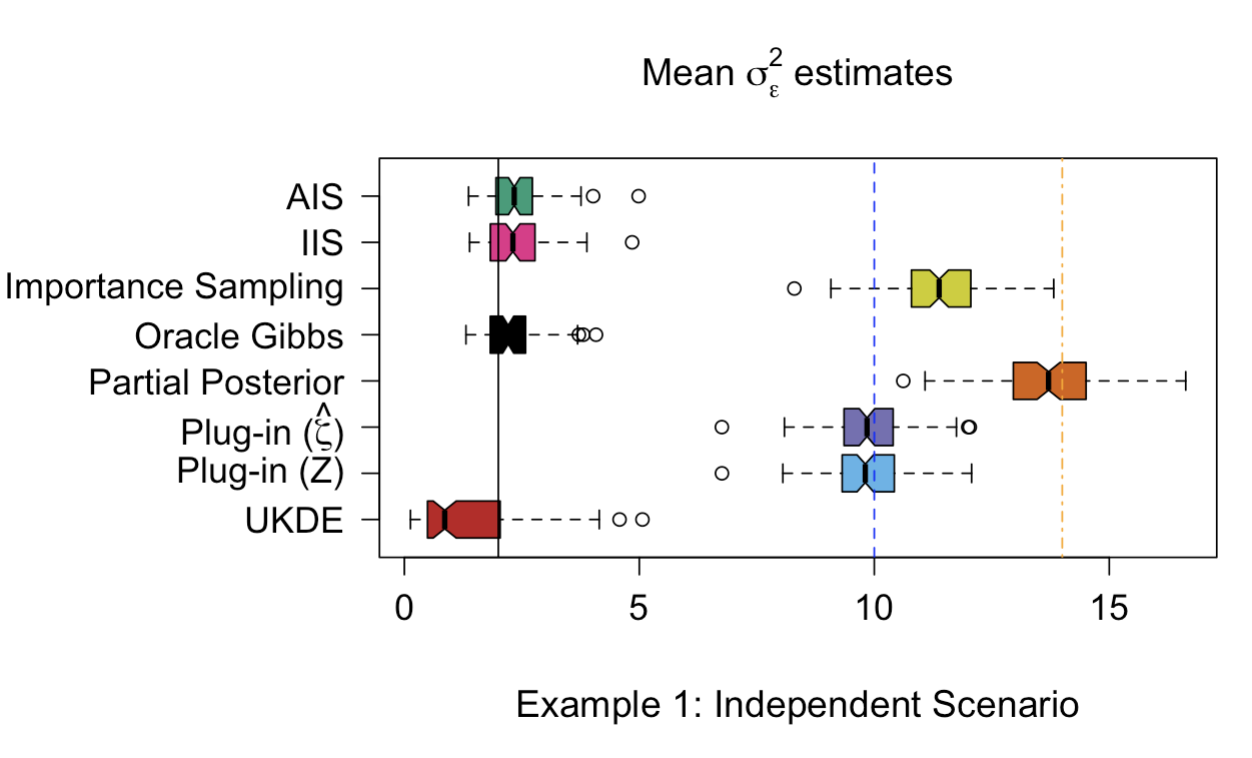}
\end{subfigure}
\begin{subfigure}{0.48\textwidth}
\includegraphics[width=\linewidth]{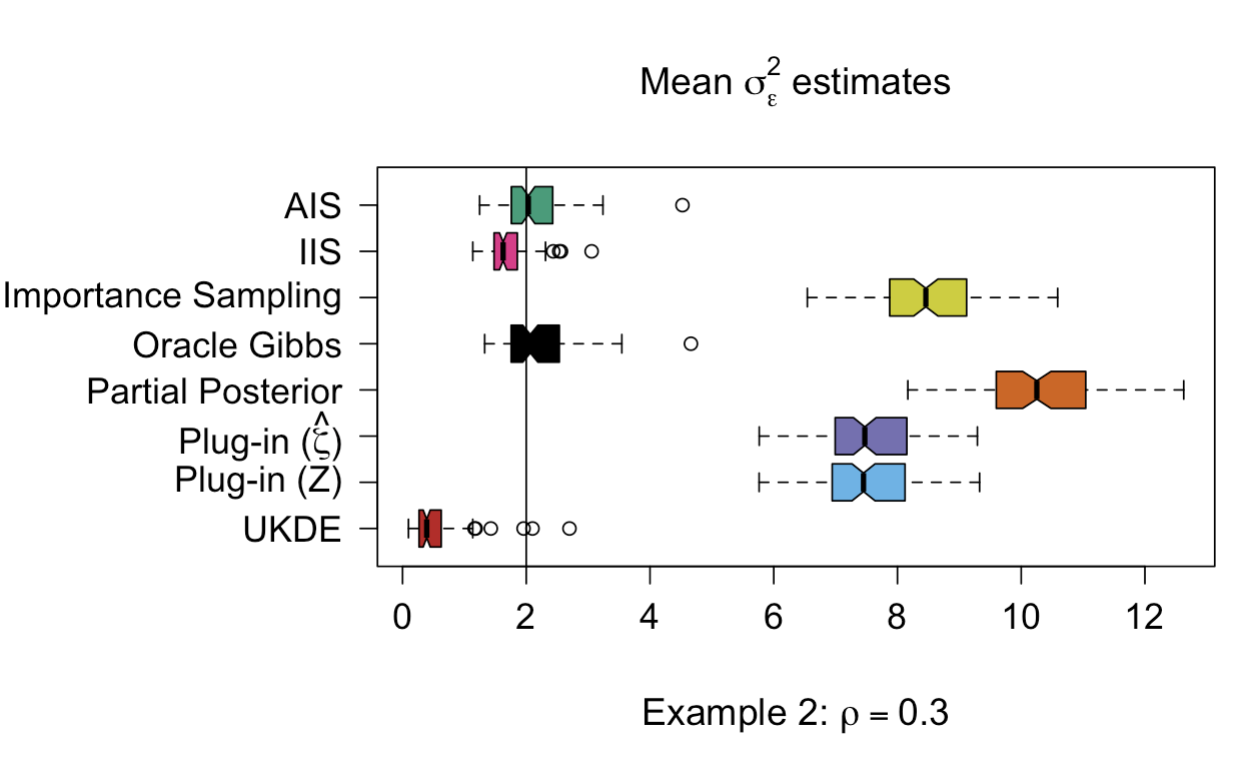}
\end{subfigure}
\caption{Posterior means for $\sigma^2_\epsilon$ in independent (left) and correlated (right) settings. The true value of $\sigma_\epsilon^2$ is indicated by a black line; for the independent example, the attenuated versions from \eqref{variance} are marked for plug-in (blue dashed line) and partial posterior (orange dash-dotted line) approaches. AIS and IIS are accurate and closely match Oracle Gibbs. Competing methods tend to overestimate $\sigma_\epsilon^2$; UKDE underestimates $\sigma_\epsilon^2$ in the correlated case.} \label{fig:sigma_sim}
\end{figure}

Next, we investigate the potential for error variance overestimation indicated by \eqref{variance}. Figure~\ref{fig:sigma_sim} shows the posterior means for $\sigma^2_\epsilon$ across the 100 simulated data sets for all methods. Both IIS and AIS consistently perform well and closely align with Oracle Gibbs. UKDE underestimates $\sigma_\epsilon^2$ in the independent setting and especially in the correlated setting. The remaining competing methods substantially overestimate $\sigma_\epsilon^2$, as predicted by \eqref{variance}. Notably, this group includes importance sampling, which highlights the problems of weight degeneracy. 

The error variance plays a critical role in uncertainty quantification for the regression effects. This is summarized by the interval widths  for $\theta_\zeta$ (Figure~\ref{fig:1}, bottom). For the simple linear regression model \eqref{eq:slr_two}, the posterior variance of  $\theta_\zeta$ is determined by 1) the error variance, 2) the covariate variance, and 3) the prior variance. In our setting, competing methods produce different values of 1) and 2); all use the same prior variance. Thus, the overestimation of $\sigma^2_\epsilon$ observed in Figure~\ref{fig:sigma_sim} contributes to the substantially wider intervals for $\theta_\zeta$ for both Plug-in ($\hat\zeta$) and the partial posterior approach. By comparison, Plug-in (Z) has much narrower intervals despite similarly overestimating $\sigma^2_\epsilon$. This occurs because Plug-in (Z) substitutes the corrupted observations $\bm z$ as the covariates in \eqref{eq:slr_two}, which overinflates the covariate variance and thus (improperly) decreases the posterior variance of $\theta_\zeta$. 



As an aggregate measure of posterior inference, Table~\ref{Table1} reports the 2-Wasserstein distances for the posterior distributions for $\theta_\zeta$ and $\sigma_\epsilon^2$ for each method compared to Oracle Gibbs. These results are averaged over all simulations in each example. Clearly, AIS dominates all competitors for posterior inference. IIS is also highly accurate in Example 1 (independent case) and performs moderately well even under stage-one posterior dependencies (Example 2). 


\begin{table}[h]
\begin{tabular}{  p{3cm}p{5cm}p{3cm}p{3cm}  }
 \hline
 \multicolumn{4}{|c|}{2-Wasserstein Distance to Oracle Gibbs Posterior for Examples 1 and 2} \\
 \hline
 Data Settings & Method& $\theta_\zeta$ &$\sigma^2_\epsilon$\\
 \hline
 Example 1:  &AIS   & {\bf 0.013}    & {\bf 0.106}\\
Independent & IIS&   {\bf 0.016}  & {\bf 0.121}  \\
 &Importance Sampling &1.455 & 9.144\\
 &Plug-in ($\hat\zeta$)   & 0.108 & 7.579\\
&Plug-in (Z) &   1.973  & 7.572\\
 &Partial Posterior& 1.958  & 11.500   \\
& UKDE& 0.085  & 1.092\\ \hline
Example 2:  &AIS   & {\bf 0.025}    & {\bf 0.070}\\
$\rho =0.3$&  IIS&   0.338  & {\bf 0.569} \\
& Importance Sampling &1.430 & 6.317\\
& Plug-in ($\hat\zeta$)   & {\bf 0.101} & 5.332\\
&Plug-in (Z) &   1.977  & 5.325\\
 &Partial Posterior& 1.959  & 8.086   \\
& UKDE& 0.211  & 1.716\\ \hline
 
\end{tabular}
\caption{2-Wasserstein distance to the posterior distributions for $\theta_\zeta$ and $\sigma_\epsilon^2$ under Oracle Gibbs. The two most accurate methods for each case are in bold. AIS performs best for both parameters in both cases.}
\label{Table1}
\end{table}



\subsection{Out-of-sample prediction}
\label{sec:4_3_post_pred}

\begin{figure}[h] 

\begin{subfigure}{0.48\textwidth}
\includegraphics[width=\linewidth]{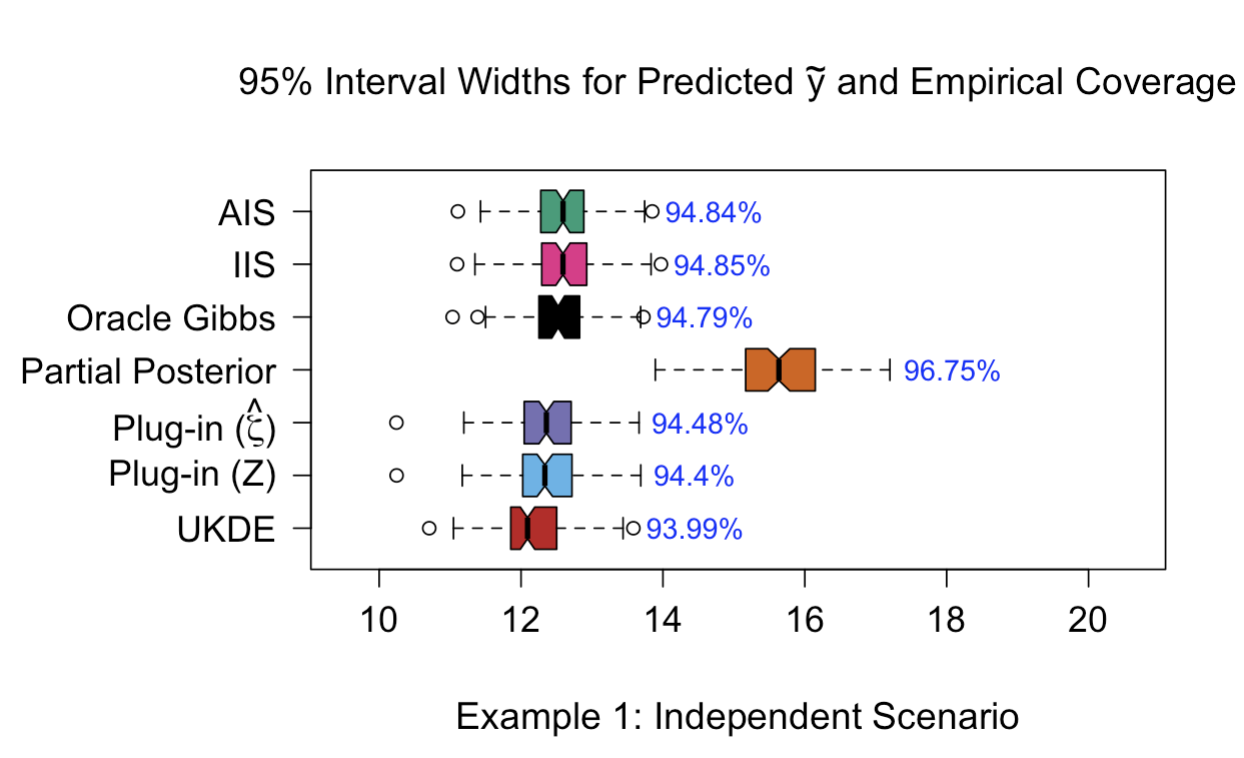}
\end{subfigure}
\begin{subfigure}{0.48\textwidth}
\includegraphics[width=\linewidth]{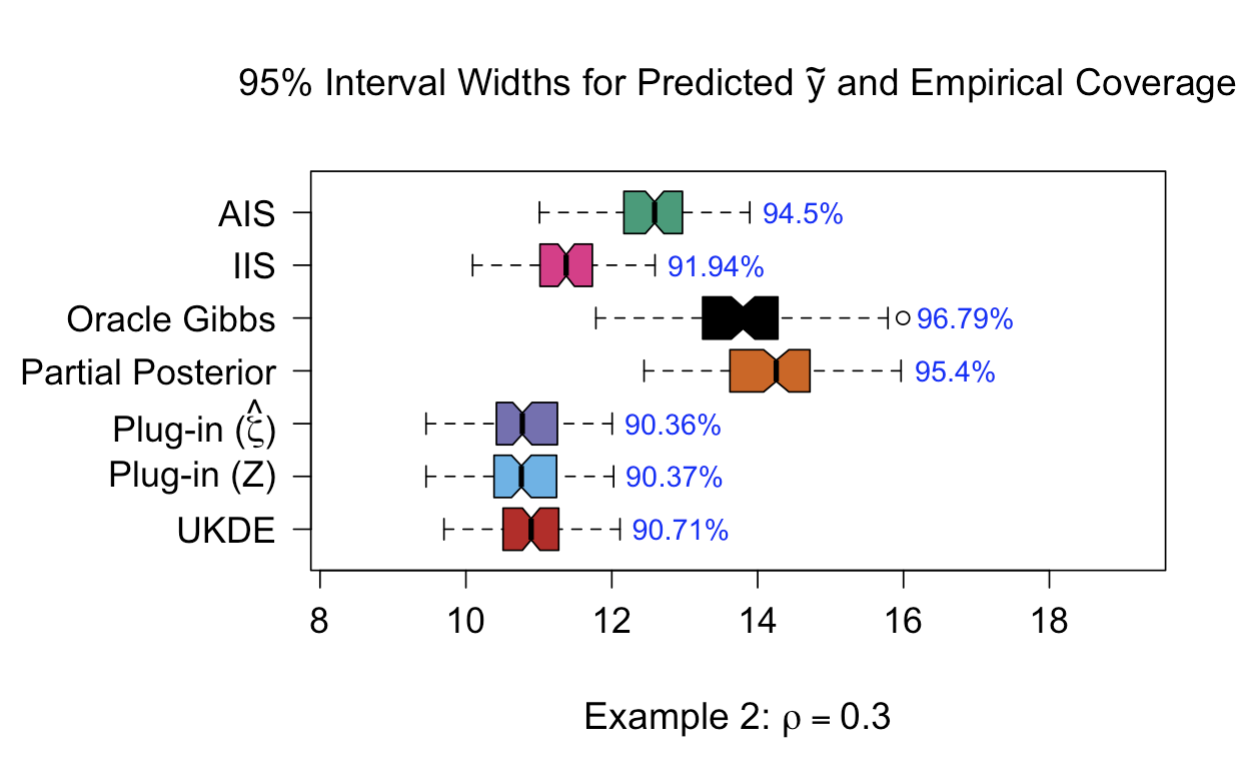}
\end{subfigure}

\caption{Interval widths (boxplots) and empirical coverage (annotations) for 95\% posterior prediction intervals on $n_{test}=1000$ out-of-sample testing points across 100 simulated datasets in independent (left) and correlated (right) settings. AIS performs well in both settings, with intervals that are narrow yet close to the nominal coverage.} \label{fig:pred_res}
\end{figure}

We evaluate the 95\% posterior prediction intervals on out-of-sample testing data. Most competing methods uses the posterior predictive simulation strategy described in Section~\ref{sec:Back_intro} based on \eqref{eq:postpred}.  Plug-in (Z) fixes $\bm{\tilde \zeta}$ at the observed value $\bm{\tilde z}$ and Plug-in ($\hat\zeta$) fixes $\bm{\tilde \zeta}$ at the partial posterior mean $\mathbb{E}(\bm{\tilde z} \mid \bm{\tilde \zeta})$. 
 The interval widths and empirical coverages are averaged over $n_{test}=1000$ out-of-sample testing points and summarized across simulations and designs in Figure~\ref{fig:pred_res}. We exclude importance sampling due to poor performance. For the independent scenario, all methods perform well and similarly, although the partial posterior intervals are excessively wide. In the correlated scenario, many of the competing methods suffer from undercoverage. 
 AIS and the partial posterior approach are the exceptions and both produce (nearly) calibrated intervals, with AIS moderately more precise.

\section{PM$_{2.5}$ Exposure and Mortality in the USA}
\label{sec:5_empmain}

We investigate the association between PM$_{2.5}$ exposure and mortality.  Specifically, PM$_{2.5}$ exposure is estimated using a Bayesian nonparametric ensemble (BNE) model  (Section~\ref{sec:5_2_model1}), while mortality is quantified as the annual age-adjusted mortality rates by county (Section~\ref{sec:5_3_model2}). Additional county-level covariates are also included (Section~\ref{sec:5_3_model2}).

\subsection{Stage One: PM$_{2.5}$ Modeling}
\label{sec:5_2_model1}

The stage-one model is a BNE model for 
PM$_{2.5}$ exposure (\citealp{benavides_bner_2025}). BNE aggregates multiple air pollution models using weights based on predictive accuracy. The authors of \cite{benavides_bner_2025} shared $S=100$ draws from the BNE posterior predictive distribution at 0.125°×0.125° resolution in the continental USA with 52,385 total grid points. Crucially, we \emph{only} have access to these partial posterior draws \eqref{eq:ppost}, not the underlying inputs (input models or data $\bm z$). For our analysis, these spatially-referenced draws are then aggregated to the county level to align with our health outcomes and covariates (Section~\ref{sec:5_3_model2}).

\begin{figure}[h]
\centering
\includegraphics[width=.49\textwidth]{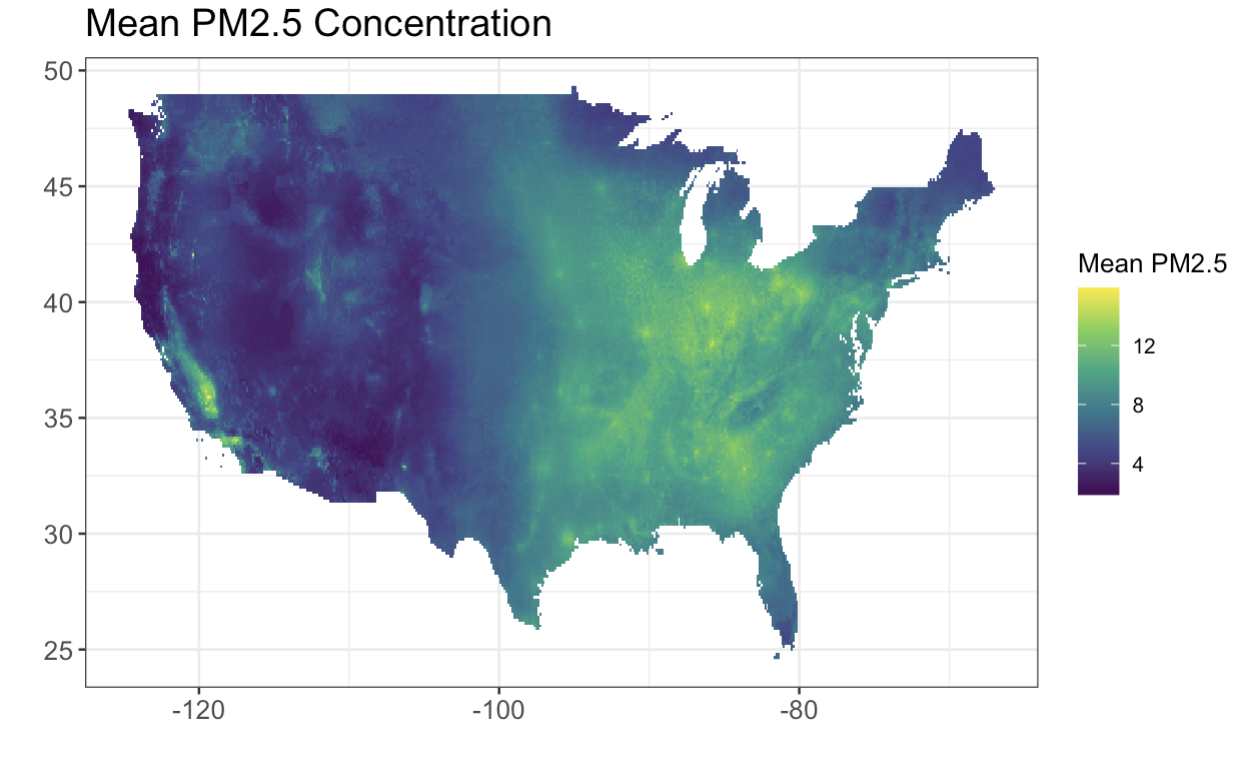}
\includegraphics[width=.49\textwidth]{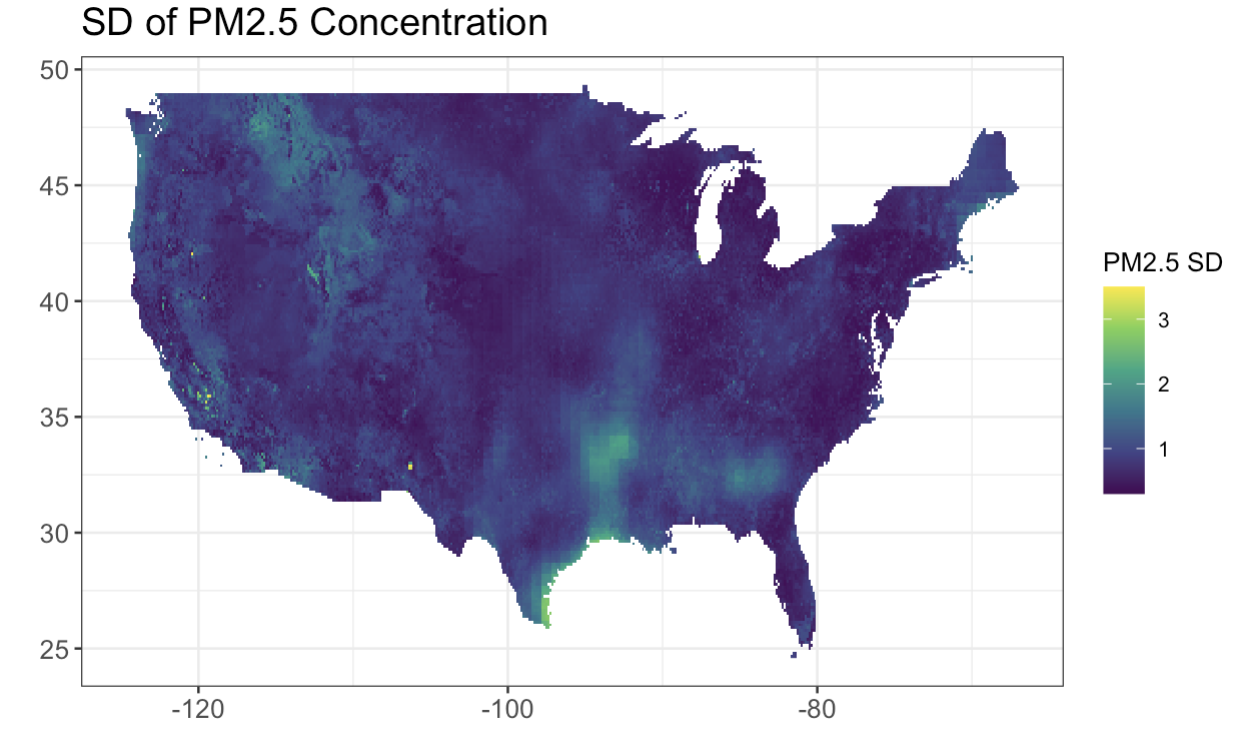}
\includegraphics[width=.49\textwidth]{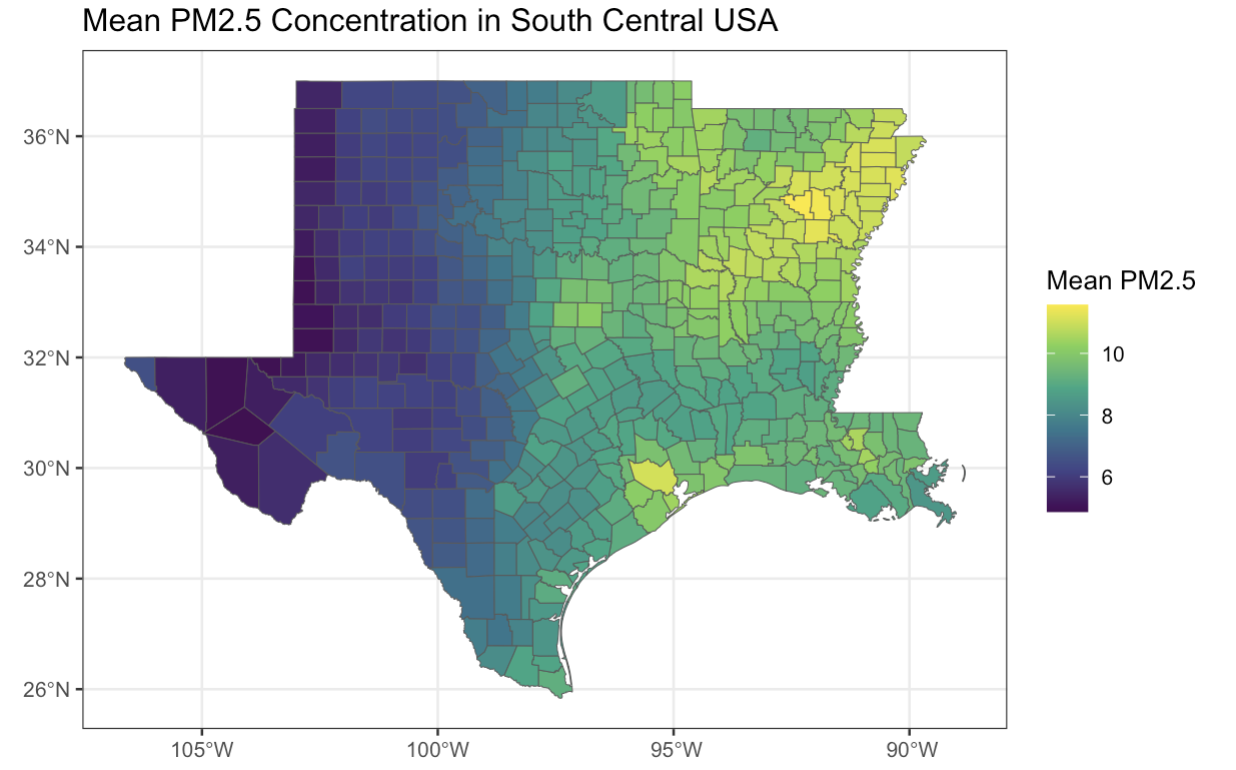}
\includegraphics[width=.48\textwidth]{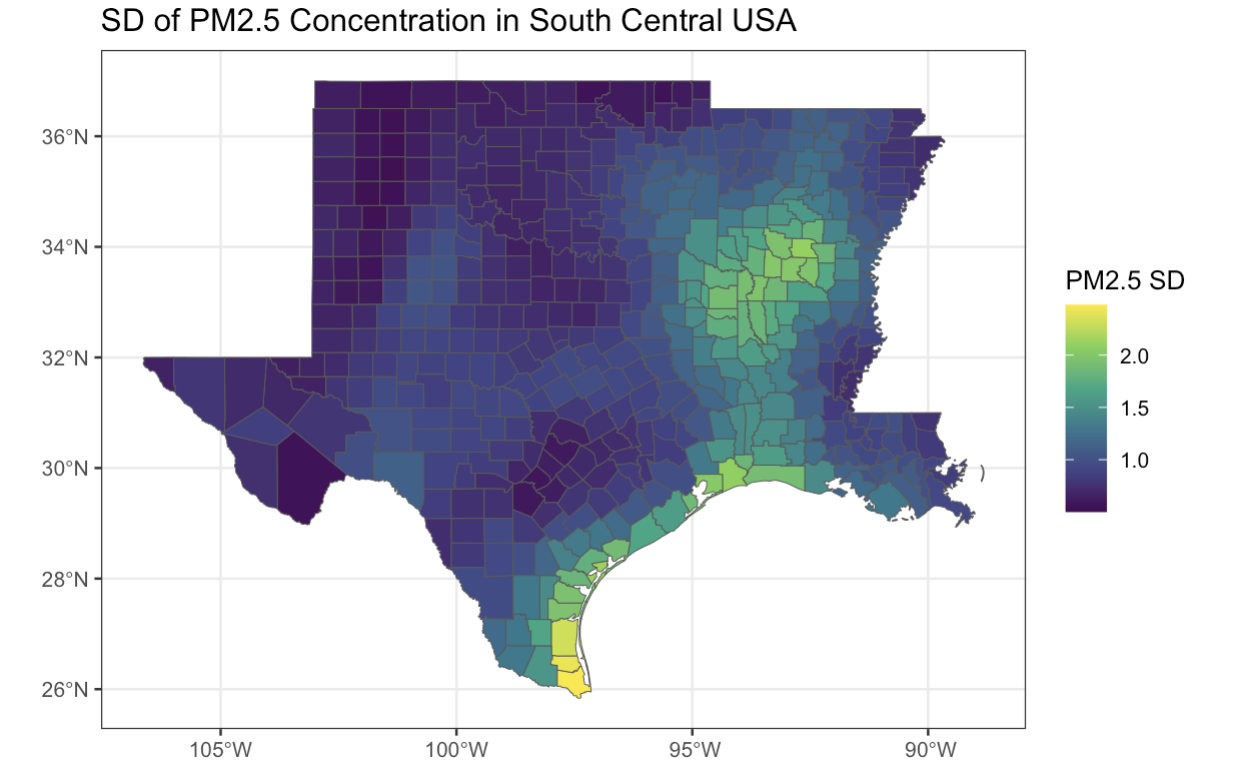}
\caption{Estimates (left) and standard deviations (right) of PM$_{2.5}$ exposure from the stage-one BNE model across the continental USA at 0.125°×0.125° resolution (top) and aggregated to the county level in the South-Central USA (bottom). The South-Central region has both higher estimates and higher uncertainties.}
\label{fig:pm}
\end{figure}

Figure~\ref{fig:pm} illustrates the BNE stage-one model output. We include both the fine-gridded output for the continental USA (top) along with the aggregated county-level output for the South-Central region (bottom). Notably, the South-Central region has higher estimates and higher uncertainties. 


\subsection{Stage Two: Modeling Mortality}
\label{sec:5_3_model2}

\begin{figure}[h]
    \centering
    \includegraphics[width=0.6\textwidth]{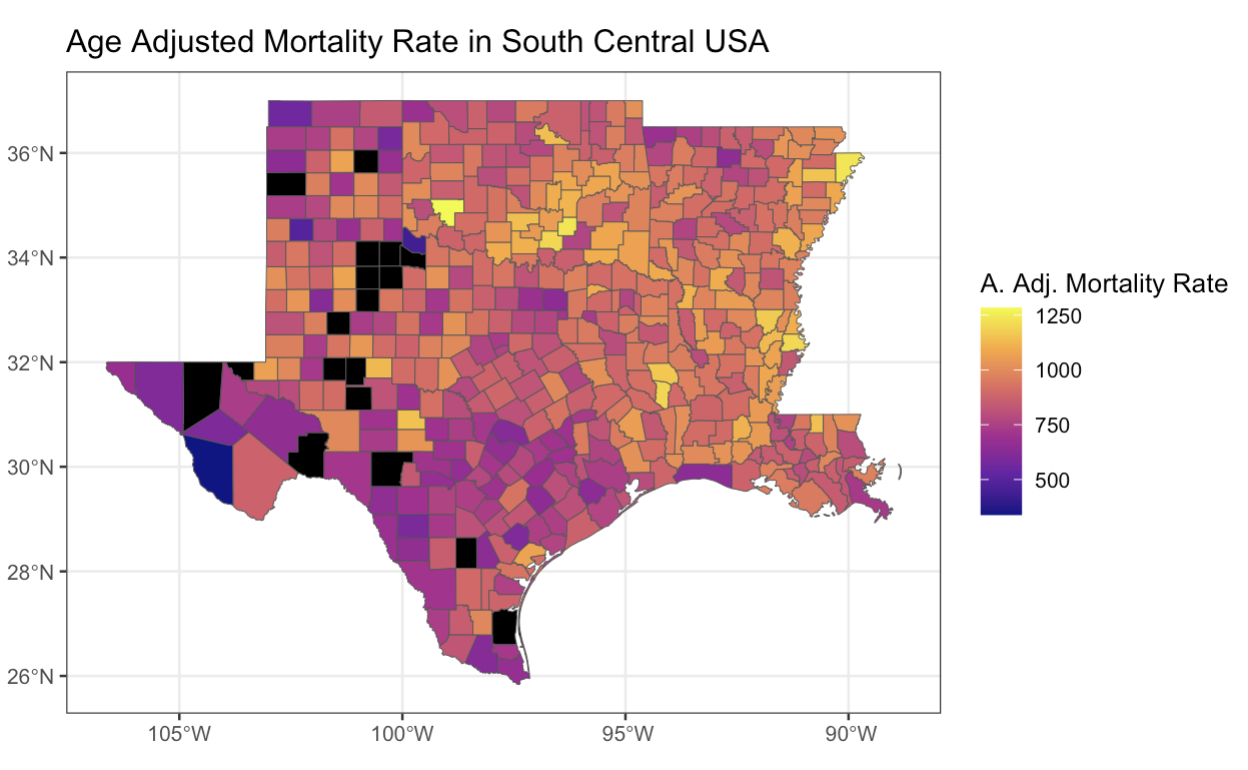}
    \caption{Age-adjusted mortality rates by county in the South-Central USA. Data for low population counties (black) are redacted for privacy.}
    \label{fig:death_rate_socent}
\end{figure}

The health outcome $y_i$ is the annual age-adjusted mortality rates for each county $i$ in the South-Central USA (Arkansas, Louisiana, Oklahoma, Texas) using data from the National Center for Health Statistics. Figure~\ref{fig:death_rate_socent} shows these age-adjusted mortality rates for $n=452$ counties in 2010. In addition, we collected covariate data $\bm x_i$ on each county $i$: median household income ($x_1$), metro/non-metro indicator based on Rural-Urban Continuum Codes ($x_2$), proportion of non-Hispanic Black residents ($x_3$), and proportion of adults without a completed high school education ($x_4$). Data on county's Rural-Urban Continuum Codes for 2013 as the closest year to 2010 was obtained from the Economic Research Service (ERS) at the USDA (\citealp{ers_rur_urb}). County-level education data for the 2008-2012 period was also obtained from ERS. The county median household income and racial composition data for 2010 were obtained from the US Census Bureau. Finally, our covariates also include an indicator of the state ($x_5,x_6, x_7$ are indicators for Louisiana, Oklahoma, and Texas, respectively), which incorporates some spatial information and adjusts for state-specific policies. 

The stage-two model is a log-linear Gaussian regression model that links the  age-adjusted mortality rates  with the stage-one PM$_{2.5}$ exposure estimates from BNE and the known covariates: 
\begin{equation}\label{mort}
    \log(y_i) = \beta_0 + \theta_\zeta\zeta_i + \bm x_i^\top  \bm \beta + \epsilon_i, \quad \epsilon_i \stackrel{iid}{\sim}N(0,\sigma_\epsilon^2), \quad i=1,\ldots,n.
\end{equation}
We use a ridge prior for the coefficients, $(\beta_0, \theta_\zeta, \bm \beta)^\top \sim MVN(\bm 0, \sigma_\theta^2 \bm I_9)$, and learn the shrinkage parameter, $\sigma_\theta \sim U(0, 1000)$.  The prior for the error variance is $\sigma^2_\epsilon \sim IG(0.01, 0.01)$.

The goal is to perform estimation and inference for the PM$_{2.5}$ effect $\theta_\zeta$. We compare the proposed IIS and AIS ($R=500$) approaches with Plug-in ($\hat\zeta$), UKDE, and the partial posterior approach. Since we do not have access to the auxiliary data $\bm z$, there is no analog for Plug-in (Z). Importance sampling is noncompetitive and thus omitted. 
For each method, we use the Gibbs sampling version of Algorithm~\ref{alg:ideal}, which substitutes each method's treatment of $\bm \zeta$ into the first step and adopts the same sampling steps for the remaining parameters $(\beta_0, \theta_\zeta, \bm \beta, \sigma_\epsilon, \sigma_\theta)$ otherwise. Each method uses all of the $S=100$ available draws from the stage-one BNE partial posterior. 
We take 10,000 samples after discarding the first 2,000. After visually inspecting traceplots of individual parameters, we do not see any obvious signs of non-convergence for any of the models. 

\subsection{Results}
\label{sec:5_4_results}

Estimates and inference for the PM$_{2.5}$ effect $\theta_\zeta$ and the error variance $\sigma_\epsilon^2$ are summarized in Figure~\ref{fig:emp_res}. Similar plots for the remaining coefficients $\bm \beta$ are in the supplementary material. First, there is broad agreement among nearly every method: there is a significant, adverse effect of PM$_{2.5}$  on age-adjusted mortality  (95\% credible intervals are above zero). The lone exception is the partial posterior approach: it reports intervals for $\theta_\zeta$ that are excessively wide and include zero. Notably, AIS and Plug-in ($\hat\zeta$) are nearly identical, with slightly smaller intervals for AIS. Comparatively, the point estimates for IIS, UKDE, and partial posterior are all slightly closer to zero. Finally, the estimates and inferences for $\sigma_\epsilon^2$ are mostly in agreement across the competing methods, with the notable exception of the partial posterior approach and its exceedingly large estimate of $\sigma_\epsilon^2$.

\begin{figure}[h]
\begin{subfigure}{0.48\textwidth}
\includegraphics[width=\linewidth]{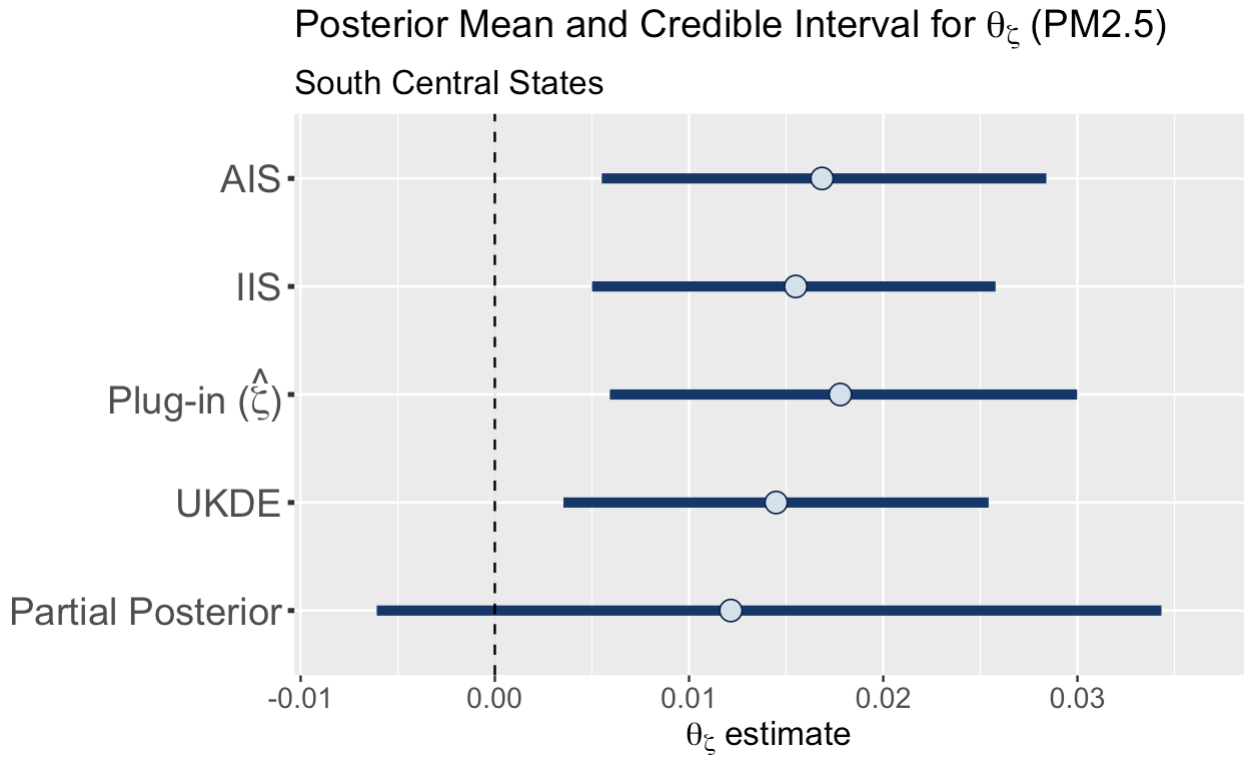}
\end{subfigure}
\begin{subfigure}{0.48\textwidth}
\includegraphics[width=\linewidth]{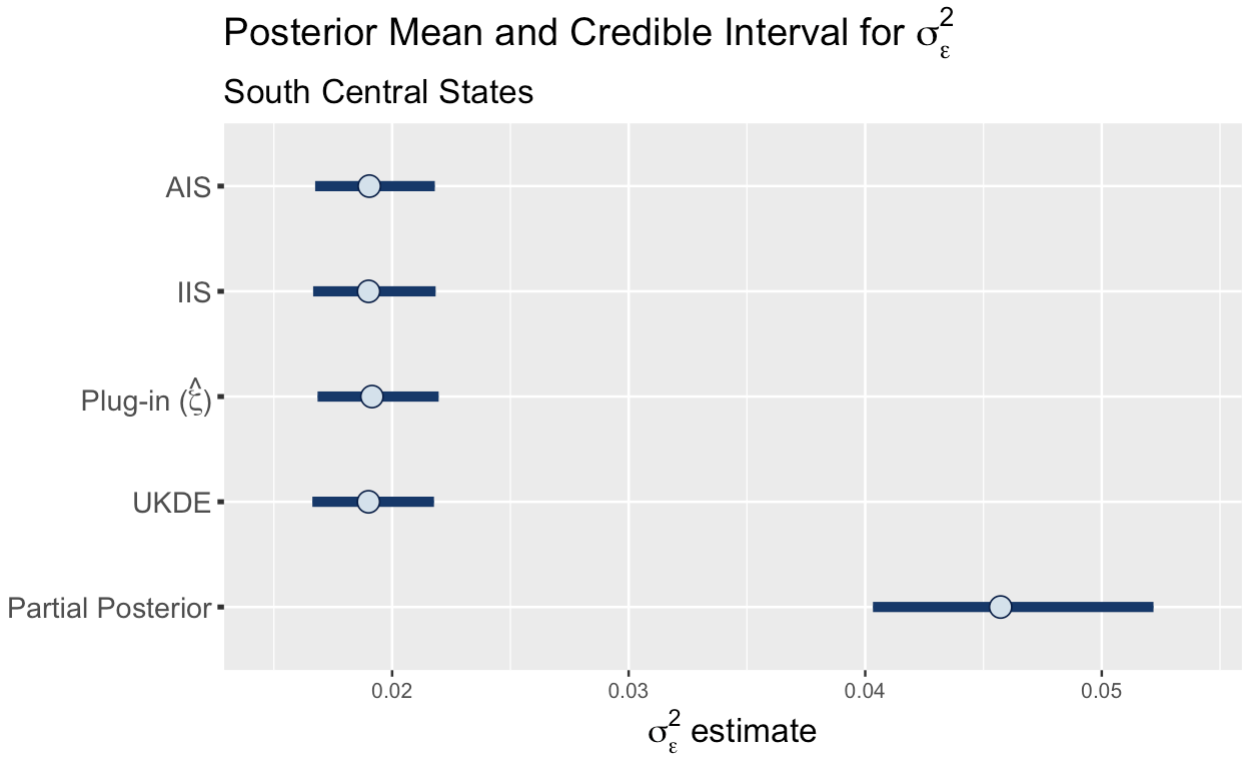}
\end{subfigure}
\caption{Posterior means (points) and 95\% credible intervals (lines) for the PM$_{2.5}$ effect $\theta_\zeta$ (left) and $\sigma_\epsilon^2$ (right) for each approach. Estimates and inference for AIS and Plug-in ($\hat\zeta$) are nearly identical, while the other methods estimate slightly smaller effects. The partial posterior intervals for $\theta_\zeta$ are excessively wide and include zero (marked by the dashed black line) and its $\sigma_\epsilon^2$ estimates are far larger than every other method.}
\label{fig:emp_res}
\end{figure}

Compared to our simulation results (Section~\ref{sec:4_Sim}), we find points of  agreement and disagreement. First, the partial posterior approach indeed appears to suffer from miscalibrated inference, poor estimation of $\sigma_\epsilon^2$, and perhaps some attenuation bias. Thus, this intuitive approach to ``propagate" uncertainty from the stage-one model into the stage-two model is misguided. However, we do not see the overconservative interval estimates for $\theta_\zeta$ (Figure~\ref{fig:1}, bottom) or the poorly-estimated $\sigma_\epsilon^2$ (Figure~\ref{fig:sigma_sim}) for Plug-in ($\hat\zeta$). Similarly for UKDE: 
we do not see the bias (Figure~\ref{fig:1}, top right), miscalibrated inference (Figure~\ref{fig:1}, bottom right), or underestimated error variance (Figure~\ref{fig:sigma_sim}). One potential explanation is that the magnitude of the PM$_{2.5}$ effect is smaller here than in the simulation studies. Notably, this term appears prominently in both the bias \eqref{bias} and the conditional variance \eqref{variance}, and the gaps between these quantities and their ground truth values disappear as $\vert \theta_\zeta\vert \to 0$. 

\subsection{Hybrid Simulated Data}
\label{sec:5_5_hybrid}

To further investigate how the behavior of these competing methods depends on the signal strength $\vert \theta_\zeta\vert$---while maintaining the features of the real data as much as possible---we construct a hybrid simulated dataset. First, we construct draw-by-draw residuals 
\[
    \epsilon_i^{t} = \log(y_i) - (\beta_0^{t} + \theta_\zeta^{t}\zeta_i^{t} + \bm x_i^\top  \bm \beta^t)
\]
where ($\beta_0^t, \theta_\zeta^t, \bm \zeta^t, \bm \beta^t$) are drawn from the joint \emph{full} data posterior distribution for $t=1,\ldots,T$. Because of its strong performance in all settings (Section~\ref{sec:4_Sim}), we use the draws from AIS. Next, we generate synthetic outcome data according to \eqref{mort}, but instead using 1) the residuals $\{\epsilon_i^t\}$ in place of the Gaussian errors $\epsilon_i$, 2) a chosen value $\theta_\zeta^*$ of the PM$_{2.5}$ effect, and 3) the previously-sampled values of the other parameters $(\beta_0^t, \bm \zeta^t, \bm \beta^t)$: 
\begin{align}
    \log(\tilde y_i^{t}) &=  \beta_0^t + \theta_\zeta^* \zeta_i^t + \bm x_i^\top \bm \beta^t + \epsilon_i^t \\ \label{eq-hybrid}
    &= \log(y_i) + (\theta_\zeta^{*} - \theta_\zeta^{t})\zeta_i^{t}
\end{align}
We consider larger values $\theta_\zeta^* \in \{0.05, 0.2\}$ for the PM$_{2.5}$ effects compared to those estimated from real data  (Figure~\ref{fig:emp_res}). 

\begin{figure}
\begin{subfigure}{0.45\textwidth}
\includegraphics[width=\linewidth]{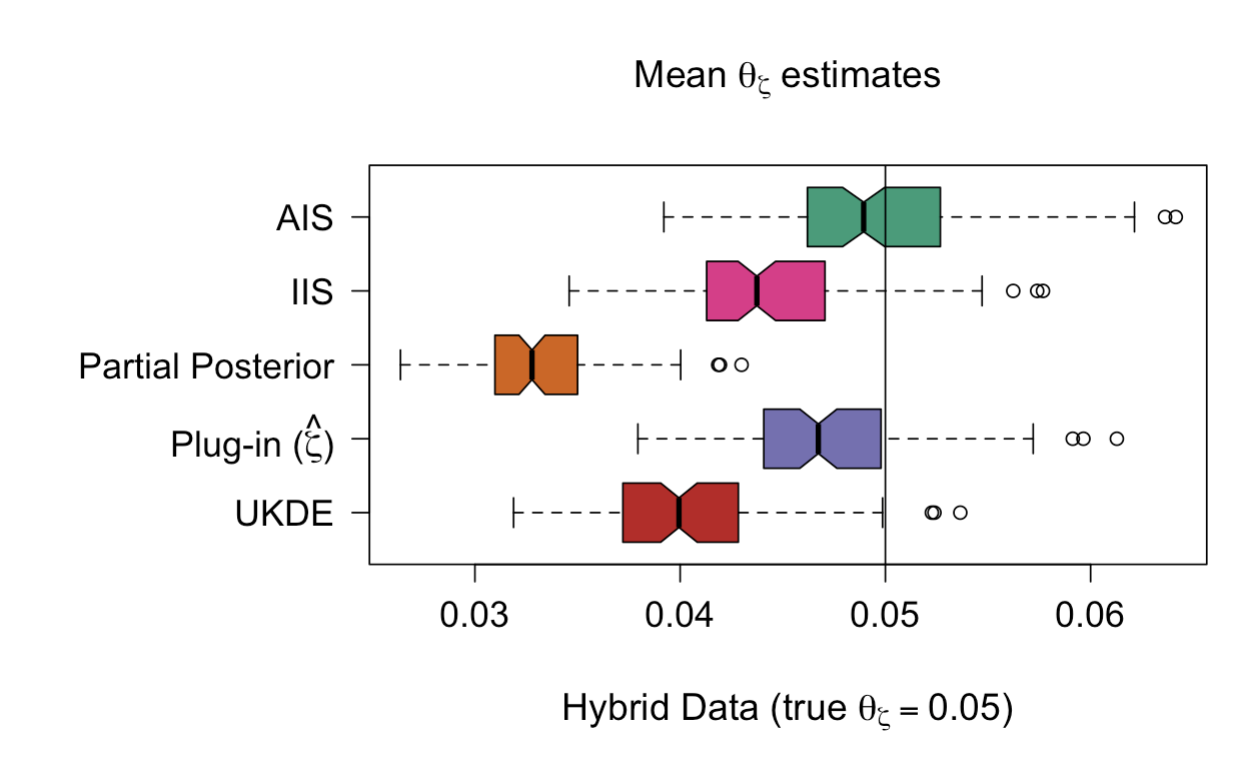}
\end{subfigure}
\begin{subfigure}{0.45\textwidth}
\includegraphics[width=\linewidth]{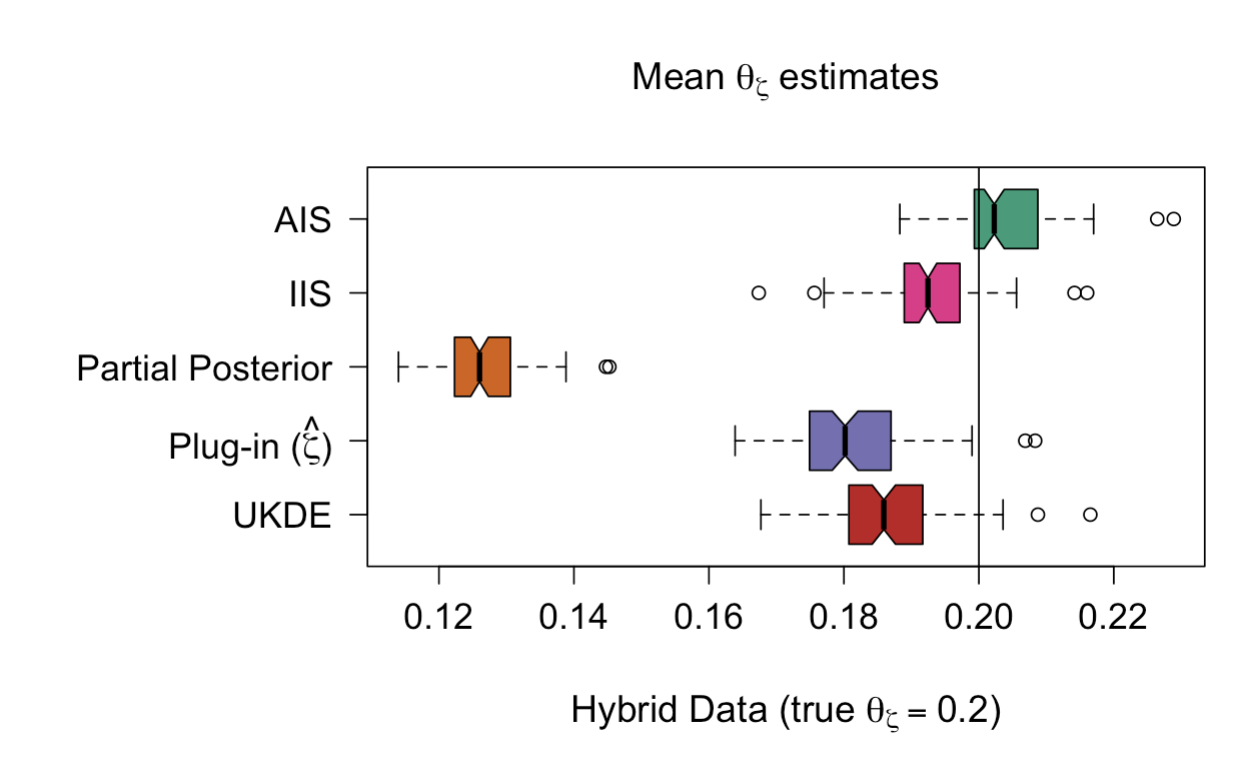}
\end{subfigure}
\begin{subfigure}{0.5\textwidth}
\includegraphics[width=\linewidth]{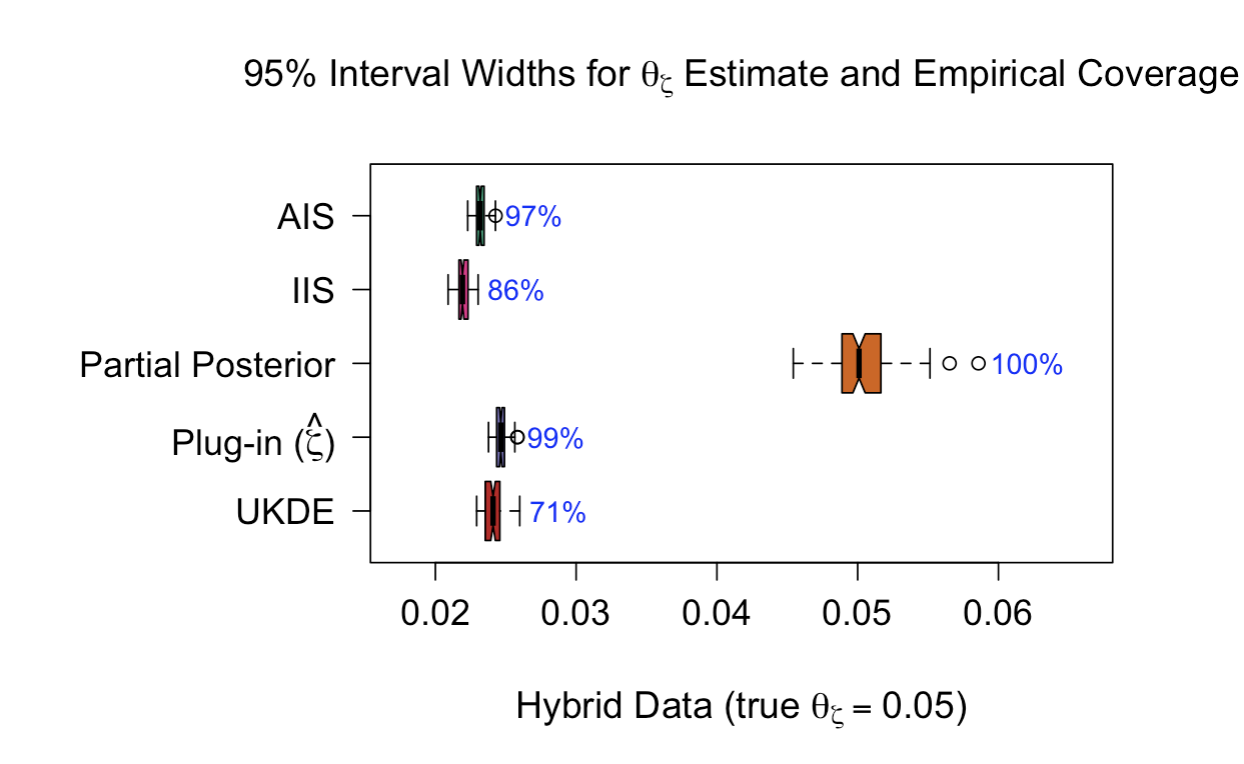}
\end{subfigure}
\begin{subfigure}{0.5\textwidth}
\includegraphics[width=\linewidth]{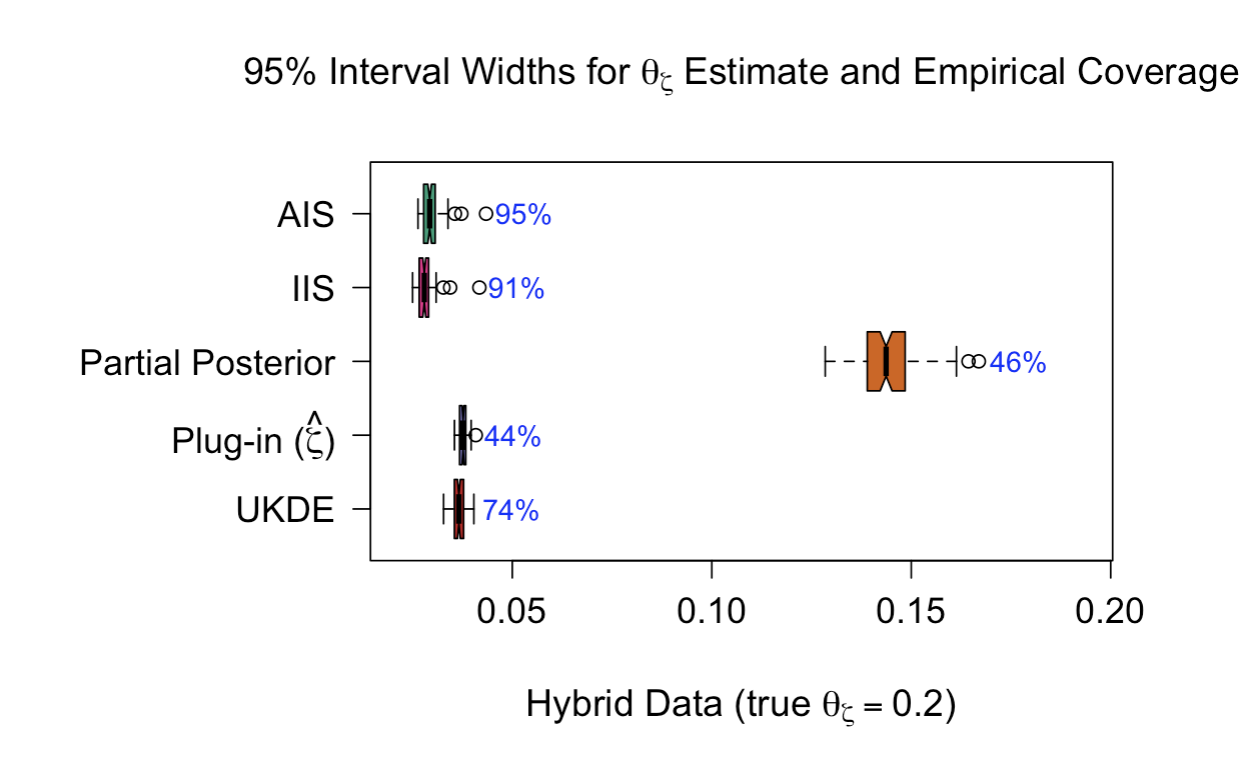}
\end{subfigure}
\caption{Posterior means (top) and 95$\%$ credible interval widths and coverage (bottom) for $\theta_\zeta$ when $\theta_\zeta^* = 0.05$ (left) and $\theta_\zeta^* = 0.2$ (right), with $\theta_\zeta^*$ marked on the plots with a solid black line. As $\vert\theta_\zeta^*\vert$ increases, the attenuation bias for competing methods is larger and the partial posterior intervals become comparatively wider.}
\label{fig:hybrid_theta100}
\end{figure}

 
 In Figure~\ref{fig:hybrid_theta100}, we summarize the posterior means and credible intervals for $\theta_\zeta$ for each method across 100 hybrid simulated datasets $\{\tilde y_i^t\}_{i=1}^n$. 
 As $\vert \theta_\zeta^*\vert$ grows, the attenuation biases become more clear, with greater separation among the methods (top). AIS maintains accurate estimation and precise inference regardless of $\vert \theta_\zeta^*\vert$, while all other competing methods are biased toward zero. The implications for uncertainty quantification are significant (bottom): AIS achieves the nominal coverage with narrow intervals, while competing methods suffer from substantial undercoverage as  $\vert \theta_\zeta^*\vert$ increases. 

\begin{figure}
\begin{subfigure}{0.48\textwidth}
\includegraphics[width=\linewidth]{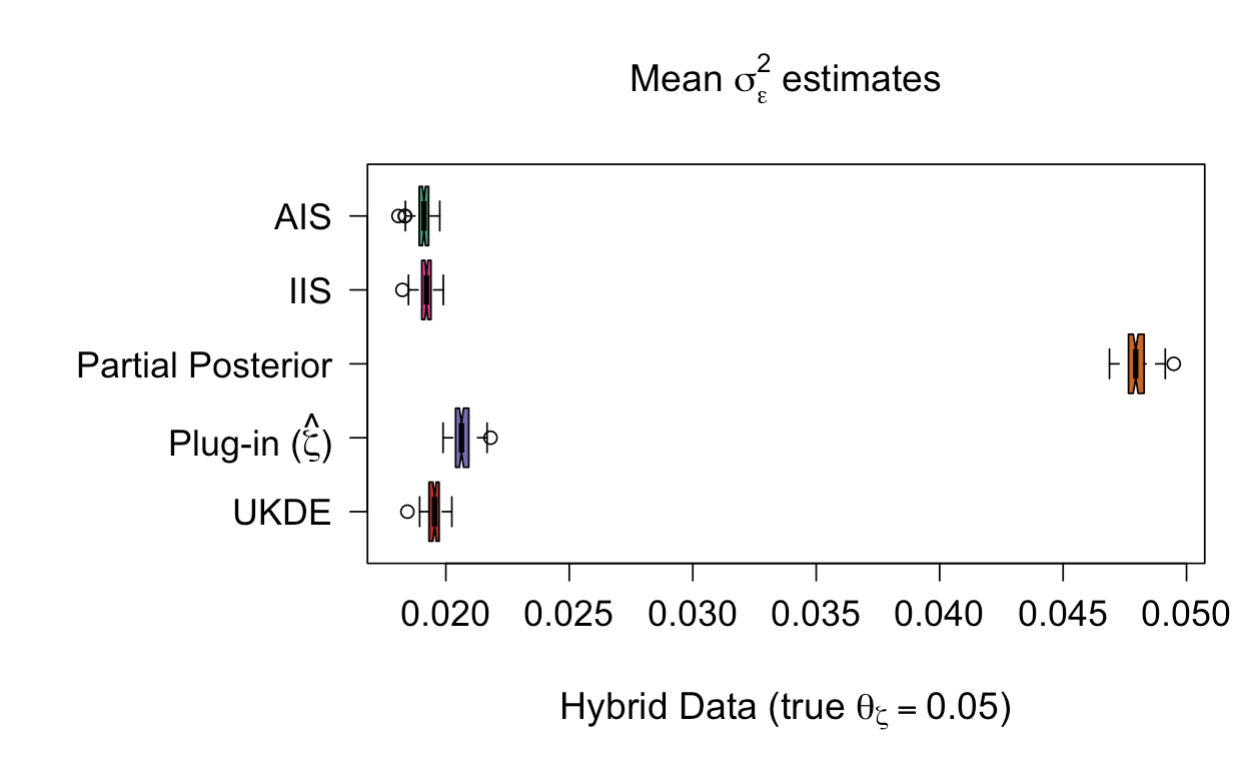}
\end{subfigure}
\begin{subfigure}{0.5\textwidth}
\includegraphics[width=\linewidth]{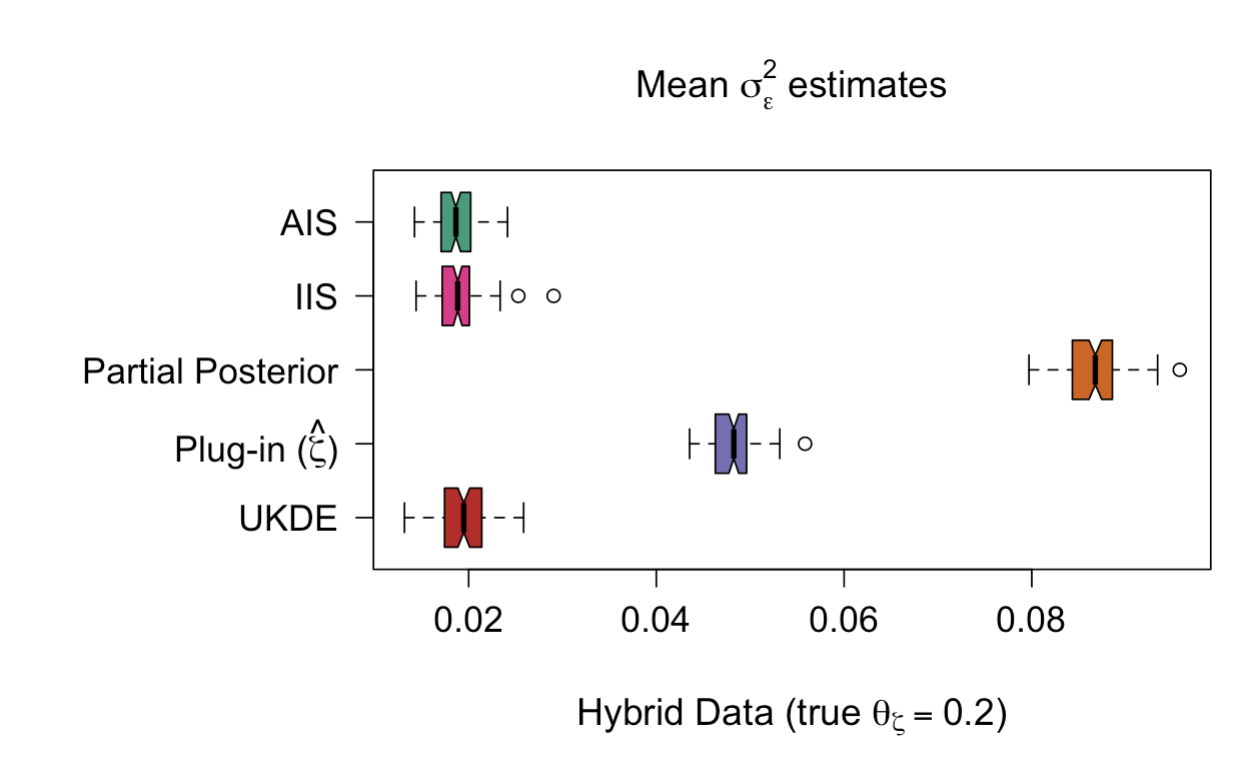}
\end{subfigure}
\caption{Posterior means and 95$\%$ credible intervals for $\sigma^2_\epsilon$ when $\theta_\zeta^* = 0.05$ (left) and $\theta_\zeta^* = 0.2$ (right). These results more closely resemble the patterns from the simulated data (Figure~\ref{fig:sigma_sim}).}
\label{fig:hybrid_sigma100}
\end{figure}

Finally, Figure~\ref{fig:hybrid_sigma100} shows posterior means for $\sigma^2_\epsilon$. The patterns from the simulation study emerge as $\vert \theta_\zeta^*\vert$ increases: Plug-in ($\hat\zeta$) and the partial posterior approach provide much larger estimates of $\sigma_\epsilon^2$. In both cases, AIS and IIS perform similarly. Notably, the true error variance does not change across these simulation settings, so stability is preferred.

\section{Discussion}
\label{sec:6_Disc}
We introduced algorithmic strategies for efficient and joint Bayesian inference with two-stage models. We focused on environmental epidemiology applications, where the inputs to a health outcome (stage-two) model require sophisticated (stage-one) modeling and estimation, especially for air pollution exposures. Stage-one models are typically estimated separately without consideration of any particular downstream analysis, which makes joint Bayesian inference challenging. We detailed how two popular approaches---using point estimates or propagating uncertainty from the stage-one model without feedback from the stage-two model---are statistically flawed, with biases and miscalibrated inferences. We proposed two solutions by modifying importance sampling: one approach (IIS) designed for independent stage-one models, and a corrected version (AIS) for more general settings. We verified the substantial empirical gains offered by these approaches using simulated data.  Finally, we documented the agreements and disagreements among competing approaches in the analysis of county-level mortality rates and PM$_{2.5}$ exposures, and found broad agreement for a significant positive (adverse) effect.

Because of its general applicability and strong performance across many different settings, we recommend AIS as a default approach. However, if we are exclusively interested in point estimation rather than inference or prediction, fixing the air pollution exposures at their stage-one model estimates (Plug-in ($\hat\zeta$)) appears to be adequate. In other contexts, we may have less information than the partial posterior draws \eqref{eq:ppost}, for example, only the (marginal) estimates and standard deviations from the stage-one model. In that case, IIS or UKDE may be preferred, since there is little information about the stage-one posterior dependencies. 



There are many promising directions for future work. First, the two-stage model setting is becoming increasingly common, with machine learning or deep learning models pre-trained on auxiliary data and subsequently fed as inputs into various downstream modeling tasks. Like in our setting, it is untenable to re-fit such complex models for each downstream model, and may not be plausible due to restricted access to  auxiliary training data. Our algorithms (AIS and IIS) may be adapted and applied to such problems. Second, while we focused on log-linear Gaussian models, our methods apply for more general stage-two models, including nonlinear, non-Gaussian, and non-iid cases. Finally, AIS used a simple normal approximation for the ratio of a joint density to a product of its marginals. Alternative estimation strategies, such as nonparametric copula models \citep{feldman_nonparametric_2024}, may further improve the performance of AIS.

\section{Software}
\label{sec5}

Software in the form of \texttt{R} code, together with a sample
input data set and complete documentation is available on
request from the corresponding author (konstantin.larin@rice.edu).

\section*{Acknowledgments}

The authors thank Marianthi-Anna Kioumourtzoglou, Robbie Parks, Jaime Benavides, Vijay Kumar, and Michael Boehnke for providing access to the PM$_{2.5}$ estimates from their BNE model.

\section*{Funding}
Kowal gratefully acknowledges funding from the National Science Foundation under award number SES-2435310.

\section*{Conflict of Interest}

None declared.

\bibliographystyle{biorefs}
\bibliography{references}

\end{document}


\title{Supplement to ``Efficient Bayesian inference for two-stage models in environmental epidemiology"}

\author{Konstantin Larin$^{1,\ast}$, Daniel R. Kowal$^2$ \\[4pt]
\text{$^1$Department of Statistics, 
Rice University,
6100 Main St, Houston, TX 77005, United States} \\
\text{
$^2$Department of Statistics and Data Science, Cornell University,
127 Hoy Rd, Ithaca, NY 14853, United States}
\\[2pt]
{konstantin.larin@rice.edu}}

\markboth%
{K. Larin and D.R. Kowal}
{Efficient Bayesian inference for two-stage models}

\maketitle

\footnotetext{To whom correspondence should be addressed.}
\appendix

Supplementary materials include full derivations of the results in Section 2.2 (Section \ref{sec:app_insuf}),  a visualized example of weight degeneracy (Section \ref{sec:app_imsam}), plots of root mean squared error (RMSE) for results in Section 4 (Section \ref{sec:app_rmse}), additional simulation results (Section \ref{sec:app_sim}), additional analyses of the South-Central United States mortality data (Section \ref{sec:app_other_theta_emp}), and additional results for hybrid simulated data (Section \ref{sec:app_hybrid_mean}).

\section{Full Derivation for Equations in Section 2.2}
\label{sec:app_insuf}

In this section we present full derivations for the analytical results from Section 2.2 for the estimates of $\theta_\zeta$ and $\sigma^2_\epsilon$ across different methods in our tractable two-stage model setting. 

We start by restating the model setting from (2.6) and (2.7). Let the stage-one model be a Gaussian measurement error model,
\begin{equation}
      \bm z = \bm \zeta + \bm u, \quad \bm u \sim  MVN(\bm 0, \bm \Sigma_u)\label{eq:slr_one}
\end{equation}      
where each vector is $n$-dimensional, $\bm \Sigma_u$ is the $n \times n$ error covariance matrix, and $\bm \zeta \sim MVN(\bm 0, \bm \Sigma_\zeta)$ is the prior. 
Let the stage-two model be a Gaussian linear model,
\begin{equation}\label{eq:slr_two}
    y_i = \beta_0 + \theta_\zeta\zeta_i + \epsilon_i, \quad \epsilon_i \stackrel{iid}{\sim}N(0,\sigma_\epsilon^2), \quad i=1,\ldots,n
\end{equation}
For simplicity, we omit known covariates $\bm x$ and assume that $\bm \Sigma_u = \sigma_u^2 \bm I_n$ and $\bm \Sigma_\zeta = \sigma_\zeta^2 \bm I_n$. 

We look at analytical results for three approaches.  Plug-in (Z) takes $\bm{\hat \zeta} = \bm z$, using the corrupted observations in place of  $\bm \zeta$. Plug-in ($\hat\zeta$) takes $\hat{\bm\zeta}=\mathbb{E}(\bm\zeta\mid \bm z) = \lambda \bm z$, which corresponds the the partial posterior mean.  Finally, Partial Posterior method substitutes partial posterior draws $[\bm \zeta \mid \bm z]$, which have the distribution $[\bm \zeta \mid \bm z] \sim MVN(\lambda \bm z, \bm \Sigma_{\zeta \mid z})$, where $\lambda = \sigma_\zeta^2/(\sigma_\zeta^2 + \sigma_u^2)$ and $\bm \Sigma_{\zeta \mid z} = \lambda \sigma_u^2 \bm I_n$.

 First, we analyze the implied regression coefficients in each case given the different ``covariates" used in place of $\bm\zeta$. Specifically, the regression coefficient is defined as the ratio of the covariance between $y$ and that covariate to the variance of that covariate.  If we use this substitution for Plug-in (Z), we see that the estimand is
 \begin{align*}
    \theta_{z} &\coloneq \frac{\mbox{Cov}( y,z)}{\mbox{Var}( z)}\\
  &= \frac{\theta_\zeta\sigma^2_\zeta}{\sigma^2_\zeta + \sigma^2_u} \\
  &= \lambda\theta_\zeta.
\end{align*}
Similarly, for Plug-in ($\hat\zeta$), we use the fact that $\hat\zeta = \lambda z$ to calculate that the estimand is
\begin{align*}
    \theta_{\hat{\zeta}} &\coloneq\frac{\mbox{Cov}(y, \lambda z)}{\mbox{Var}(\lambda z)}\\ 
    &= \frac{\lambda\theta_\zeta\sigma^2_\zeta}{\lambda^2\sigma^2_\zeta + \lambda^2\sigma^2_u } \\
    &= \frac{\theta_\zeta\sigma^2_\zeta}{\{\sigma_\zeta^2/(\sigma_\zeta^2 + \sigma_u^2)\}(\sigma_\zeta^2 + \sigma_u^2)}\\
    &= \theta_\zeta
\end{align*}
showing that the estimand corresponds to the ground truth with no attenuation bias. 

For the Partial Posterior method, we use the definition of $\lambda= \sigma_\zeta^2/(\sigma_\zeta^2 + \sigma_u^2)$ to see that the Partial Posterior estimand has the same attenuation bias as the results of the Plug-in (Z) method:
\begin{align*}
    \theta_{\zeta\mid z} &\coloneq \frac{\mbox{Cov}( y,[\zeta\mid  z])}{\mbox{Var}( \zeta\mid  z)}\\
  &= \frac{\lambda\theta_\zeta\sigma^2_\zeta}{\lambda^2\sigma^2_\zeta + \lambda^2\sigma^2_u + \lambda\sigma^2_u} \\
    &= \frac{\theta_\zeta\sigma^2_\zeta}{\frac{\sigma^2_\zeta(\sigma^2_\zeta + \sigma^2_u) }{\sigma^2_\zeta + \sigma^2_u}+\sigma^2_u} \\
    &= \lambda \theta_\zeta.
     \end{align*}

 

Next, we obtain estimands for the conditional variance of $y$ given the corresponding covariate for each of the three methods. Specifically, we first use the law of total variance, $Var(Y) = E[Var(Y|\zeta)]+ Var(E[Y|\zeta])$, and the formula for the variance of the conditional expectation for simple linear models, $ Var(E[Y|\zeta]) = Var(\beta_0 +\theta_\zeta\zeta) = \theta_\zeta^2Var(\zeta)$,  to define the conditional variance in this case as the difference between the marginal variance Var($y$) and the product of the square of the slope coefficient and variance of the covariate.  For Plug-in (Z), we use our previous derivation for $\theta_z$ to show that this formula simplifies to
\begin{align*}
    \mbox{Var}( y \mid z) &\coloneq \mbox{Var}( y) - \theta_z^2\mbox{Var}(z) \\
    &= \mbox{Var}( y) - \frac{\mbox{Cov}( y, z)^2}{\mbox{Var}( z)}\\
  &= \theta^2_\zeta\sigma^2_\zeta + \sigma^2_\epsilon - \frac{\theta^2_\zeta (\sigma^2_\zeta)^2}{\sigma^2_\zeta + \sigma^2_u}\\
    &= \sigma^2_\epsilon + \theta^2_\zeta ( \frac{\sigma^2_\zeta \sigma^2_u}{\sigma^2_\zeta + \sigma^2_u} ) \\
    &= \sigma^2_\epsilon + \theta^2_\zeta \lambda \sigma^2_u
\end{align*}
demonstrating that conditional variance under the Plug-in (Z) approach includes an additive term $\theta^2_\zeta \lambda \sigma^2_u$ that inflates the variance.

We use an equivalent process to write down a formula for the estimand of conditional variance for the Plug-in ($\hat\zeta$) approach. Then, we simplify: 
\begin{align*}    
    \mbox{Var}( y\mid \hat{\zeta}) &\coloneq \mbox{Var}( y) - \theta_{\hat{\zeta}}^2 \mbox{Var}(\hat{\zeta}) \\
    &=\mbox{Var}( y) - \frac{\mbox{Cov}( y,\lambda z)^2}{\mbox{Var}(\lambda  z)} \\
    &= \theta_\zeta^2\sigma^2_\zeta+ \sigma^2_\epsilon  - \frac{\theta_\zeta^2(\sigma^2_\zeta)^2}{(\sigma^2_\zeta+\sigma^2_u)} \\
    &= \sigma^2_\epsilon + \theta_\zeta^2(\frac{\sigma^2_\zeta(\sigma^2_\zeta+\sigma^2_u-\sigma^2_\zeta)}{\sigma^2_\zeta + \sigma^2_u}) \\
    &=  \sigma^2_\epsilon + \theta_\zeta^2\lambda\sigma^2_u
\end{align*}
using  $\lambda= \sigma_\zeta^2/(\sigma_\zeta^2 + \sigma_u^2)$ and previous results for $\mbox{Var}(\hat{\zeta})$ and $\theta_{\hat{\zeta}}$.  We see that the estimand for conditional variance for Plug-in ($\hat\zeta$) is equal to the estimand for Plug-in (Z).

Finally, we analytically derive results for the conditional variance estimand for the Partial Posterior method, and see that it ends up different from the Plug-in methods:
     \begin{align*}
    \mbox{Var}( y\mid [{ \zeta} \mid  z]) &\coloneq \mbox{Var}( y) - \theta_{\zeta\mid z}^2 \mbox{Var}(\zeta\mid z) \\
    &= \mbox{Var}( y) - \frac{\mbox{Cov}( y, [ \zeta\mid  z])^2}{\mbox{Var}(  \zeta\mid  z)}\\
    &= \theta^2_\zeta\sigma^2_\zeta + \sigma^2_\epsilon - \frac{\theta_\zeta^2\lambda^2(\sigma^2_\zeta)^2}{\lambda^2\sigma^2_\zeta + (\lambda+1)\lambda \sigma^2_u} \\
    &= \sigma^2_\epsilon + \theta^2_\zeta\left\{\sigma^2_\zeta - \frac{\lambda (\sigma^2_\zeta)^2}{\lambda\sigma^2_\zeta + (\lambda+1) \sigma^2_u}\right\} \\
    &= \sigma^2_\epsilon +  \theta^2_\zeta(1+\lambda)\lambda\sigma^2_u
     \end{align*} 
Specifically, the estimand for the conditional variance for the Partial Posterior is even more inflated than for the other methods. Thus, we confirm that the two Plug-in methods lead to the estimands of conditional variance that are inflated by the same amount, while the Partial Posterior approach has even greater variance inflation. In addition, all the additive terms depend on the signal strength $\theta_\zeta$, attenuation factor $\lambda$, and measurement error $\sigma^2_u$.

\section{Example of regular importance sampling weights}
\label{sec:app_imsam}
\begin{figure}
    \centering
    \includegraphics[width=0.6\textwidth]{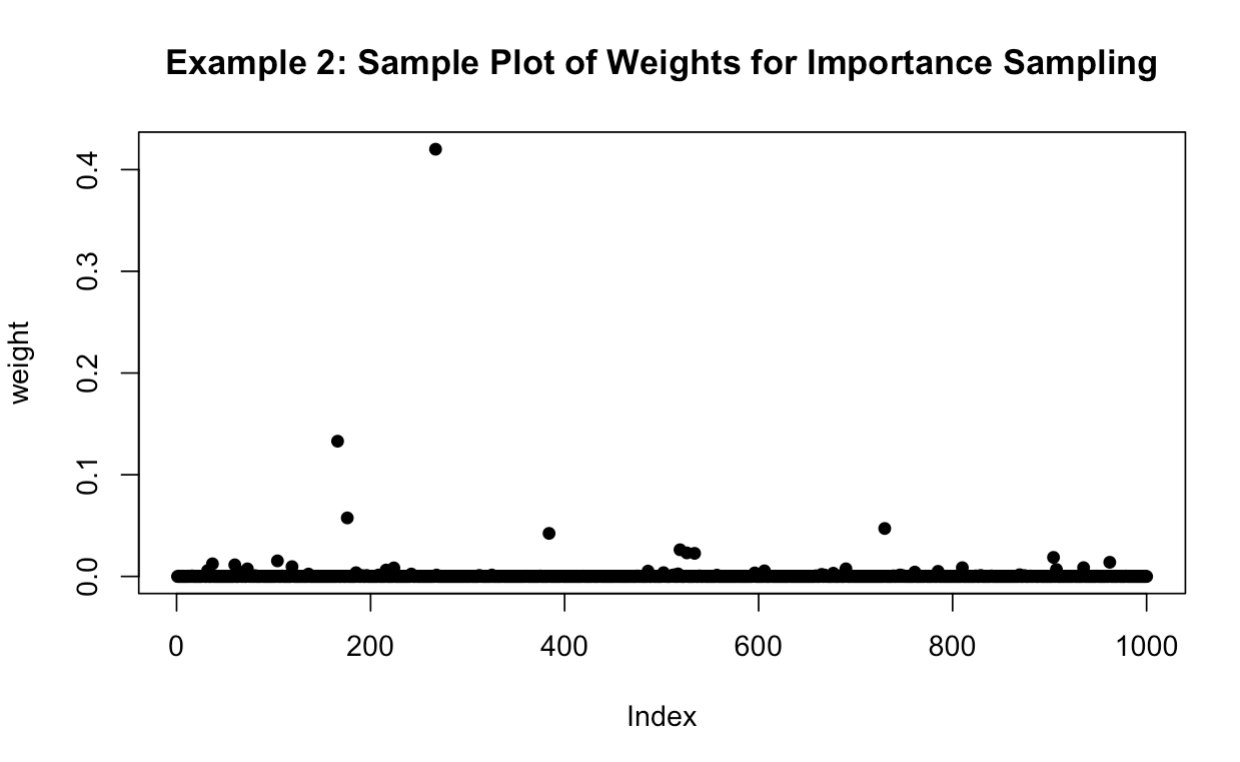}
    \caption{Normalized weights during a single step of the importance sampling algorithm in one of the simulations for Example 2 ($\rho=0.3$). The plot clearly shows weight degeneracy, with a single weight out of 1000 with the value of over 0.4.}
    \label{fig:is_weight}
\end{figure}

We illustrate the weight degeneracy problem that occurs when importance sampling is used for higher dimensional data using one of the simulations from Section 4 as an example. Figure \ref{fig:is_weight} shows the values of normalized weights during one of the steps of the importance sampling algorithm for Example 2 (correlated scenario with $\rho=0.3$). The plot shows that, among 1000 weights, a weight for a single observation has the value equal to almost half of the total weight. A few other observations also have disproportionately large weights, and the remaining observations all have weights near zero. This weight degeneracy means that only a small fraction of the partial posterior draws are viable candidates for SIR sampling, which leads to the poor performance of importance sampling observed in  Section 4.2.

\section[RMSE of y and predicted y-tilde for Examples 1 and 2]{RMSE of \(\bm y\) and out-of-sample \(\bm y\) for Examples 1 and 2}
\label{sec:app_rmse}

\begin{figure}[t!] 
\begin{subfigure}{0.45\textwidth}
\includegraphics[width=\linewidth]{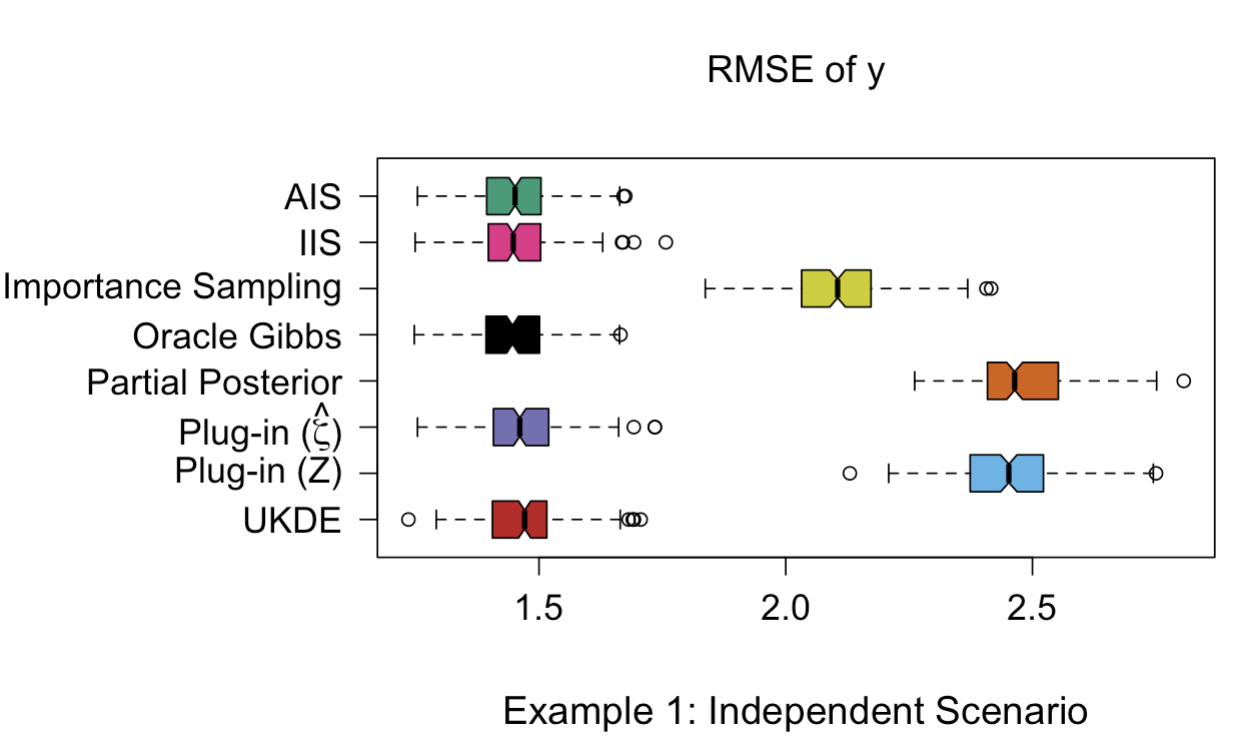}
\end{subfigure}
\begin{subfigure}{0.45\textwidth}
\includegraphics[width=\linewidth]{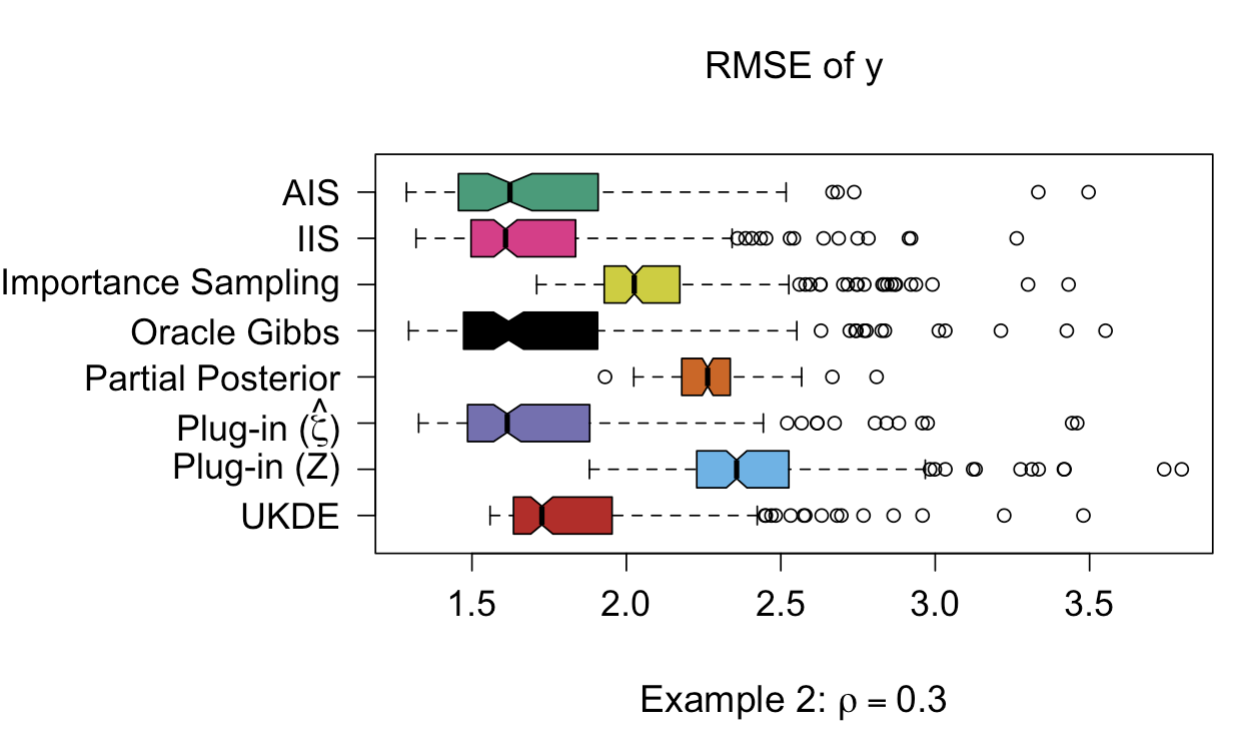}
\end{subfigure}
\caption{ RMSE of $\bm y$ in independent (left) and correlated (right) settings. AIS and IIS perform comparably to Oracle Gibbs in all cases. } \label{fig:rmse}
\end{figure}

We augment the main text by providing additional results from the simulations in Section 4. Figure \ref{fig:rmse} shows root mean square error (RMSE) for $\bm y$ across 100 simulations from Section 4.3 for all methods for Examples 1 and 2. Although there is  more variability in RMSE across simulations for Example 2, comparative performance of different methods stays the same. AIS, IIS, and Plug-in ($\hat\zeta$) perform similarly  with RMSE matching the results of Oracle Gibbs. UKDE performs slightly worse in the correlated case with a slightly higher mean RMSE. Importance sampling, Plug-in(Z), and Partial Posterior all perform substantially worse.

\begin{figure}[t!] 

\begin{subfigure}{0.48\textwidth}
\includegraphics[width=\linewidth]{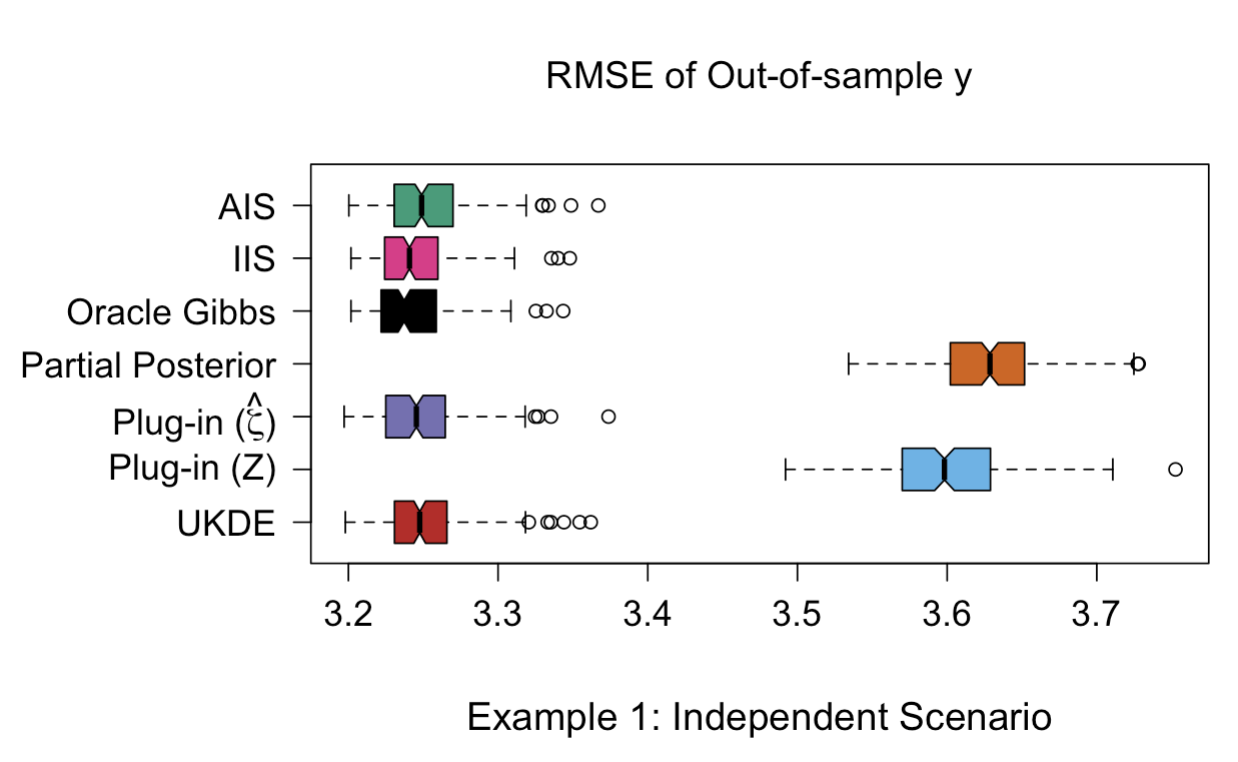}
\end{subfigure}
\begin{subfigure}{0.48\textwidth}
\includegraphics[width=\linewidth]{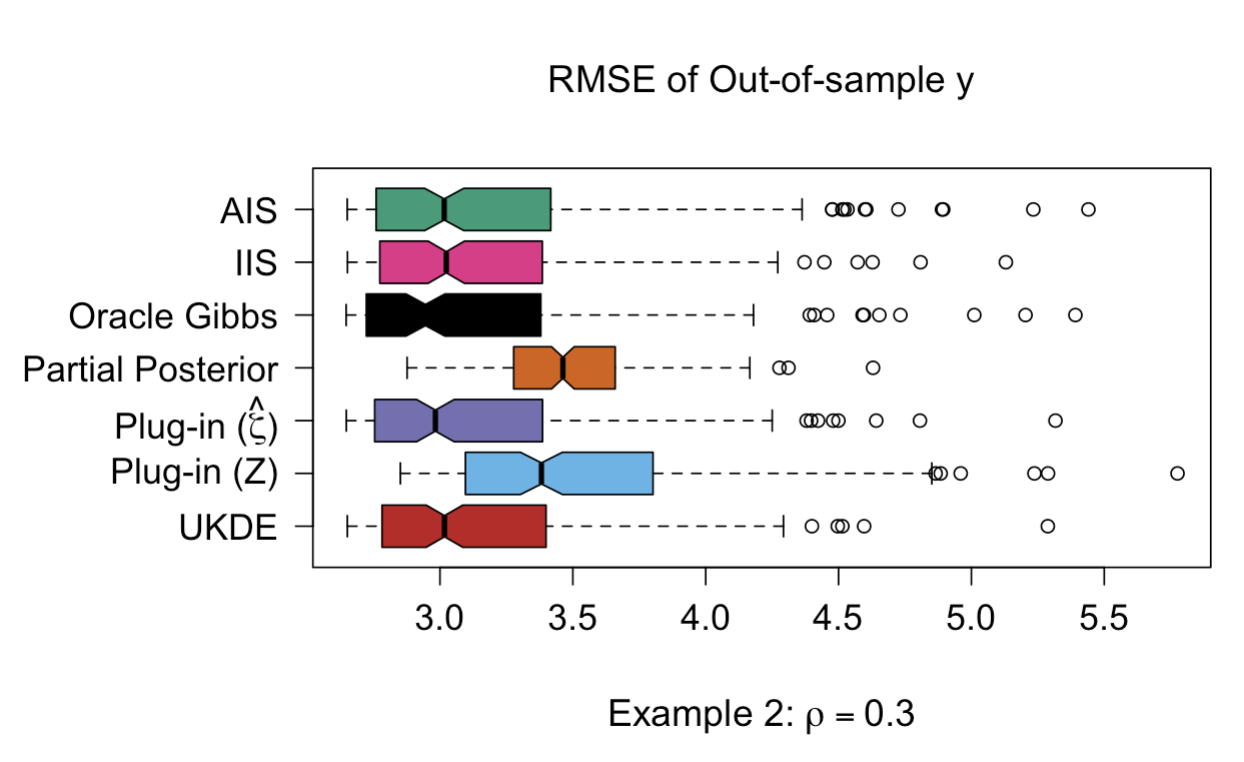}
\end{subfigure}

\caption{The plots shows RMSE for out-of sample $\bm{ y}$ values  in independent (left) and correlated (right) settings. Partial Posterior and Plug-in (Z) perform the worst, but the other methods perform on par with Oracle Gibbs.} \label{fig:pred_rmse}
\end{figure}

Figure \ref{fig:pred_rmse} shows RMSE for out-of-sample $\bm{ y}$ for the simulations from Section 4.4. Importance sampling is not included due to its poor performance in the simulations. While Plug-in (Z) and Partial Posterior have higher RMSE, all the other methods perform similarly.

\newpage
\section{Further Simulation Results}
\label{sec:app_sim}

Below we include results for additional simulation setting that were not included in Section 4.2, with different combinations of $\rho$, $\theta_\zeta$, and $\sigma^2_\zeta$. Importance sampling is once again excluded due to poor performance. We show that the results from Section 4.2 hold across different simulation settings: AIS consistently outperforms all competitors in all scenarios, while IIS is highly competitive  in  the independent scenarios.

\begin{figure}[t!] 
\begin{subfigure}{0.48\textwidth}
\includegraphics[width=\linewidth]{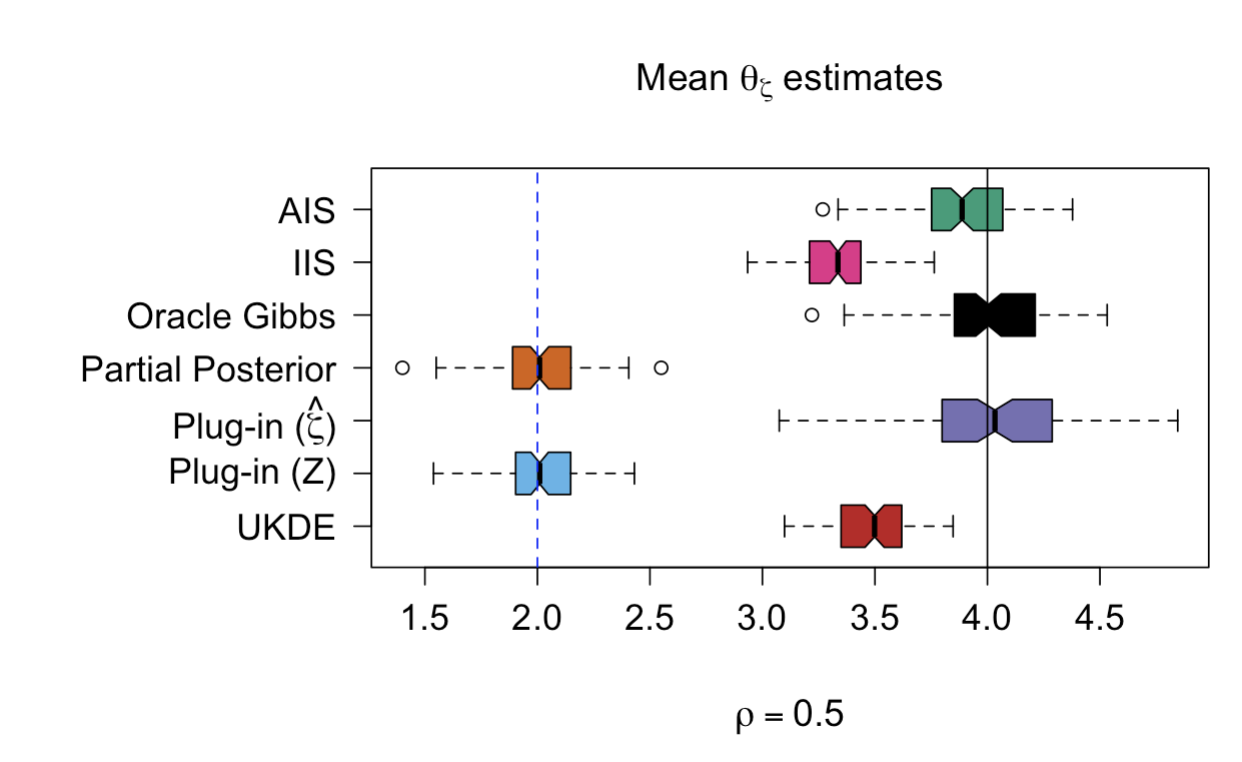}
\end{subfigure}
\begin{subfigure}{0.5\textwidth}
\includegraphics[width=\linewidth]{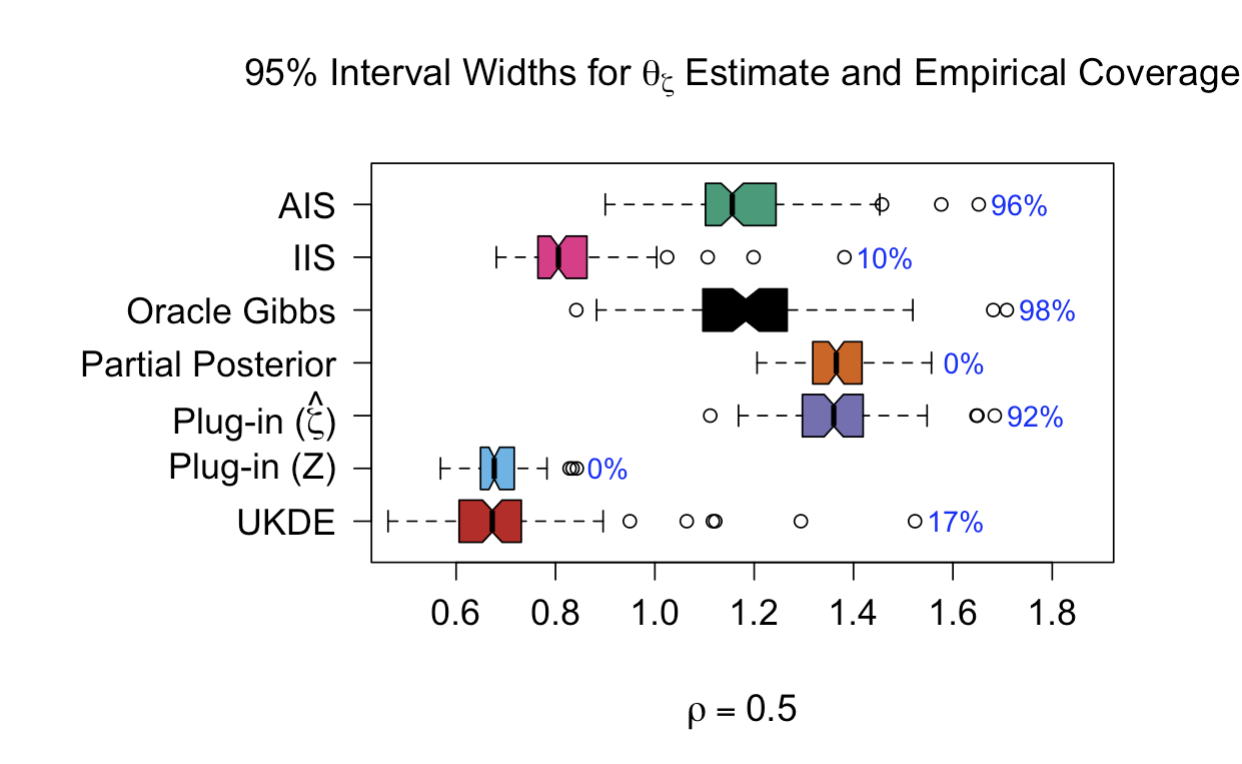}
\end{subfigure}
\caption{Posterior means (top) and 95\% credible interval widths and coverages (bottom) for $\theta_\zeta$ for each method for $\rho=0.5$. For estimation, we mark the true regression coefficient (solid black) and the attenuated version (dashed blue); for inference, the annotations give the empirical coverages. AIS performs comparably with Oracle Gibbs in both cases, with the most accurate interval widths, while IIS and UKDE fall behind. Plug-in ($\hat\zeta$) has good point estimates, but overly large interval width.} \label{fig:theta04-05}
\end{figure}

\begin{figure}[t!] 
\begin{subfigure}{0.48\textwidth}
\includegraphics[width=\linewidth]{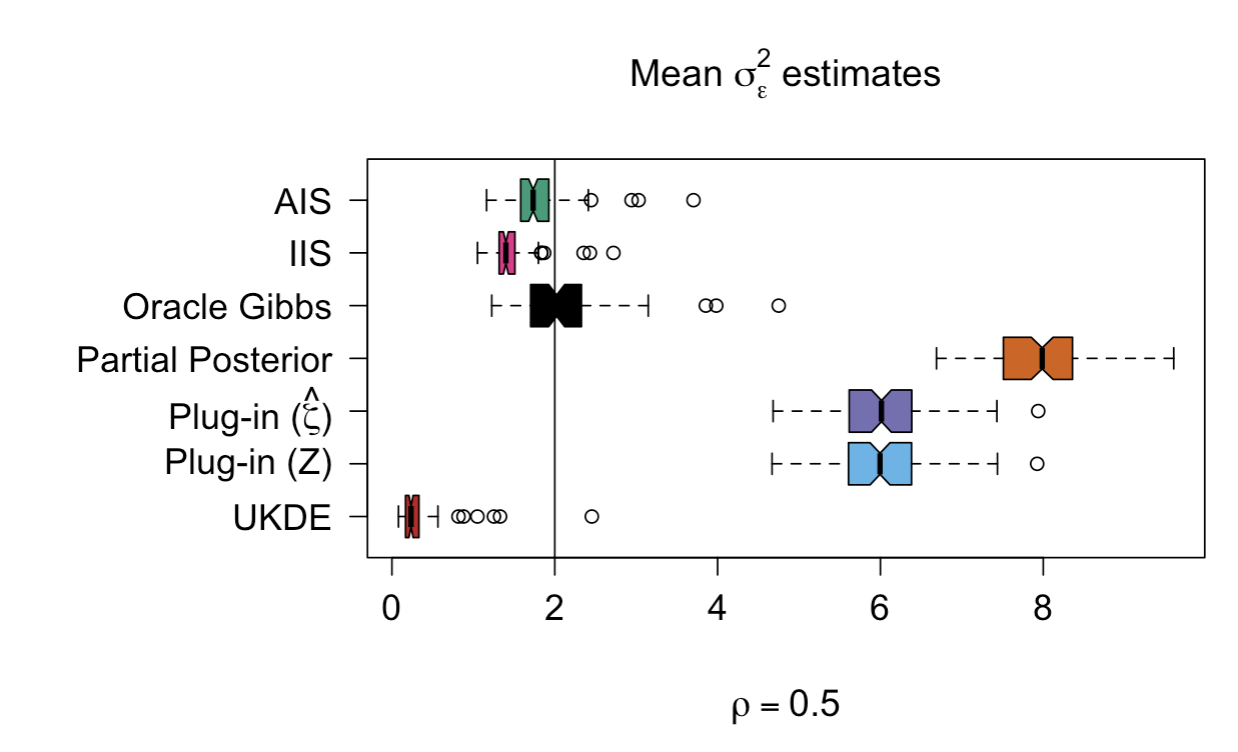}
\end{subfigure}
\caption{Posterior mean estimates for $\sigma^2_\epsilon$ for each method for $\rho=0.5$, with the true variance marked by the solid black line. AIS is accurate and closely matched Oracle Gibbs, while IIS slightly underestimates $\sigma^2_\epsilon$. UKDE greatly undestimates $\sigma^2_\epsilon$, while the other methods overestimate it.} \label{fig:sigma_04-05}
\end{figure}

First, we provide results for models with stronger dependencies within $\bm\zeta\mid \bm z$ than in Example 2, specifically $\rho=0.5$, while all other parameters stay the same as for models in Section 4.2. Figure \ref{fig:theta04-05} shows posterior means (top) and  95\% credible interval widths and coverages (bottom) for $\theta_\zeta$ across all simulations for these two settings. Figure \ref{fig:sigma_04-05} shows posterior means for $\sigma^2_\epsilon$, while Table \ref{Table0405} the 2-Wasserstein distances to the Oracle Gibbs posterior distribution for $\theta_\zeta$ and $\sigma^2_\epsilon$. We see similar results to the $\rho=0.3$ scenario, with AIS continuing to perform the best.

\begin{figure}[t!] 

\begin{subfigure}{0.48\textwidth}
\includegraphics[width=\linewidth]{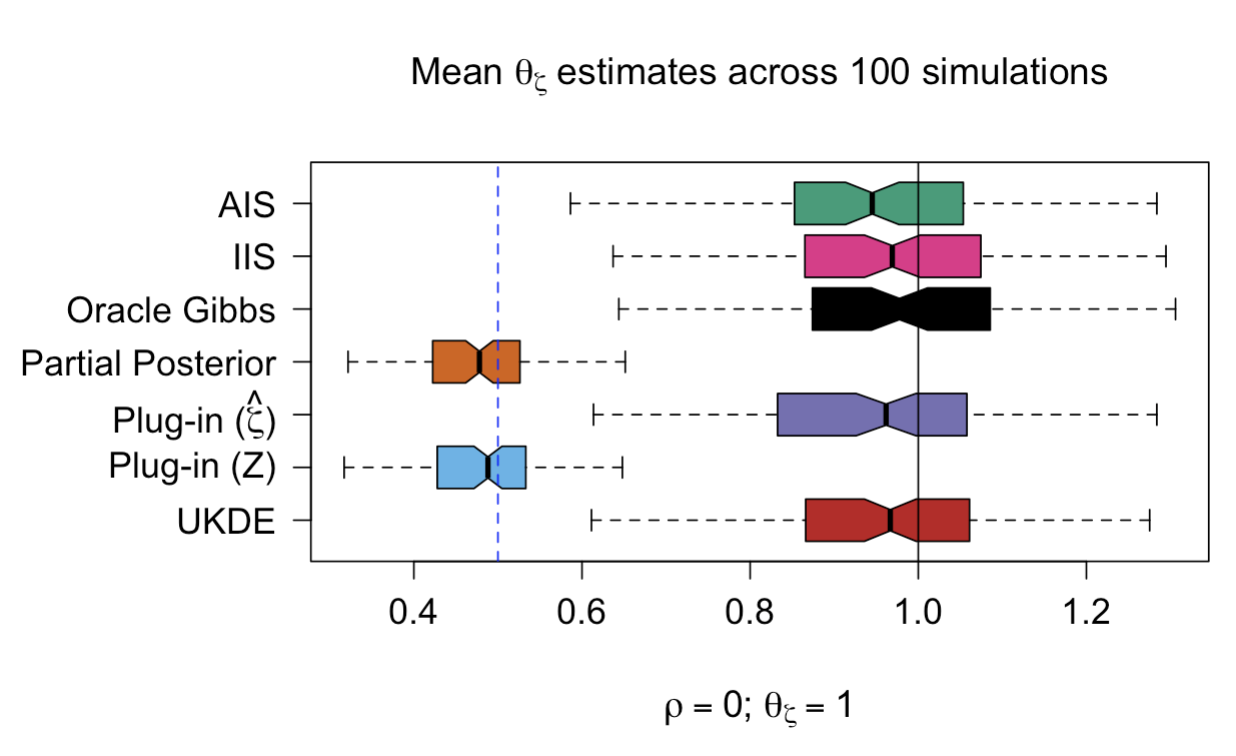}
\end{subfigure}
\begin{subfigure}{0.48\textwidth}
\includegraphics[width=\linewidth]{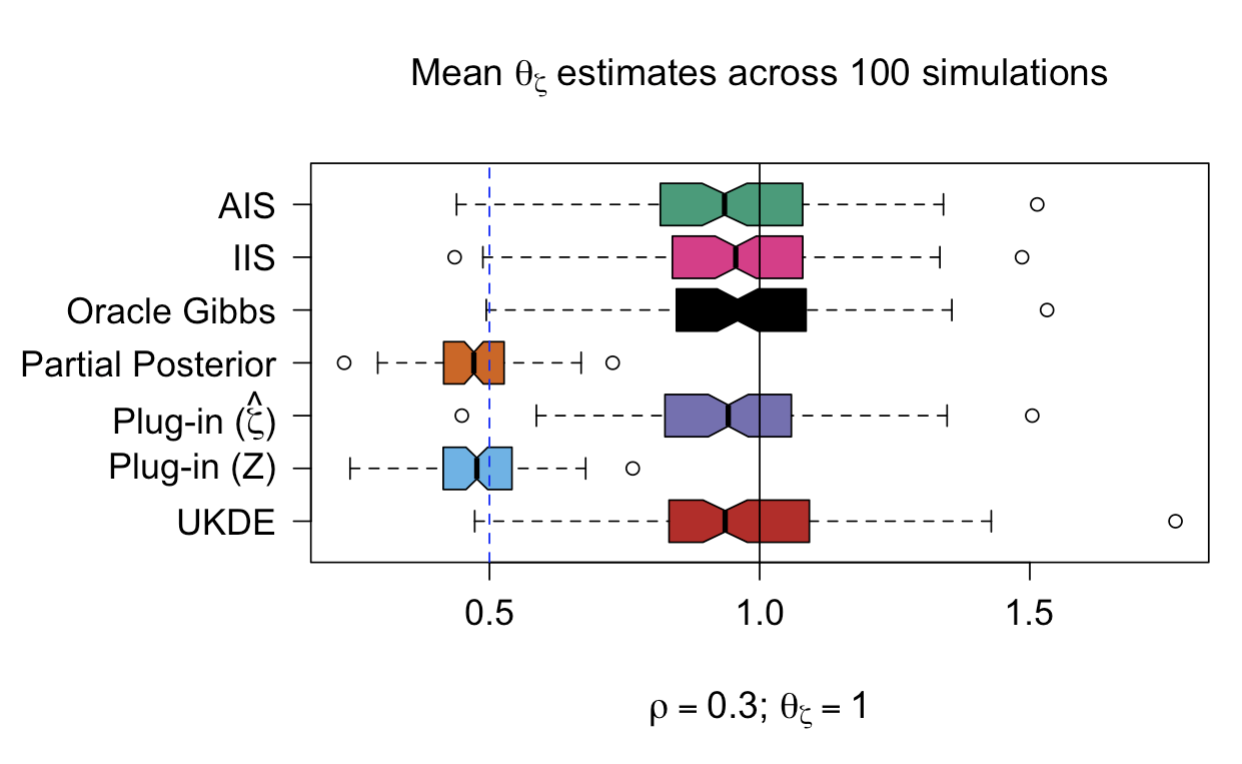}
\end{subfigure}
\begin{subfigure}{0.5\textwidth}
\includegraphics[width=\linewidth]{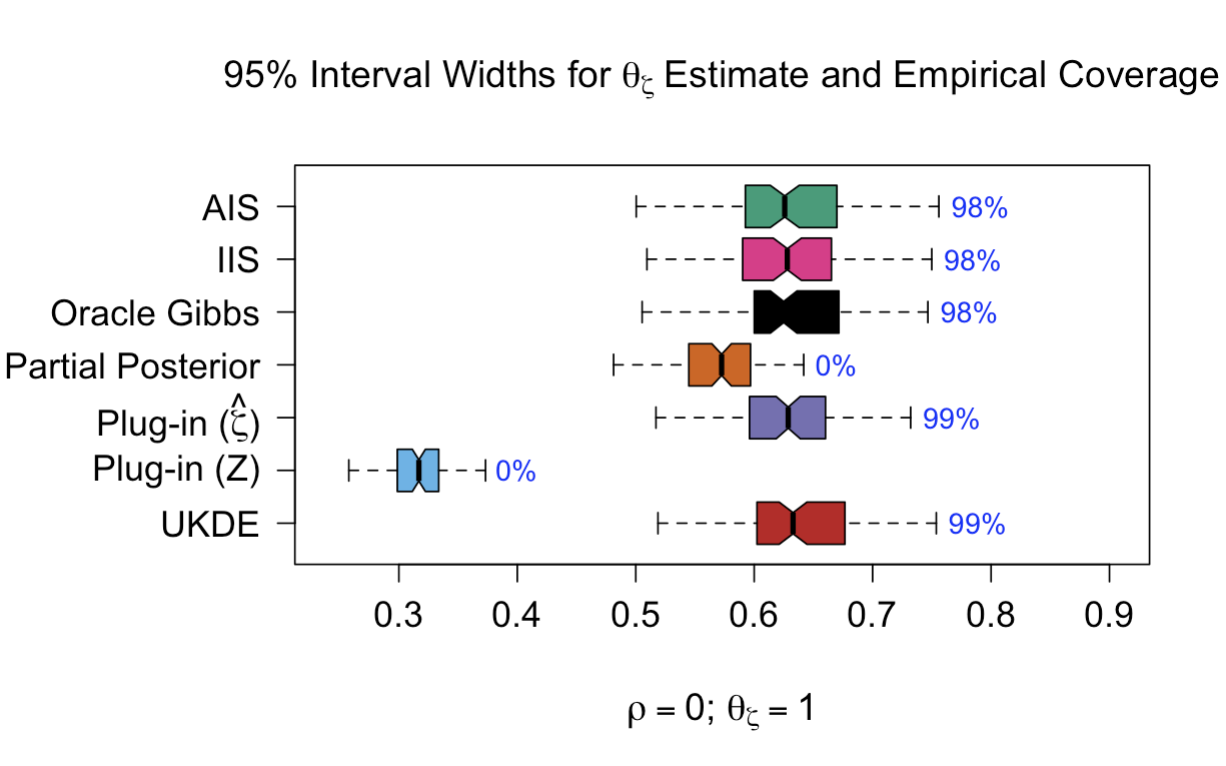}
\end{subfigure}
\begin{subfigure}{0.5\textwidth}
\includegraphics[width=\linewidth]{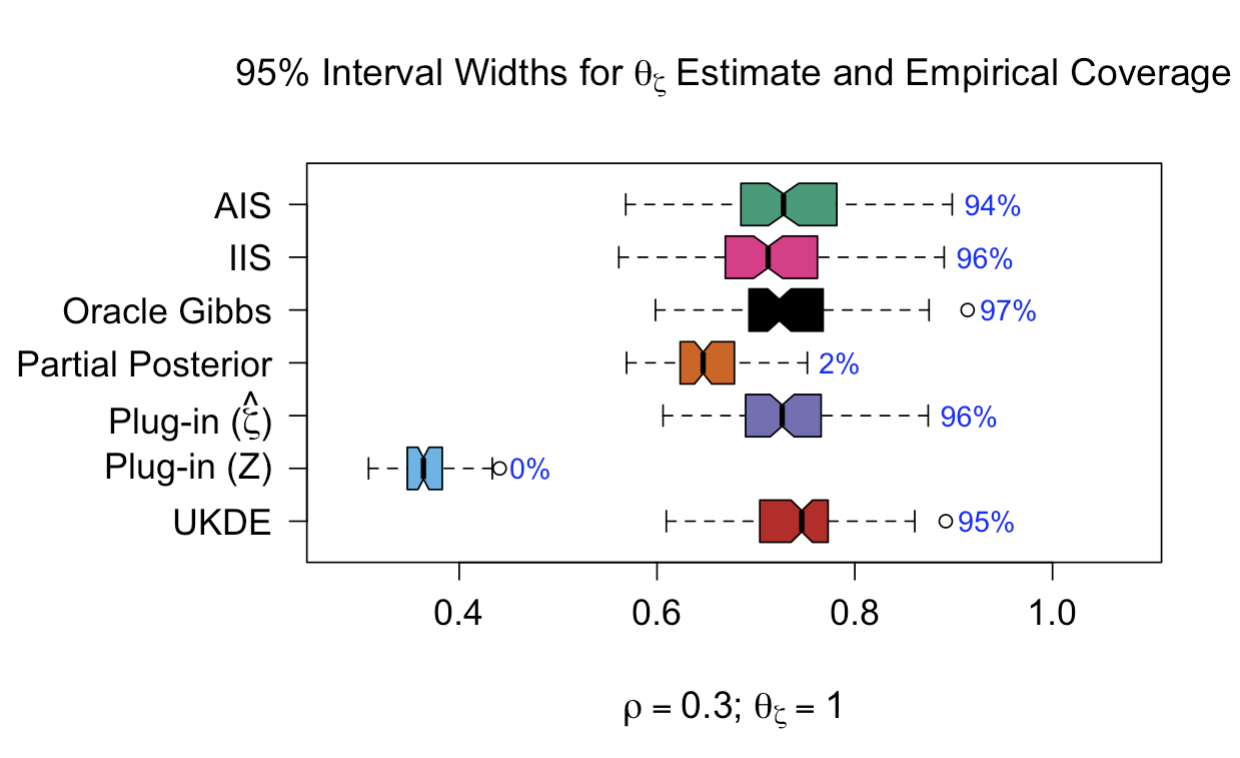}
\end{subfigure}
\caption{Posterior means (top) and 95\% credible interval widths and coverages (bottom) for $\theta_\zeta=1$ in independent $\rho=0$ (left) and correlated $\rho=0.3$ (right) settings. For estimation, we mark the true regression coefficient (solid black) and the attenuated version (dashed blue); for inference, the annotations give the empirical coverages. Here, with smaller signal strength $\theta_\zeta$, all methods aside from Partial Posterior and Plug-in (Z) seem to perform very similar visually both in terms of point estimates and interval widths.} \label{fig:theta_bz1}
\end{figure}

\begin{figure}[t!] 
\begin{subfigure}{0.48\textwidth}
\includegraphics[width=\linewidth]{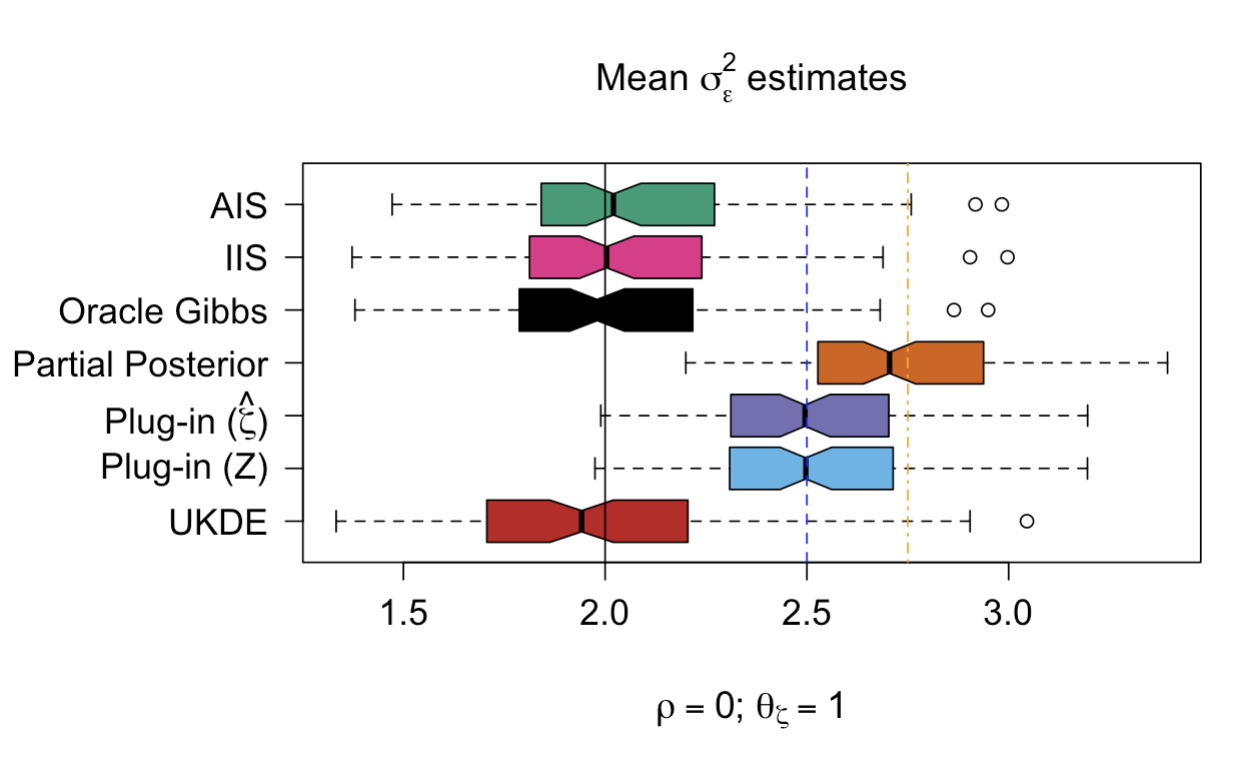}
\end{subfigure}
\begin{subfigure}{0.48\textwidth}
\includegraphics[width=\linewidth]{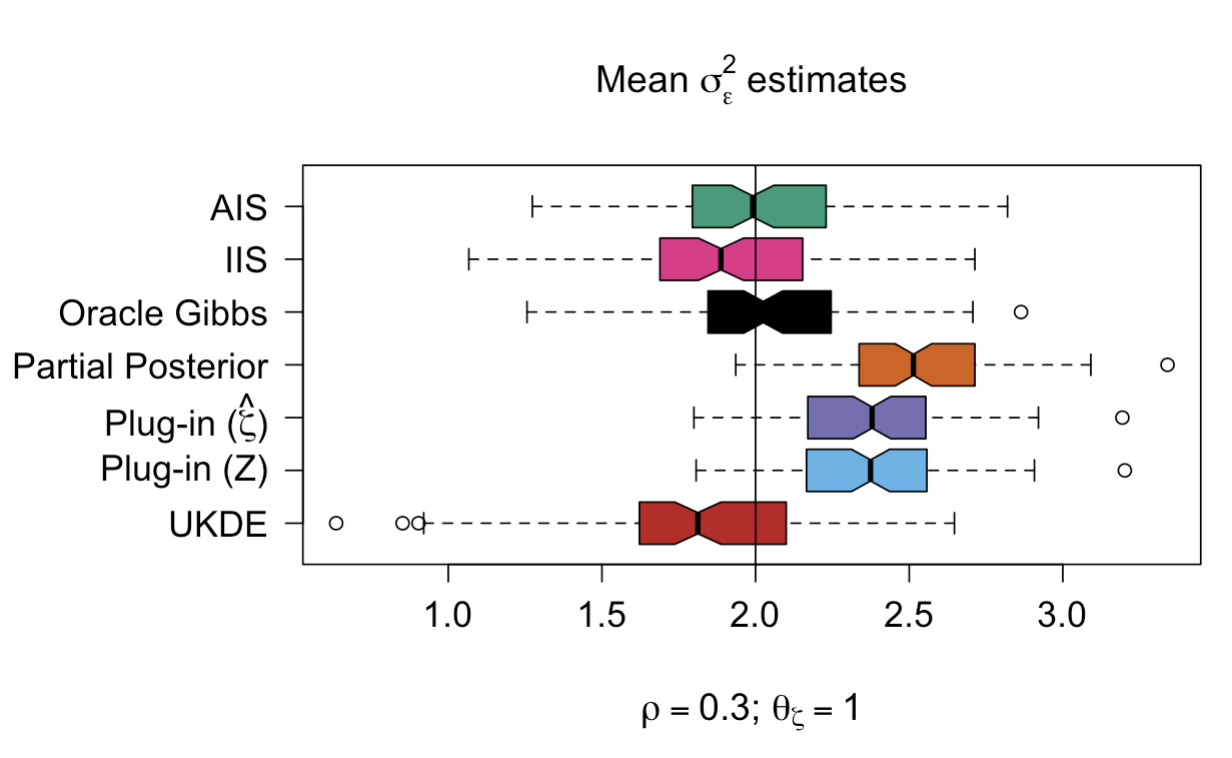}
\end{subfigure}
\caption{Posterior means for $\sigma^2_\epsilon$ in independent (left) and correlated (right) settings for $\theta_\zeta=1$; the true value of $\sigma_\epsilon^2$ is indicated by a black line; the attenuated versions for the independent setting for Plug-in (blue dashed line) and Partial Posterior (orange dash-dotted line) are also included on the left. AIS still performs the best in terms of estimating $\sigma^2_\epsilon$, but the differences with the other methods are smaller than for $\theta_\zeta=4$.} \label{fig:sigma_bz1}
\end{figure}

\begin{figure}[t!] 

\begin{subfigure}{0.48\textwidth}
\includegraphics[width=\linewidth]{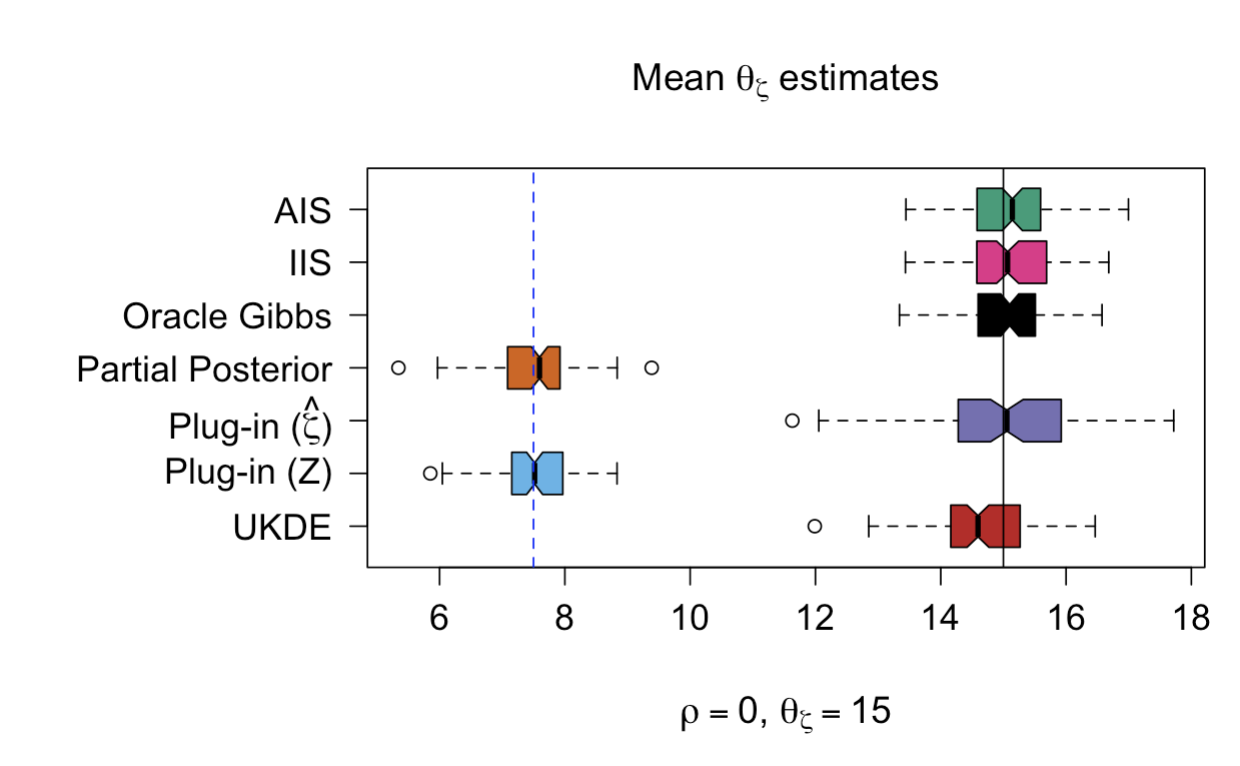}
\end{subfigure}
\begin{subfigure}{0.48\textwidth}
\includegraphics[width=\linewidth]{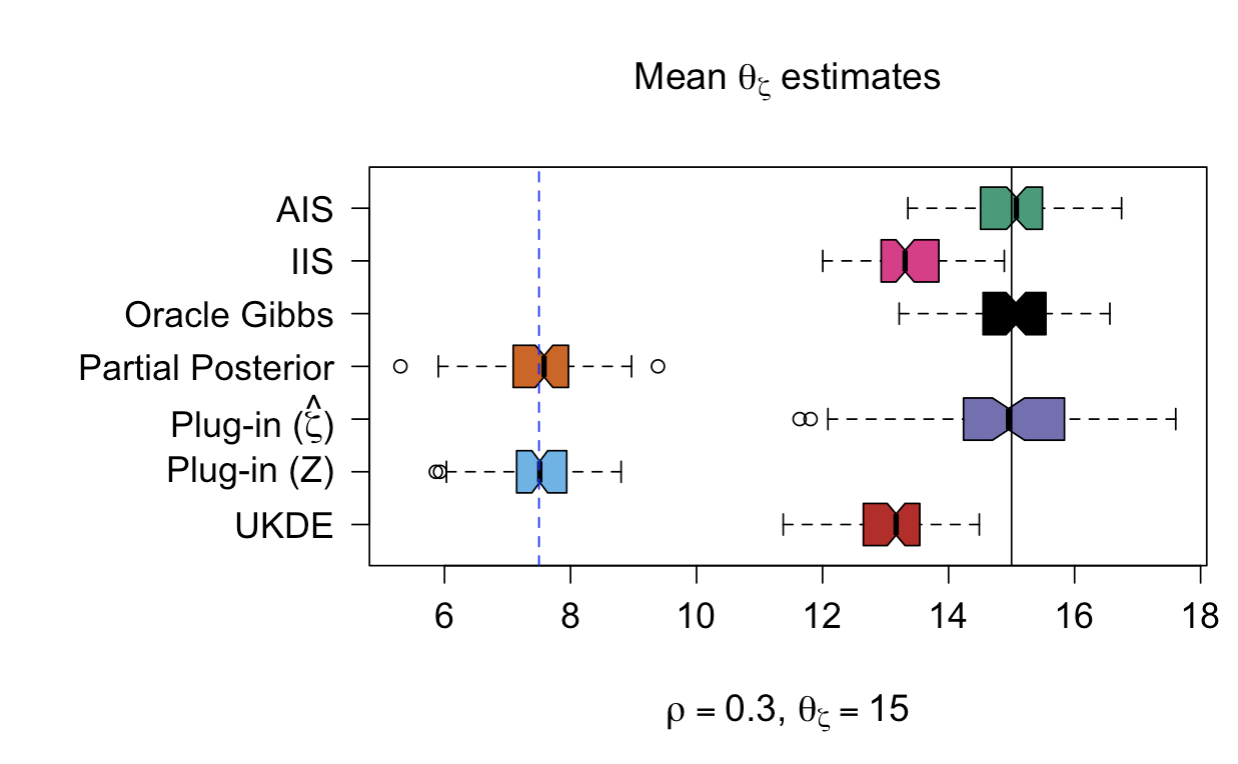}
\end{subfigure}
\begin{subfigure}{0.5\textwidth}
\includegraphics[width=\linewidth]{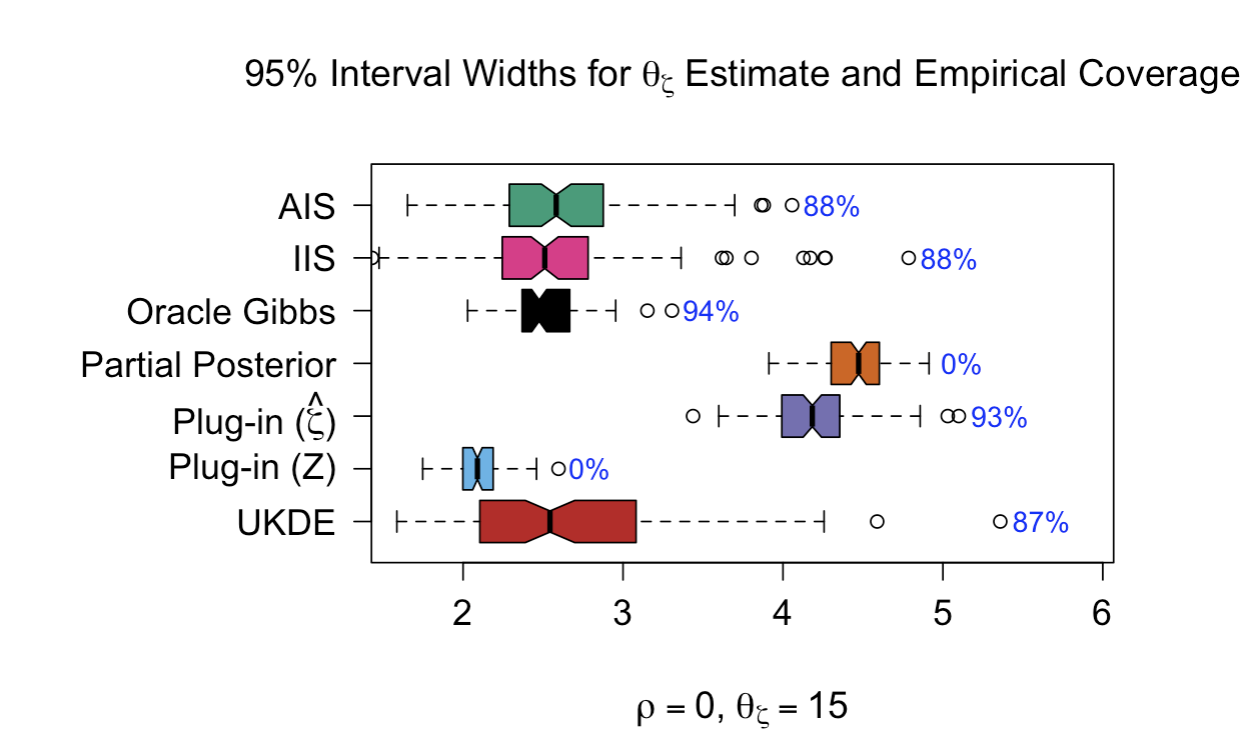}
\end{subfigure}
\begin{subfigure}{0.5\textwidth}
\includegraphics[width=\linewidth]{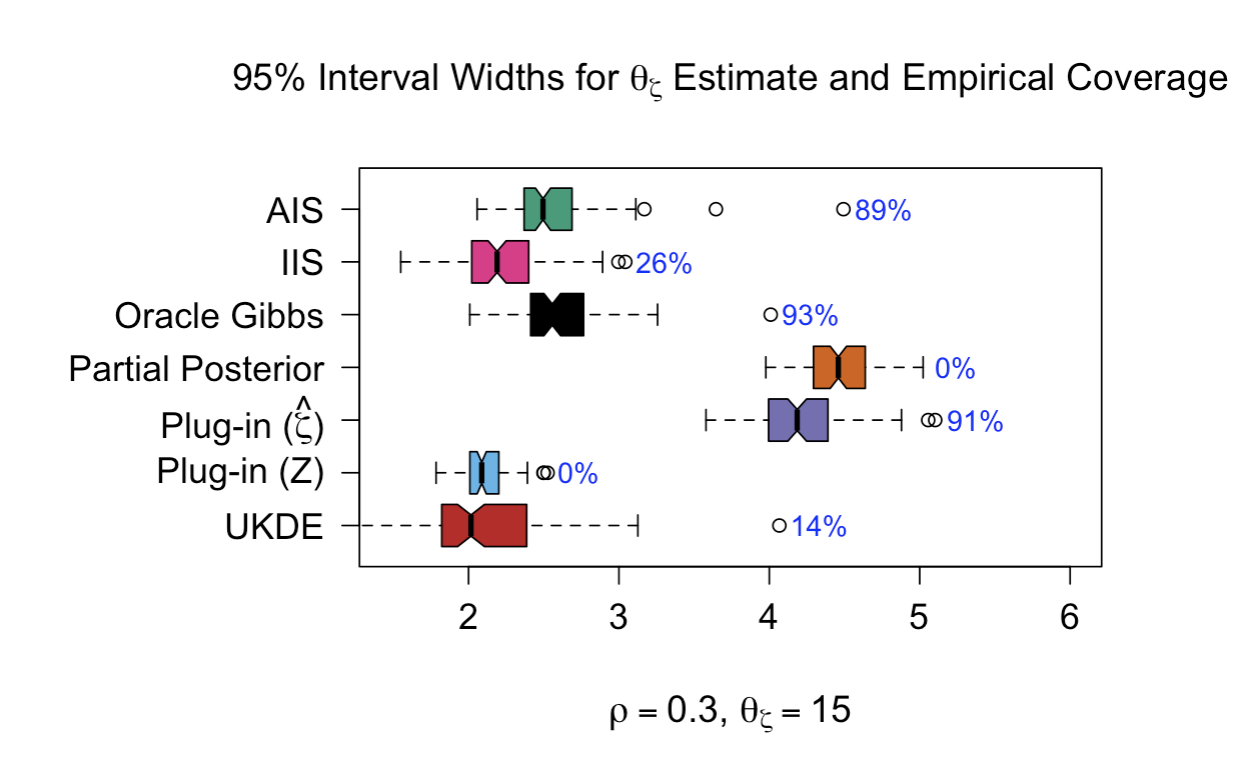}
\end{subfigure}
\caption{Posterior means (top) and 95\% credible interval widths and coverages (bottom) for $\theta_\zeta=15$ in independent $\rho=0$ (left) and correlated $\rho=0.3$ (right) settings. For estimation, we mark the true regression coefficient (solid black) and the attenuated version (dashed blue); for inference, the annotations give the empirical coverages. AIS provides accurate point estimation and the best interval width with good coverage. Plug-in ($\hat\zeta$) has good point estimates, but the interval widths are greater than for Oracle Gibbs.} \label{fig:theta_bz15}
\end{figure}

\begin{figure}[t!] 
\begin{subfigure}{0.48\textwidth}
\includegraphics[width=\linewidth]{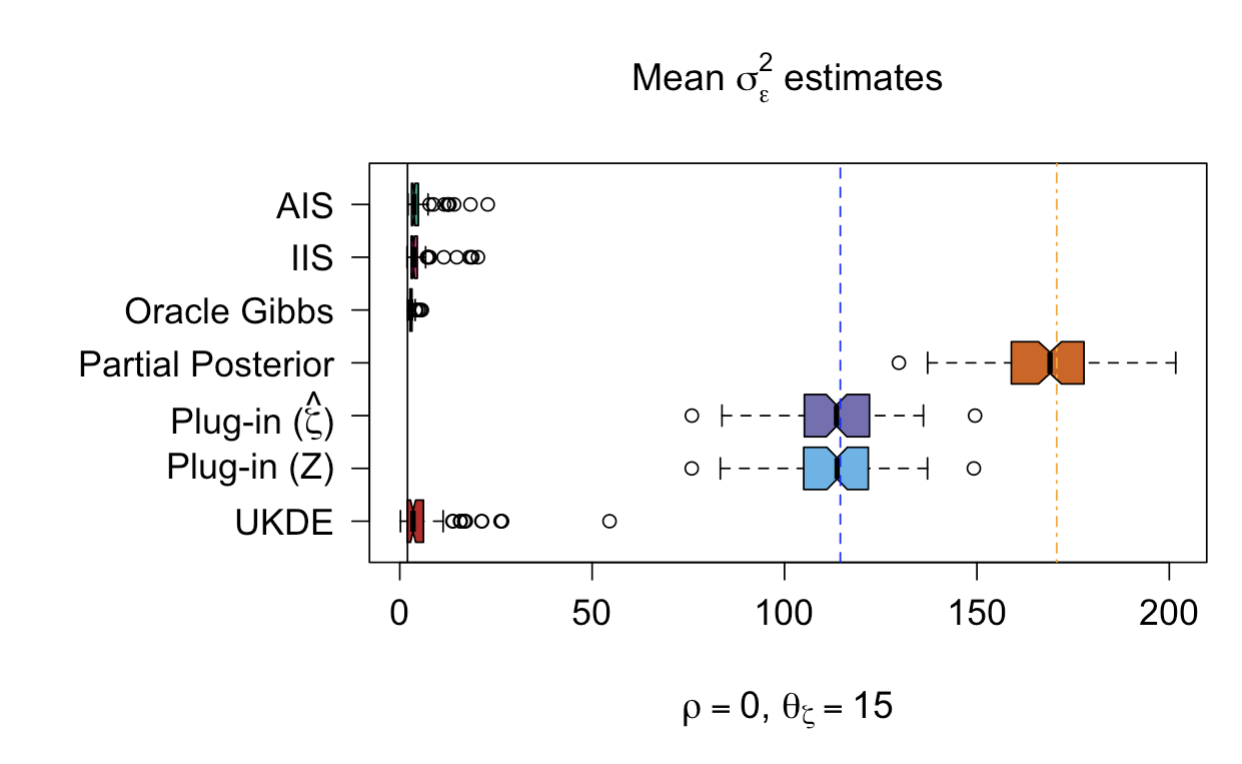}
\end{subfigure}
\begin{subfigure}{0.48\textwidth}
\includegraphics[width=\linewidth]{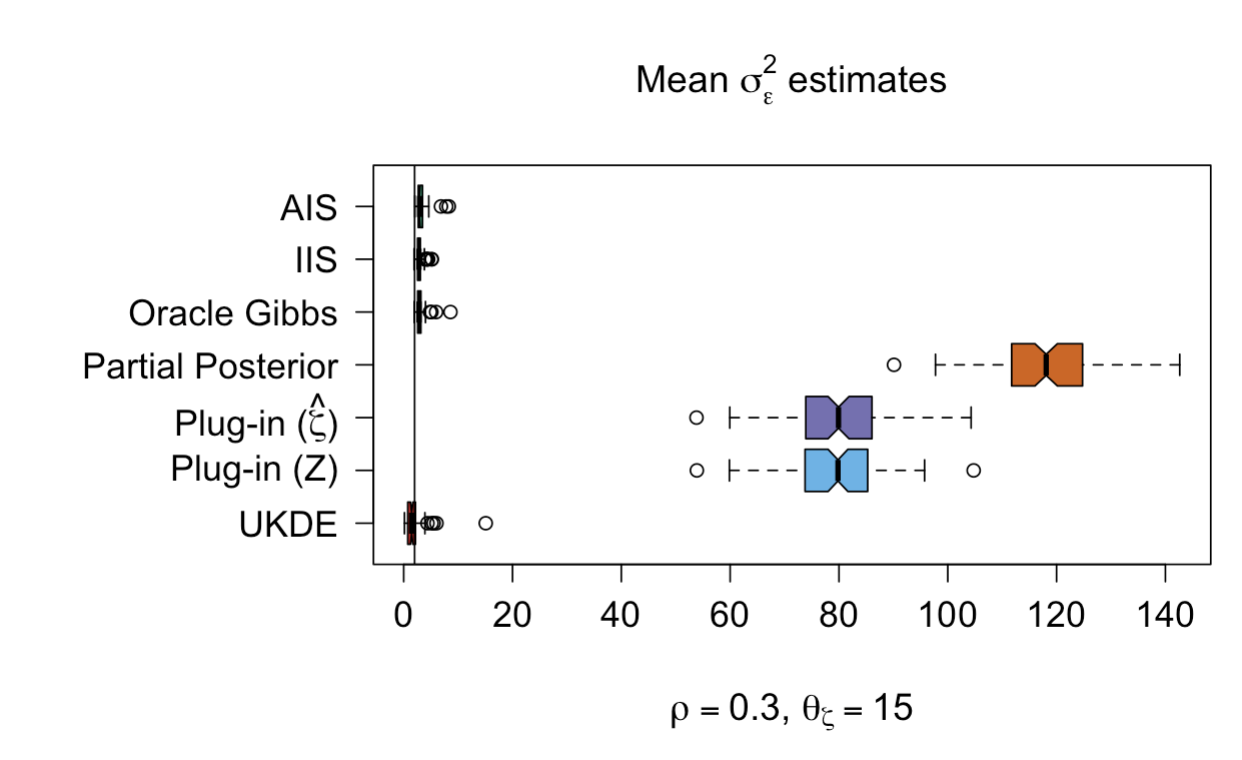}
\end{subfigure}
\caption{Posterior means for $\sigma^2_\epsilon$ in independent (left) and correlated (right) settings for $\theta_\zeta=15$; the true value of $\sigma_\epsilon^2$ is indicated by a black line; the attenuated versions for the independent setting  for Plug-in (blue dashed line) and Partial Posterior (orange dash-dotted line) are also included on the left. The large $\theta_\zeta$ leads to great variance inflation for the Plug-in and Partial Posterior methods. AIS, IIS, and UKDE seem to perform similarly well and are close to the true value and the Oracle Gibbs result.} \label{fig:sigma_bz15}
\end{figure}

Figures \ref{fig:theta_bz1} and \ref{fig:sigma_bz1}, along with Table \ref{Tabletheta1}, provide results for the case with smaller $\theta_\zeta=1$ compared to $\theta_\zeta=4$ in Section 4.2 for the independent ($\rho=0.0$) and the correlated ($\rho=0.3$) scenarios. Smaller $\theta_\zeta$ leads to smaller differences across methods, e.g., the differences between AIS and IIS and UKDE are a lot less visible in the correlated scenario. For comparison, Figures \ref{fig:theta_bz15}, \ref{fig:sigma_bz15}, and Table \ref{Tabletheta15} show results of simulations for the larger $\theta_\zeta=15$, which remain similar to the $\theta_\zeta=4$ scenario.

\begin{figure}[t!] 

\begin{subfigure}{0.48\textwidth}
\includegraphics[width=\linewidth]{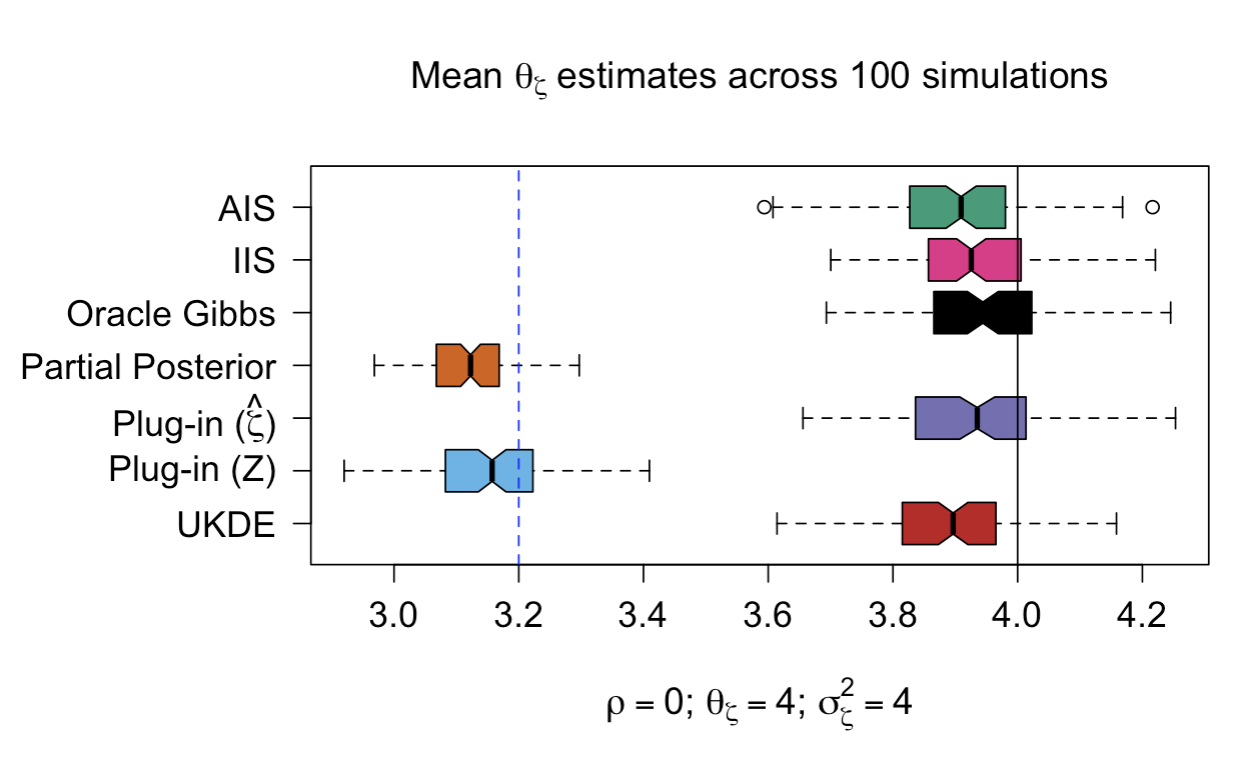}
\end{subfigure}
\begin{subfigure}{0.48\textwidth}
\includegraphics[width=\linewidth]{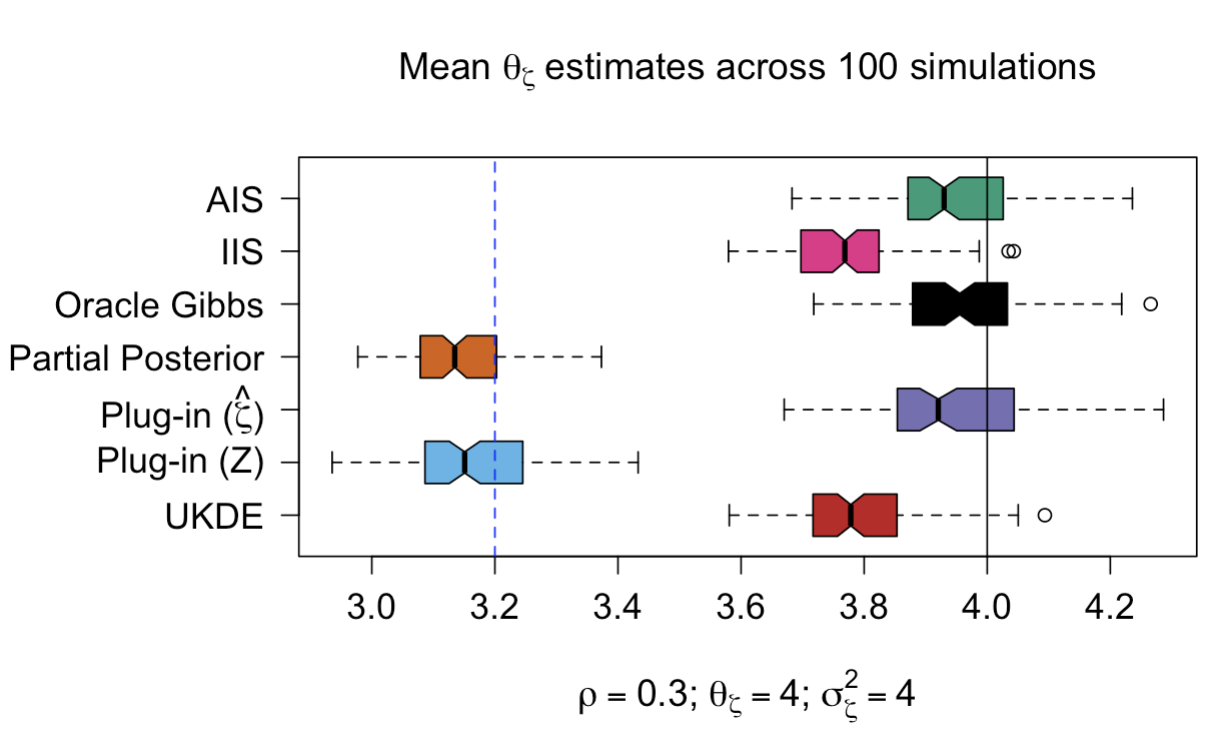}
\end{subfigure}
\begin{subfigure}{0.5\textwidth}
\includegraphics[width=\linewidth]{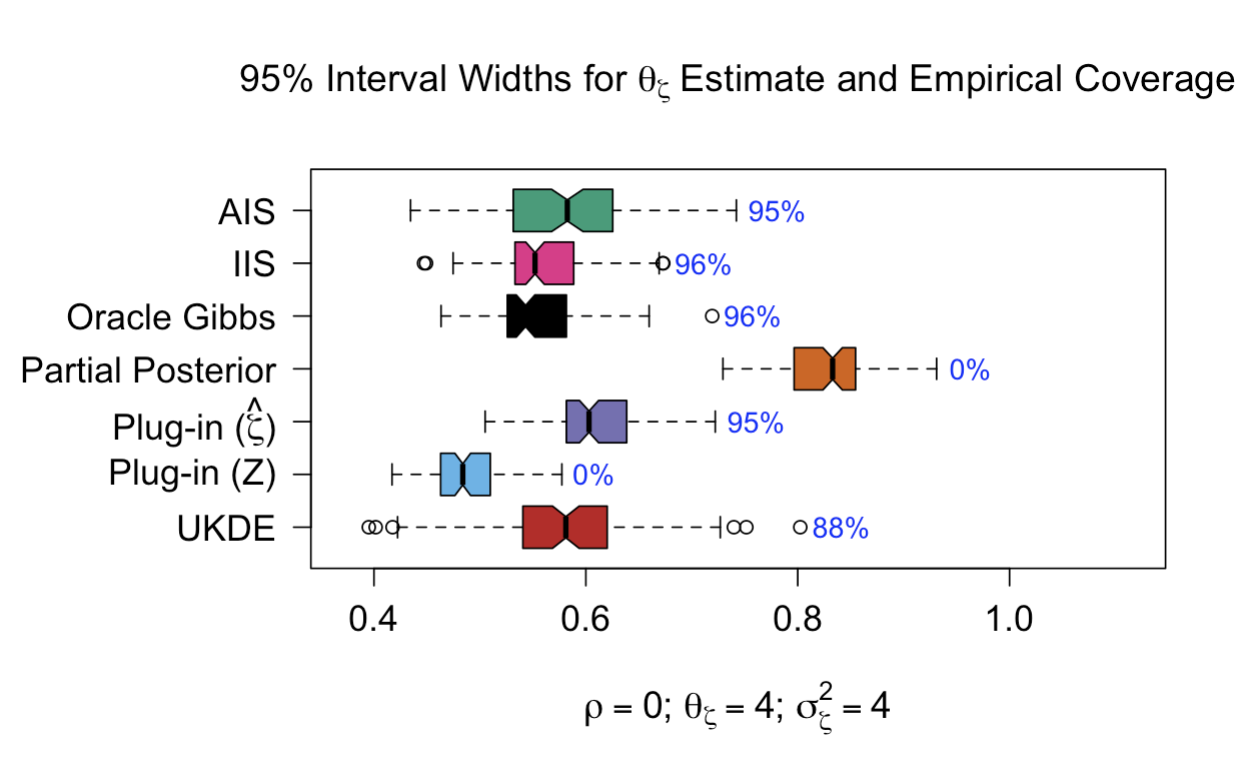}
\end{subfigure}
\begin{subfigure}{0.5\textwidth}
\includegraphics[width=\linewidth]{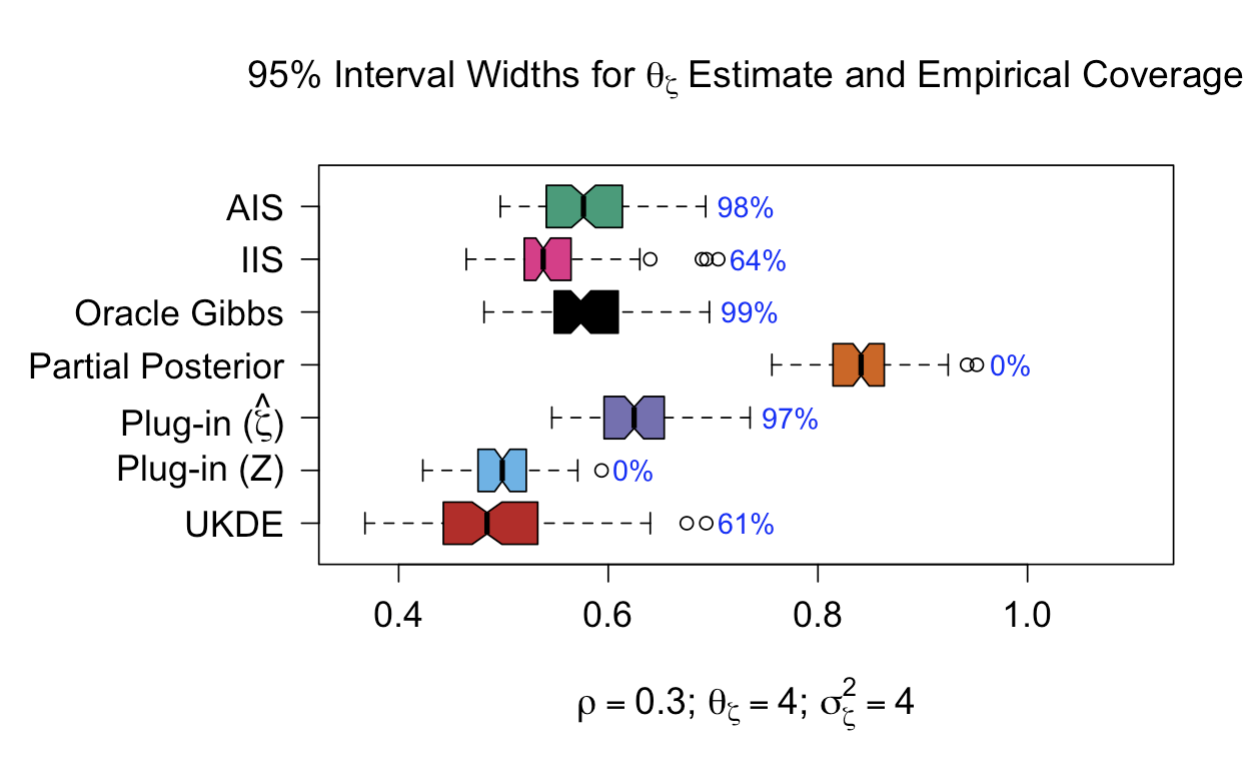}
\end{subfigure}
\caption{Posterior means (top) and 95\% credible interval widths and coverages (bottom) for $\sigma^2_\zeta=4$, $\theta_\zeta=4$ in independent $\rho=0$ (left) and correlated $\rho=0.3$ (right) settings. For estimation, we mark the true regression coefficient (solid black) and the attenuated version (dashed blue); for inference, the annotations give the empirical coverages. We see that AIS and Plug-in ($\hat\zeta$) both have accurate point estimates and  similar interval widths, though AIS has a slightly narrower one that is also closer to the Oracle Gibbs result. IIS and UKDE perform well in the independent scenario, but worse in the correlated one.} \label{fig:theta_sz2}
\end{figure}

\begin{figure}[t!] 
\begin{subfigure}{0.48\textwidth}
\includegraphics[width=\linewidth]{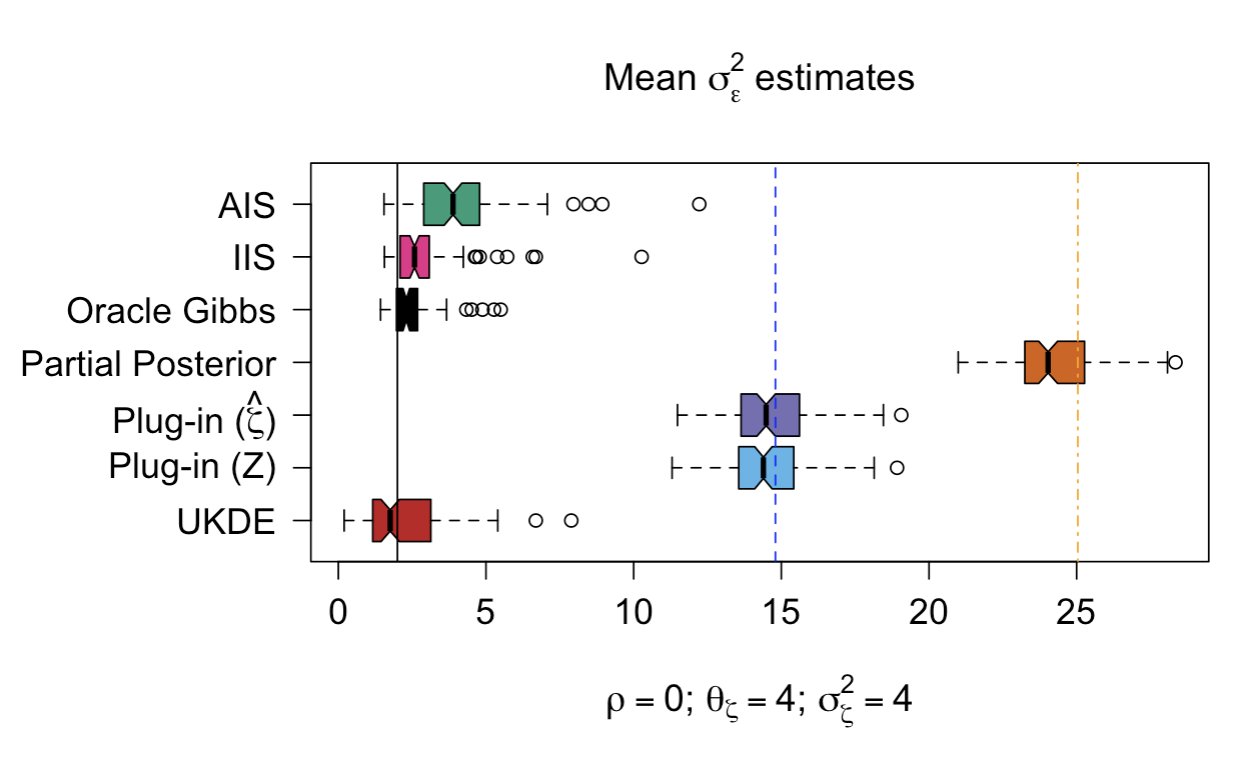}
\end{subfigure}
\begin{subfigure}{0.48\textwidth}
\includegraphics[width=\linewidth]{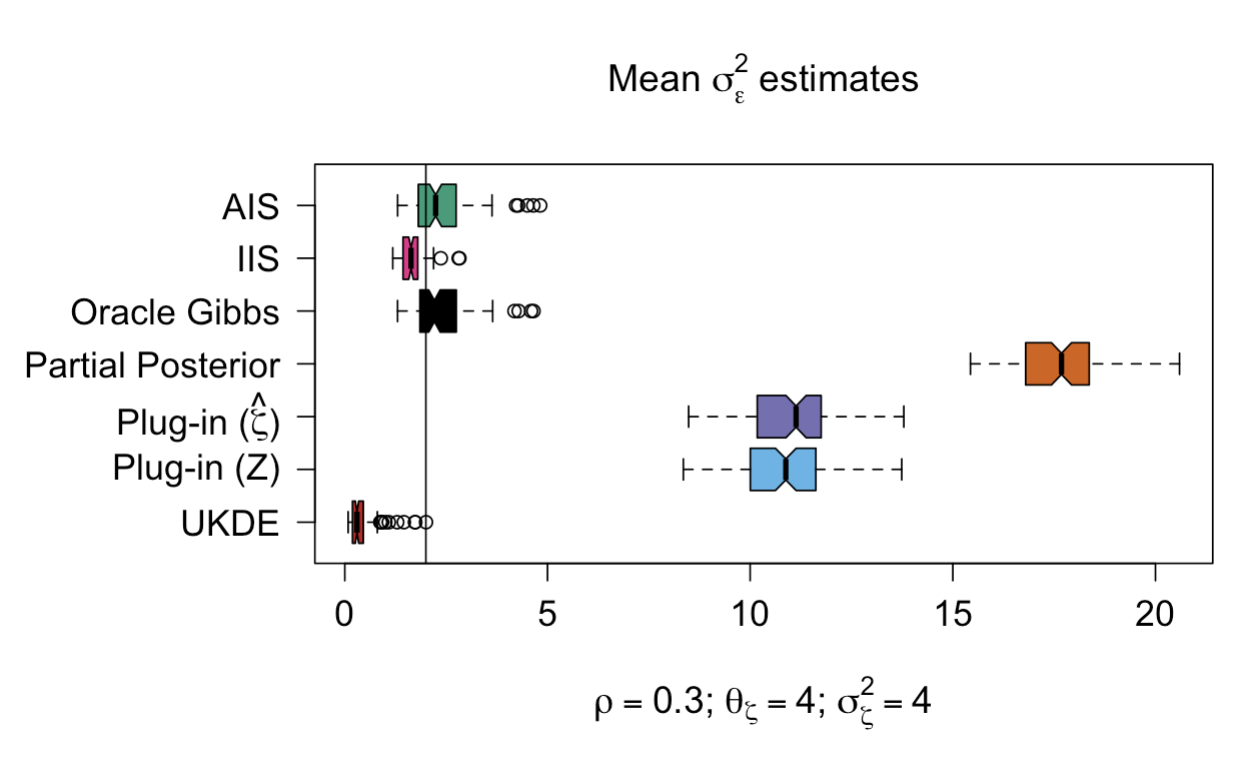}
\end{subfigure}
\caption{Posterior means for $\sigma^2_\epsilon$ in independent (left) and correlated (right) settings for $\sigma^2_\zeta=4$, $\theta_\zeta=4$; the true value of $\sigma_\epsilon^2$ is indicated by a black line; the attenuated versions for the independent setting  for Plug-in (blue dashed line) and Partial Posterior (orange dash-dotted line) are also included on the left. AIS and IIS have consistently accurate estimates that match Oracle Gibbs, UKDE underestimates $\sigma^2_\epsilon$ in the correlated scenario despite accurate estimate in the independnet scenario; all other methods consistently overestimate $\sigma^2_\epsilon$.} \label{fig:sigma_sz2}
\end{figure}

Finally, Figures \ref{fig:theta_sz2}, \ref{fig:sigma_sz2}, and Table \ref{TableSZ2} show result for simulations with true $\sigma^2_\zeta=4$ instead of $\sigma^2_\zeta=1$ as in the main text. This leads $\lambda = \frac{\sigma^2_\zeta}{\sigma^2_\zeta+\sigma^2_u}$ to be closer to 1, resulting in smaller attenuation and variance inflation for the Plug-in and Partial Posterior methods. However, AIS remains consistently the method that is both accurate and closest to the Oracle Gibbs results.

\begin{table} 
\begin{tabular}{  p{3cm}p{5cm}p{3cm}p{3cm}  }
 \hline
 \multicolumn{4}{|c|}{2-Wasserstein Distance to Oracle Gibbs Result for the $\rho=0.5$ setting} \\
 \hline
 Data Settings & Method& $\theta_\zeta$ &$\sigma^2_\epsilon$\\
 \hline
$\rho=0.5$  &AIS   & {\bf0.080}    &{\bf 0.362}\\
&  IIS&   0.653  & 0.806 \\
& Plug-in ($\hat\zeta$)   & {\bf 0.079} & 3.707\\
&Plug-in (Z) &   1.957  & 3.703\\
 &Partial Posterior& 1.971  & 5.534   \\
& UKDE& 0.602	 & 1.634\\ \hline
 
\end{tabular}
\caption{2-Wasserstein distance to the posterior distributions for $\theta_\zeta$ and $\sigma_\epsilon^2$ under Oracle Gibbs for $\rho=0.5$. AIS is consistently the best performer, though Plug-in ($\hat\zeta$) has the same distance for $\theta_\zeta$ in the latter setting.}
\label{Table0405}
\end{table}

\begin{table} 
\begin{tabular}{  p{3cm}p{5cm}p{3cm}p{3cm}  }
 \hline
 \multicolumn{4}{|c|}{2-Wasserstein Distance for $\rho=0$ (Independent) and $\rho=0.3$ settings when $\theta_\zeta=1$} \\
 \hline
 Data Settings & Method& $\theta_\zeta$ &$\sigma^2_\epsilon$\\
 \hline
  $\rho=0$  &AIS   & {\bf0.006}    &{\bf 0.012}\\
& IIS&    {\bf0.009}  &{\bf 0.014}  \\
 &Plug-in ($\hat\zeta$)   & 0.021& 0.497\\
&Plug-in (Z) &    0.505  &0.491\\
 &Partial Posterior& 0.504  & 0.713   \\
& UKDE& {\bf0.007}  &0.065\\ \hline
$\rho=0.3$  &AIS   & {\bf0.011}    & {\bf0.028}\\
&  IIS&   {\bf0.016}  & 0.130 \\
& Plug-in ($\hat\zeta$)   & 0.021 & 0.359\\
&Plug-in (Z) &   0.510  &0.354\\
 &Partial Posterior& 0.503  & 0.509   \\
& UKDE& {\bf0.012}	 & 0.233\\ \hline
 
\end{tabular}
\caption{2-Wasserstein distance to the posterior distributions for $\theta_\zeta$ and $\sigma_\epsilon^2$ under Oracle Gibbs for $\theta_\zeta=1$. For $\theta_\zeta$, AIS, IIS, and UKDE have almost the same distances and are the best performers, with Plug-in ($\hat\zeta$) performing slightly worse. For $\sigma_\epsilon^2$, AIS is clearly the best method in the correlated scenario, though IIS matches it in the independent scenario. }
\label{Tabletheta1}
\end{table}

\begin{table} 
\begin{tabular}{  p{3cm}p{5cm}p{3cm}p{3cm}  }
 \hline
 \multicolumn{4}{|c|}{2-Wasserstein Distance for $\rho=0$ (Independent) and $\rho=0.3$ settings when $\theta_\zeta=15$} \\
 \hline
 Data Settings & Method& $\theta_\zeta$ &$\sigma^2_\epsilon$\\
 \hline
 $\rho=0$  &AIS   &{\bf 0.086	}    &{\bf3.204}\\
 & IIS&    {\bf0.089}  & {\bf3.373}  \\
 &Plug-in ($\hat\zeta$)   & 0.639	& 111.349\\
&Plug-in (Z) &    7.554	  &111.320\\
 &Partial Posterior& 7.573	  & 167.410   \\
& UKDE& 0.443	  &9.679\\ \hline
$\rho=0.3$ &AIS   & {\bf0.036}    &{\bf0.233}\\
 &  IIS&   1.683  & {\bf0.774} \\
& Plug-in ($\hat\zeta$)   & {\bf0.632} & 77.328\\
&Plug-in (Z) &   7.535  &77.279\\
 &Partial Posterior& 7.552  &116.428  \\
& UKDE& 1.975	 & { 1.785}\\ \hline

\end{tabular}
\caption{2-Wasserstein distance to the posterior distributions for $\theta_\zeta$ and $\sigma_\epsilon^2$ for $\theta_\zeta=15$ under Oracle Gibbs. AIS has one of the smallest distance for $\theta_\zeta$ in both scenarios, though IIS is slightly better in the independent scenario, and Plug-in ($\hat\zeta$) is similar in the correlated scenario. AIS performs best for $\sigma_\epsilon^2$ in both scenarios, although IIS performs only a little worse.}
\label{Tabletheta15}
\end{table}

\begin{table} 
\begin{tabular}{  p{3cm}p{5cm}p{3cm}p{3cm}  }
 \hline
 \multicolumn{4}{|c|}{2-Wasserstein Distance for $\rho=0$ (Independent) and $\rho=0.3$ settings when $\sigma^2_\zeta=4$ and $\theta_\zeta=4$} \\
 \hline
 Data Settings & Method& $\theta_\zeta$ &$\sigma^2_\epsilon$\\
 \hline
$\rho=0$  &AIS   & {\bf0.009}    &{\bf0.256}\\
& IIS&   {\bf 0.011}  & {\bf0.539}  \\
 &Plug-in ($\hat\zeta$)   & 0.027& 12.250\\
&Plug-in (Z) &    0.792  & 12.141\\
 &Partial Posterior& 0.826  & 21.856  \\
& UKDE& 0.056  & 0.769 \\ \hline
$\rho=0.3$  &AIS   & {\bf0.010}    & {\bf0.019} \\
&  IIS&  0.189  & {\bf0.841} \\
& Plug-in ($\hat\zeta$)   & {\bf0.028} &8.656 \\
&Plug-in (Z) &   0.794  & 8.555 \\
 &Partial Posterior& 0.817  & 15.408   \\
& UKDE& 0.174	 & 2.054\\ \hline

\end{tabular}
\caption{2-Wasserstein distance to the posterior distributions for $\theta_\zeta$ and $\sigma_\epsilon^2$ under Oracle Gibb for $\sigma^2_\zeta=4$. AIS performs the best for both parameters in both cases.}
\label{TableSZ2}
\end{table}

\clearpage

\section{Regression coefficient estimates for the mortality rate model}
\label{sec:app_other_theta_emp}
In this section, we add to the empirical results presented in Section 5.3 by providing posterior means and credible intervals for the $\bm\beta$ coefficients for known covariates in our mortality model. Our known covariates are as follows: $x_1$ is the median household income, $x_2$ is the metro/non-metro indicator, $x_3$ is the proportion of  non-Hispanic Black residents, $x_4$ is the proportion of adults without complete high school education. and $x_5$ through $x_7$ are state indicators for Louisiana, Oklahoma, and Texas, respectively. The coefficient for the median household income ($\beta_1$) has a significant negative effect for all models, suggesting that poorer counties have worse health outcomes. For all models other than Partial Posterior, proportion of non-Hispanic Black residents ($\beta_3$) and state indicator for Oklahoma ($\beta_6$) have a significant positive effect on county-wide mortality. Interestingly, the proportion of adults without HS education ($\beta_4$) has a significant negative effect accounting for the other variables for all methods other than Partial Posterior, where it is negative but non-significant. For all methods, the 95$\%$ credible intervals include 0 for coefficients for the metro indicator ($\beta_2$), as well as the state indicators for Louisiana ($\beta_5$) and Texas ($\beta_7$). Overall, the methods have similar posterior means and credible intervals across the coefficients, with the exception of Partial Posterior, which consistently has wider credible intervals and thus fewer variables with a significant effect.


 \begin{figure}
 \centering
\begin{subfigure}{0.48\textwidth}
\includegraphics[width=\linewidth]{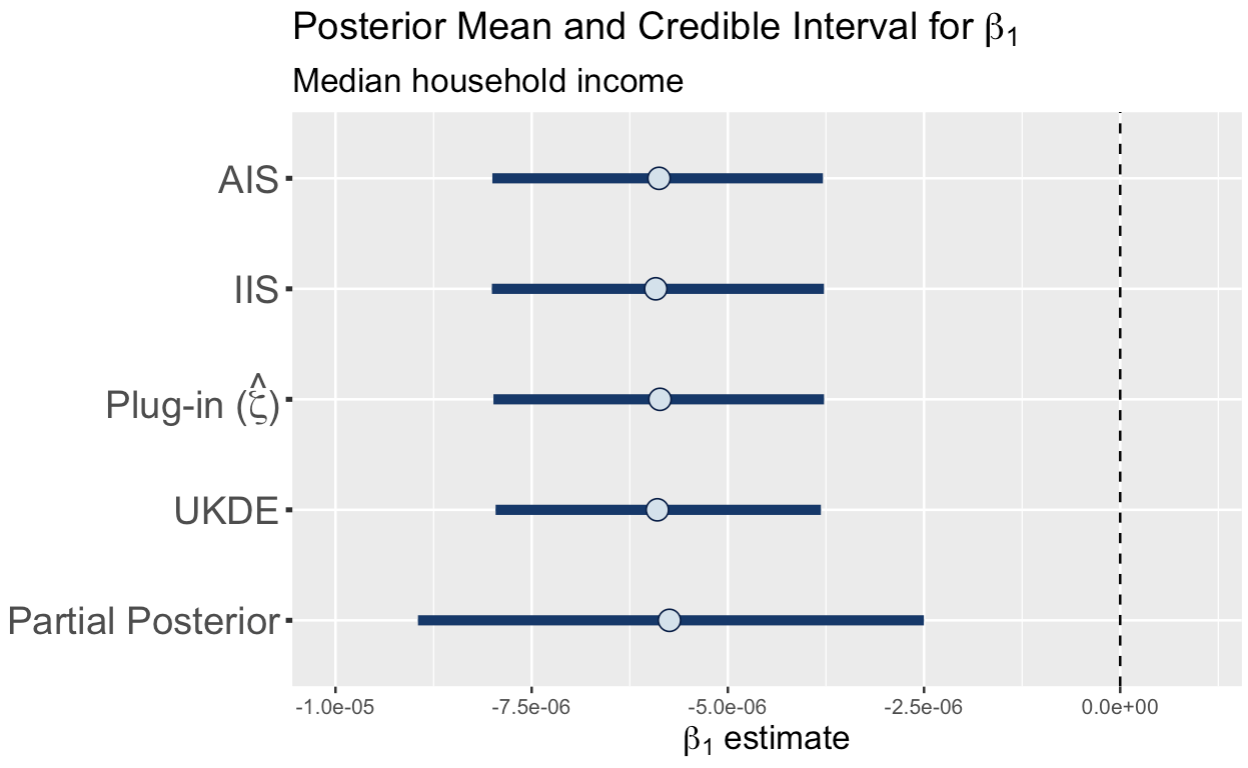}

\end{subfigure}
\begin{subfigure}{0.48\textwidth}
\includegraphics[width=\linewidth]{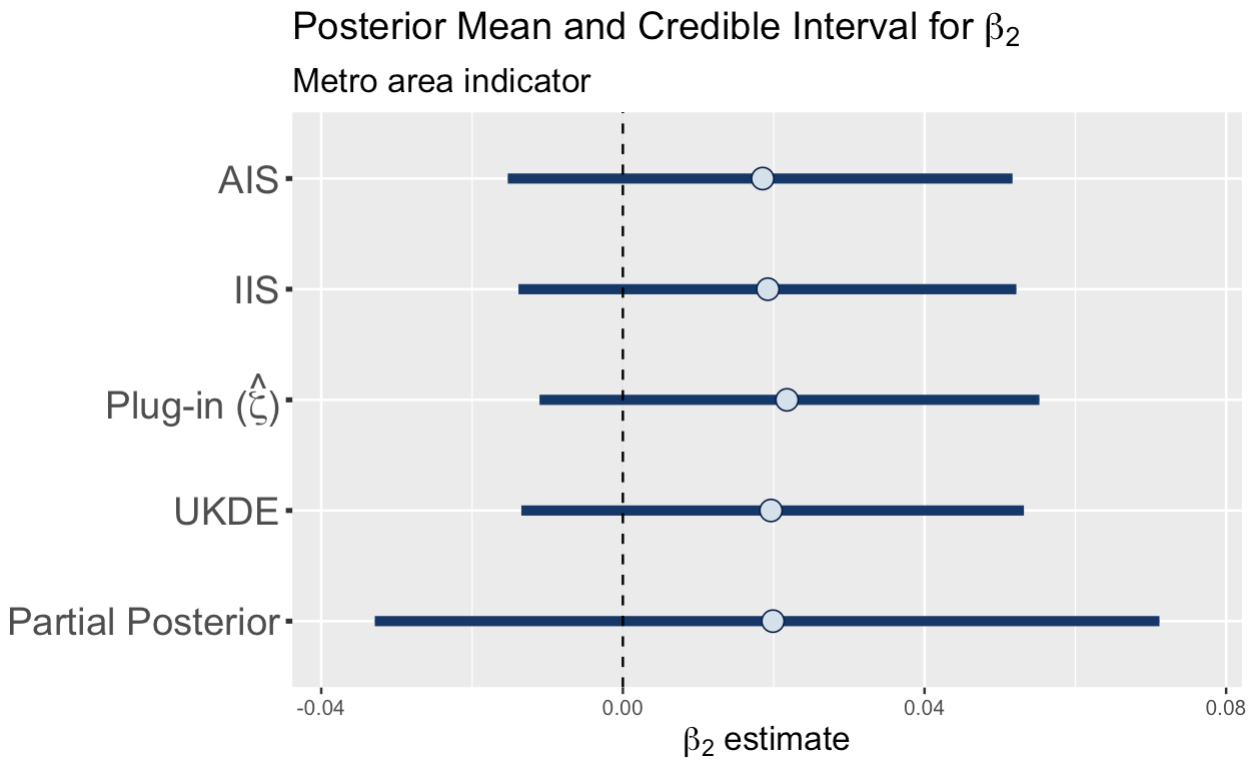}

\end{subfigure}
\begin{subfigure}{0.48\textwidth}
\includegraphics[width=\linewidth]{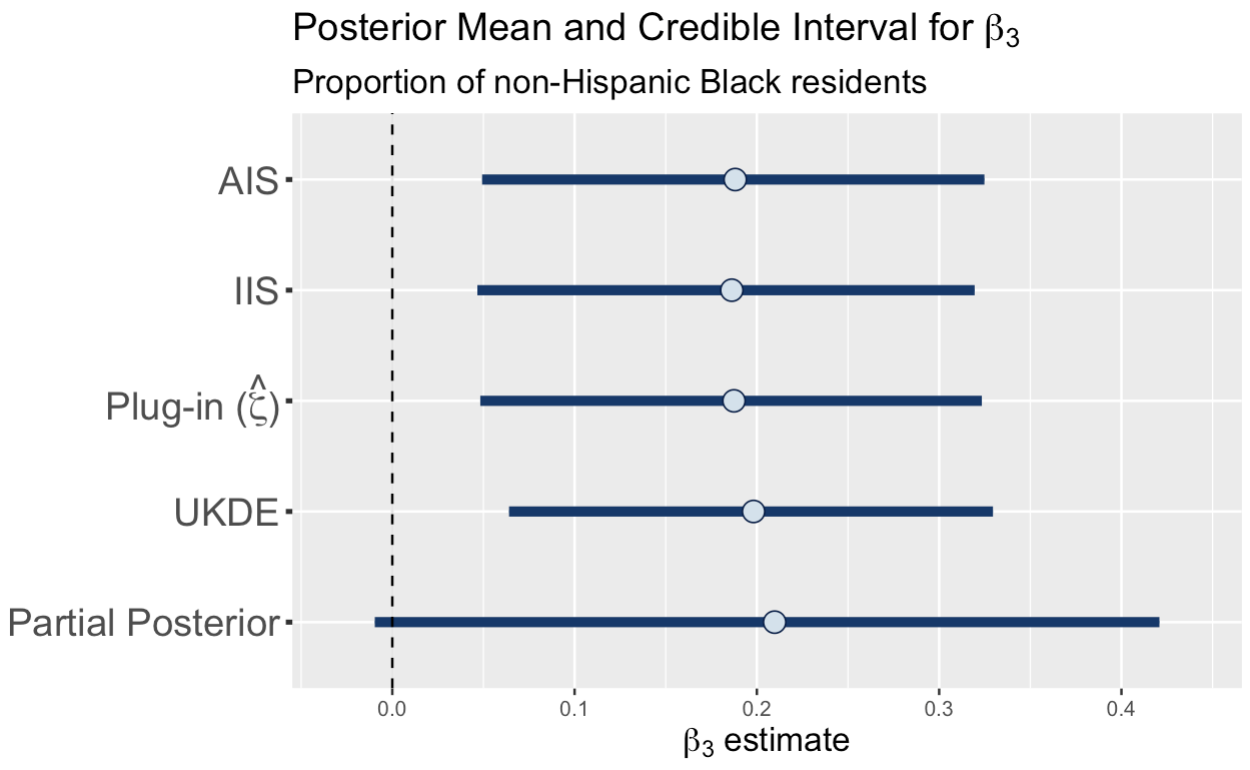}

\end{subfigure}
\begin{subfigure}{0.48\textwidth}
\includegraphics[width=\linewidth]{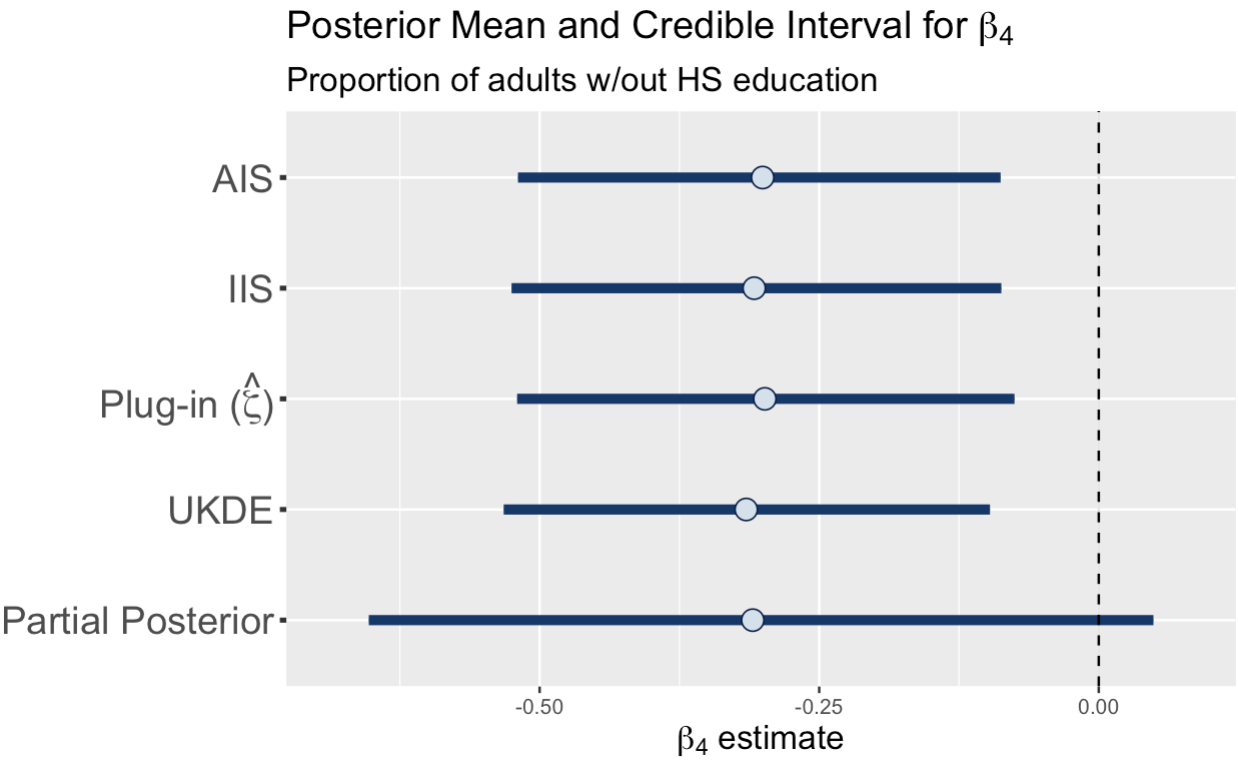}

\end{subfigure}
\begin{subfigure}{0.48\textwidth}
\includegraphics[width=\linewidth]{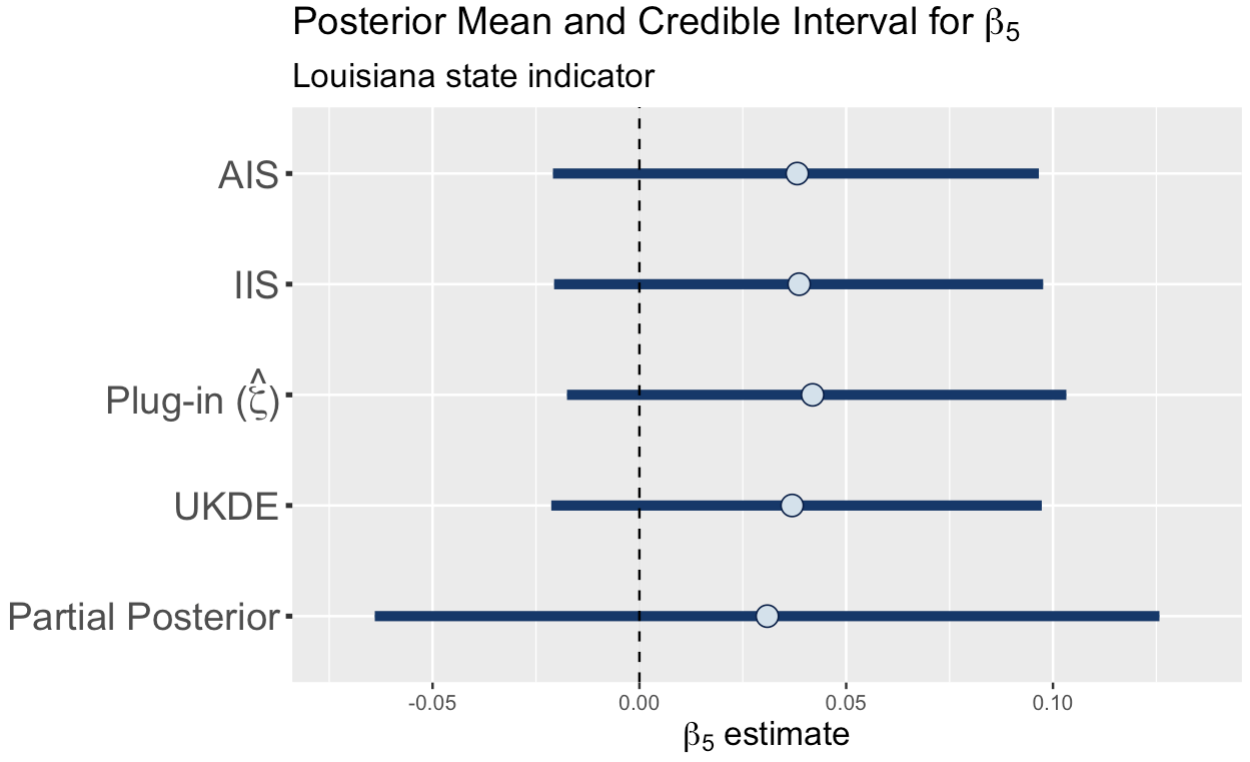}

\end{subfigure}
\begin{subfigure}{0.48\textwidth}
\includegraphics[width=\linewidth]{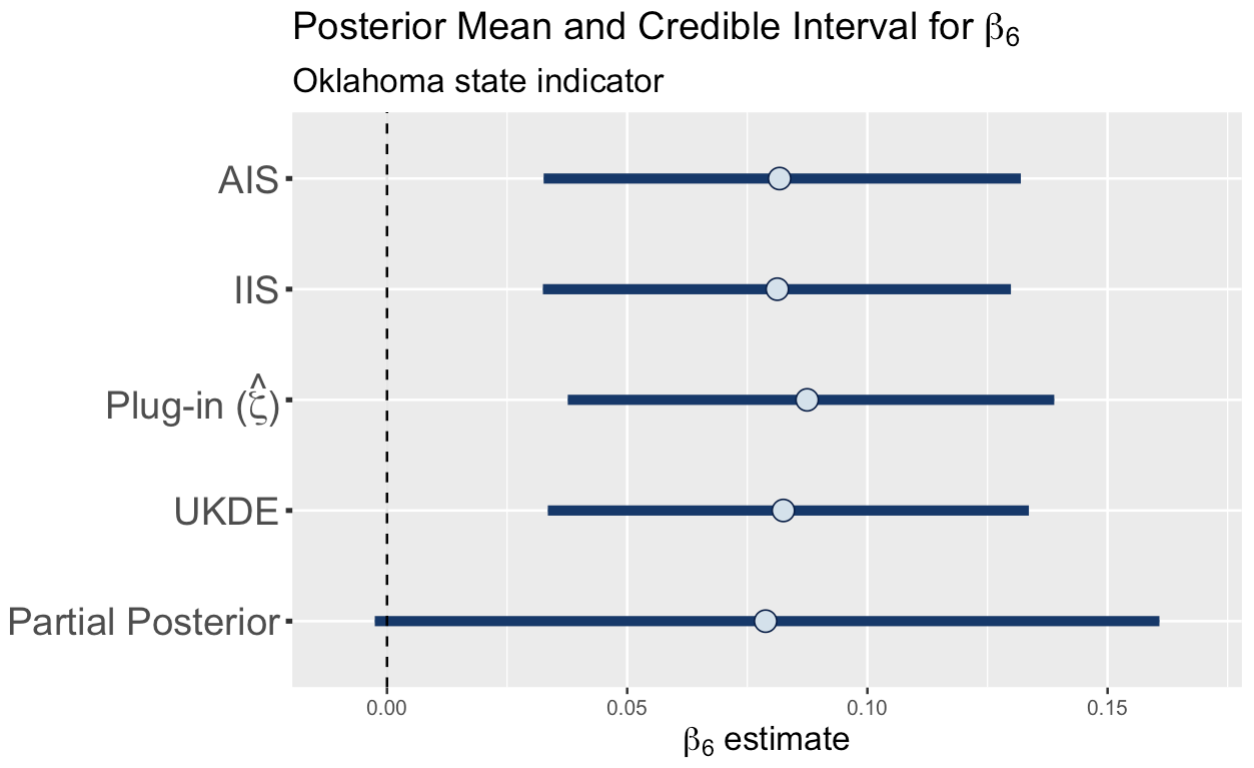}

\end{subfigure}
\begin{subfigure}{0.48\textwidth}
\includegraphics[width=\linewidth]{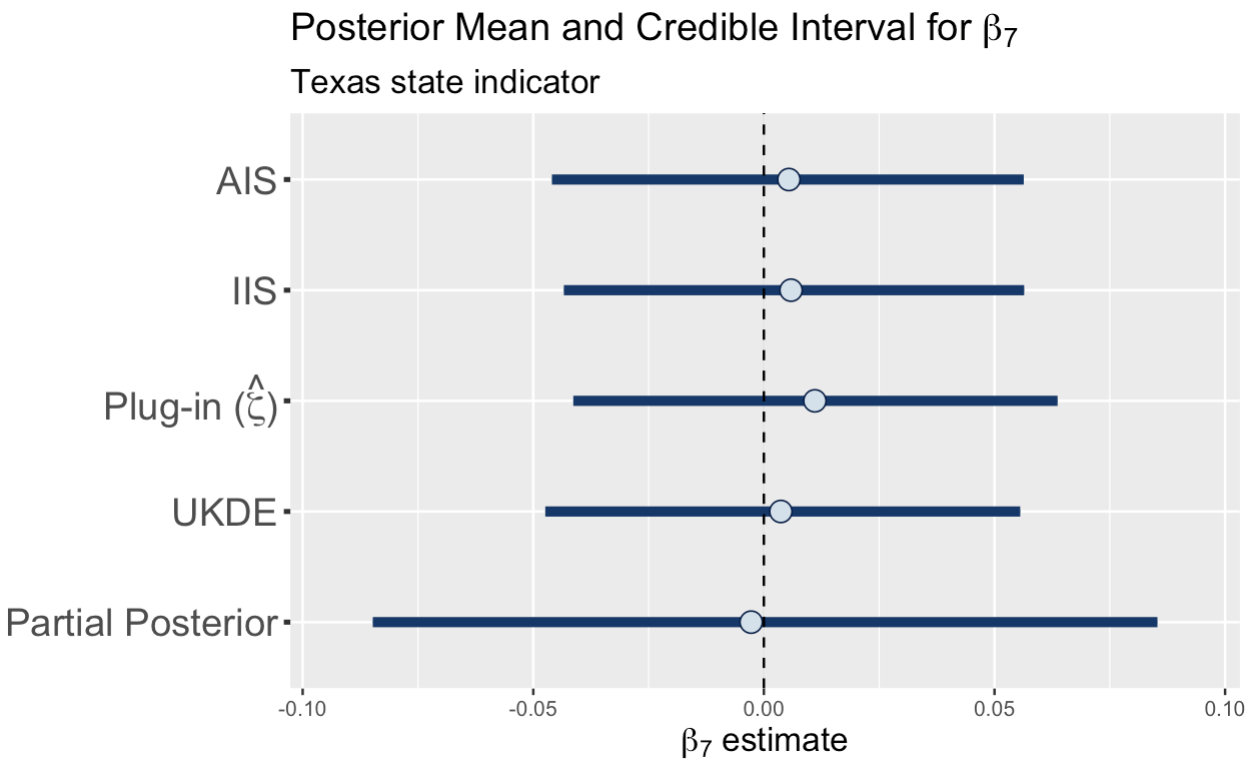}

\end{subfigure}

\caption{Posterior means and credible intervals for the $\beta_1,...,\beta_7$ regression coefficients in our age-adjusted mortality model for South Central States from Section 5.3. We see much greater uncertainty for the Partial Posterior method across all coefficients, but little difference between the other methods. The dashed black vertical line at zero helps show significance.}
\label{fig:emp_o_thetas}
\end{figure}

\section{Hybrid Simulated Data Using Sample Mean of the Draws}
\label{sec:app_hybrid_mean}
\begin{figure}
\begin{subfigure}{0.47\textwidth}
\includegraphics[width=\linewidth]{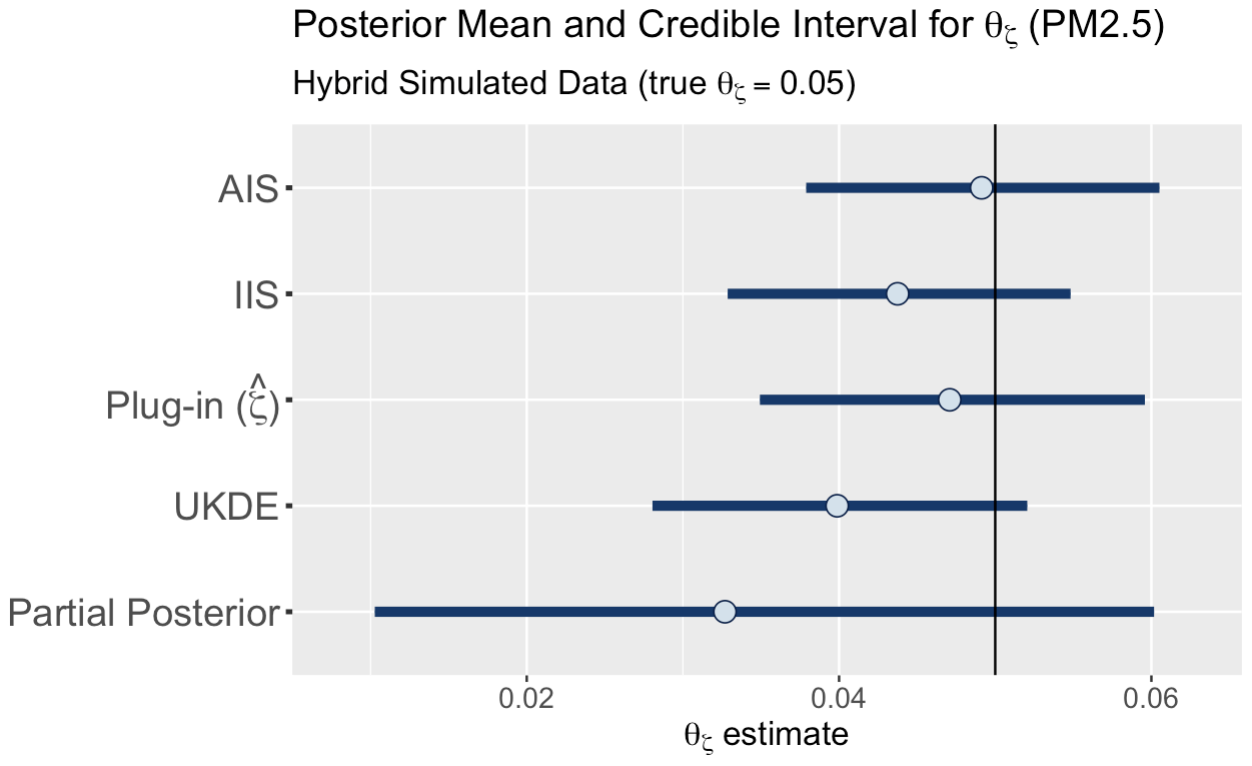}
\end{subfigure}
\begin{subfigure}{0.5\textwidth}
\includegraphics[width=\linewidth]{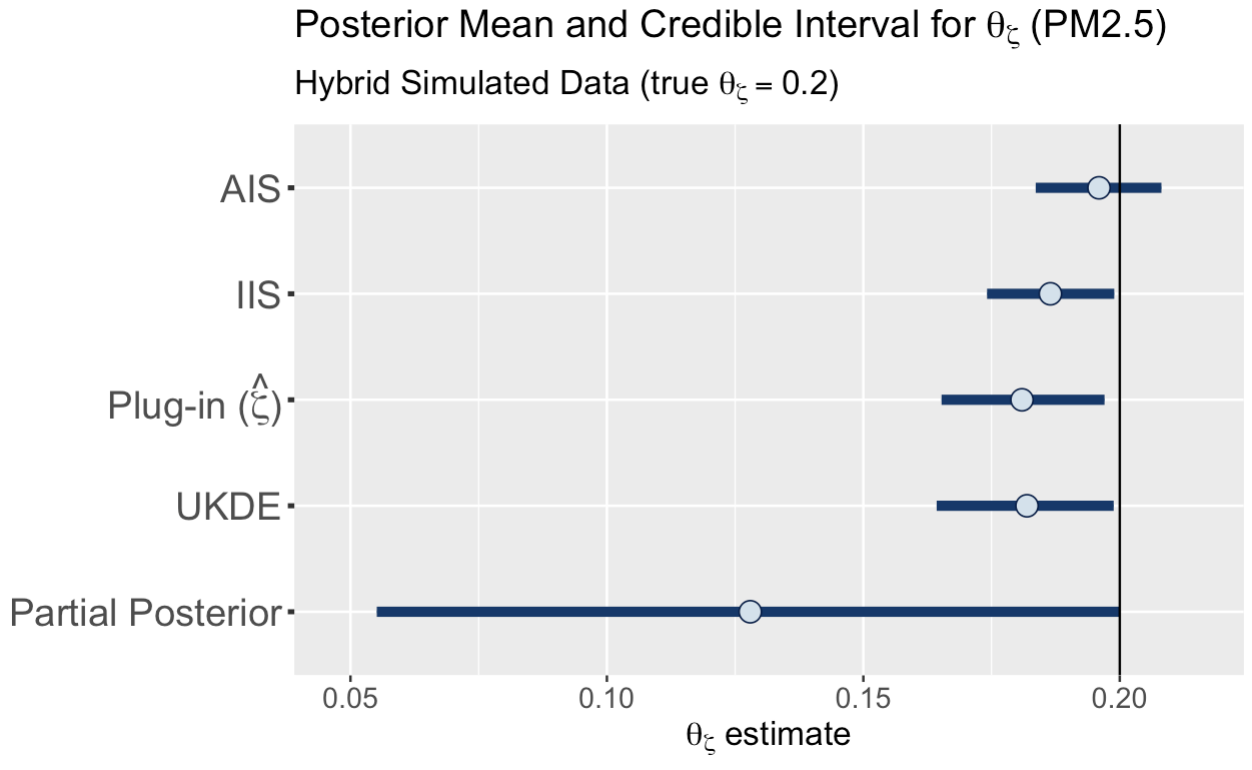}
\end{subfigure}
\caption{Posterior means and 95$\%$ credible intervals for $\theta_\zeta$ when $\theta_\zeta^* = 0.05$ (left) and $\theta_\zeta^* = 0.2$ (right), with $\theta_\zeta^*$ marked with a solid black line. As $\vert\theta_\zeta^*\vert$ increases, the attenuation bias for methods other than AIS gets larger and the partial posterior intervals become comparatively wider.}
\label{fig:hybrid_theta}
\end{figure}

\begin{figure}
\begin{subfigure}{0.48\textwidth}
\includegraphics[width=\linewidth]{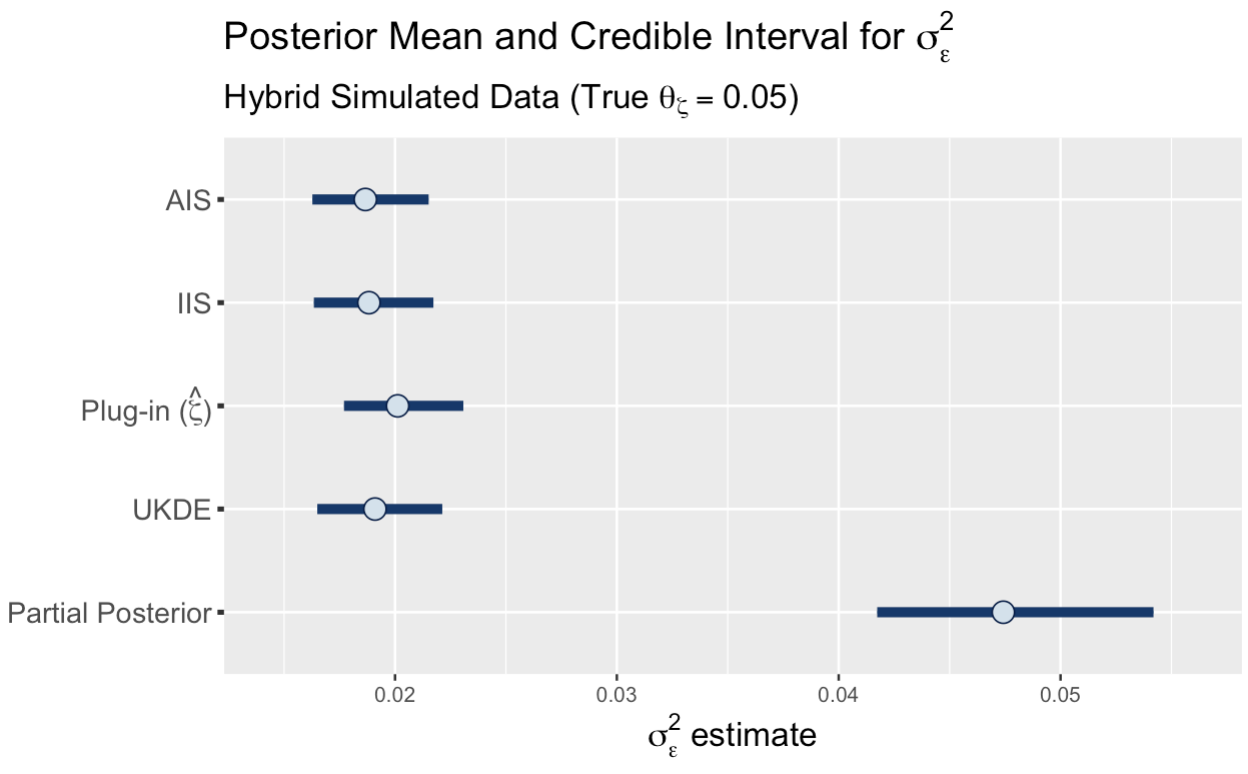}
\end{subfigure}
\begin{subfigure}{0.5\textwidth}
\includegraphics[width=\linewidth]{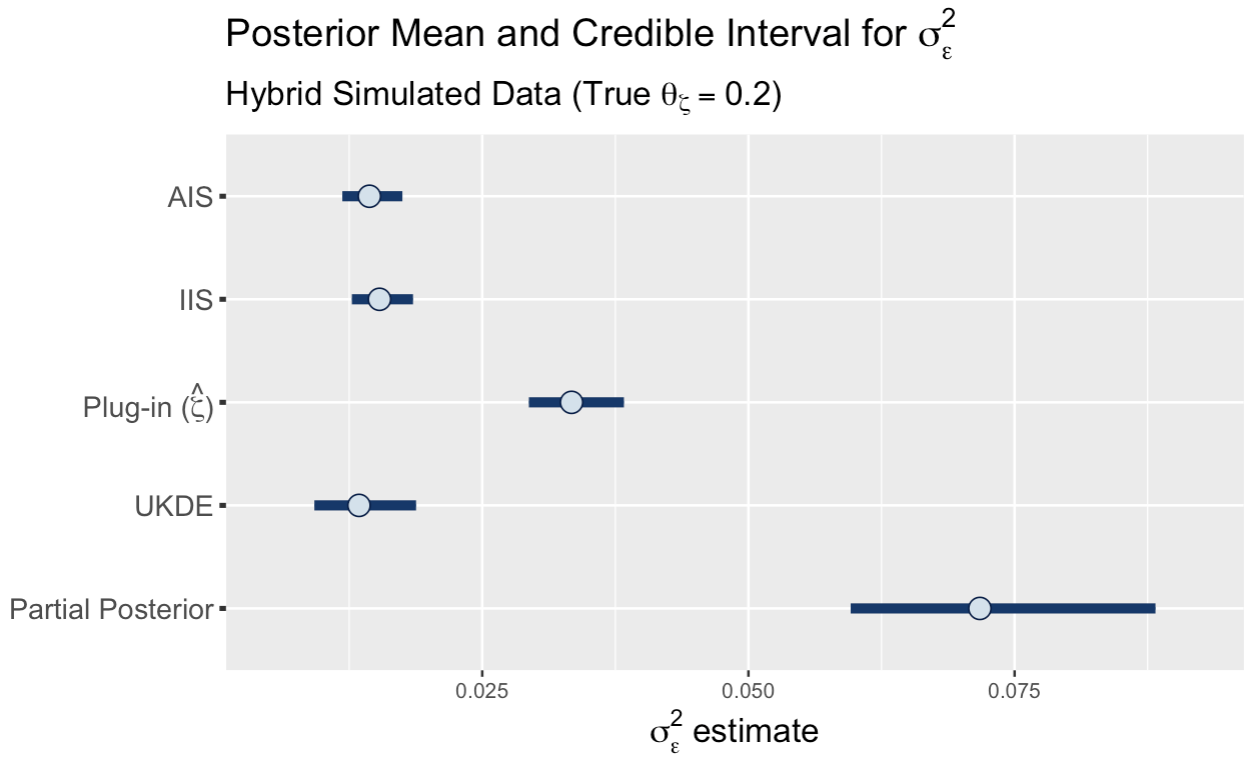}
\end{subfigure}
\caption{Posterior means for $\sigma^2_\epsilon$ when $\theta_\zeta^* = 0.05$ (left) and $\theta_\zeta^* = 0.2$ (right). We clearly see variance inflation for Partial Posterior and Plug-in ($\hat\zeta$) on the right.}
\label{fig:hybrid_sigma}
\end{figure}

In Figures ~\ref{fig:hybrid_theta} and ~\ref{fig:hybrid_sigma}, we plot the posterior means and credible intervals from the hybrid synthetic data using the sample means of $\{\tilde y_i^t\}_{i=1}^n$ across $T$ simulations. This leads to plots that are closer in presentation to results for real data as shown in Figure 7 in Section 5.3. We clearly see that the credible intervals for the Partial Posterior approach get even wider relative to the other methods as the signal strength increases, while the estimates for $\sigma^2_\epsilon$ in the Plug-in($\hat\zeta$) approach become more clearly inflated as the signal strength increases. These results agree with our conclusions from 100 simulations from $\{\tilde y_i^t\}_{i=1}^n$ in Figures 8 and 9 in Section 5.4 and confirm that greater signal strength leads to increased attenuation bias and variance inflation for methods that do not correctly account for uncertainty in the stage-one model. 

